\documentclass[accepted=2024-11-18,a4paper]{quantumarticle}
\pdfoutput=1

\usepackage{revtex}
\usepackage{comandi_tesi}

\usepackage{hyperref}
\usepackage[capitalise]{cleveref}
\usepackage{pgfplots}
\usepgfplotslibrary{groupplots}
\usepgfplotslibrary{fillbetween}
\pgfplotsset{compat=newest}

\newcommand{\vtheta}[0]{\boldsymbol{\theta}}
\newcommand{\hvtheta}[0]{\boldsymbol{\widehat{\theta}}}
\newcommand{\vx}[0]{\boldsymbol{x}}
\newcommand{\vy}[0]{\boldsymbol{y}}
\newcommand{\vsigma}[0]{\boldsymbol{\sigma}}
\newcommand{\vtau}[0]{\boldsymbol{\tau}}
\newcommand{\vlambda}[0]{\boldsymbol{\lambda}}
\newcommand{\vdelta}[0]{\boldsymbol{\delta}}
\newcommand{\sg}[1]{\text{sg}\left[#1\right]}
\newcommand{\cov}[0]{\Sigma}
\newcommand{\vu}[0]{\boldsymbol{u}}
\newcommand{\wt}[0]{\widetilde}
\def\mathunderline#1#2{\color{#1}\underline{{\color{black}#2}}\color{black}}

\begin{document}
\title{Model-aware reinforcement learning for high-performance Bayesian experimental design in quantum metrology}

\author{Federico Belliardo}
\affiliation{NEST, Scuola Normale Superiore, I-56126 Pisa, Italy}
\author{Fabio Zoratti}
\affiliation{Scuola Normale Superiore, I-56126 Pisa, Italy}
\author{Florian Marquardt}
\affiliation{Max Planck Institute for the Science of Light and Physics Department, University of Erlangen-Nuremberg, 91058 Erlangen, Germany}
\author{Vittorio Giovannetti}
\affiliation{NEST, Scuola Normale Superiore and Istituto Nanoscienze-CNR, I-56126 Pisa, Italy}

\begin{abstract}
	Quantum sensors offer flexibility in control during estimation, allowing manipulation across various parameters by the experimenter. For each sensing platform, determining the optimal controls to enhance the sensor's precision remains a challenging task. While an analytical solution may be unattainable, machine learning presents a promising approach for many systems of interest, especially considering the capabilities of modern hardware. We introduce a versatile procedure capable of optimizing a wide range of problems in quantum metrology and estimation by combining model-aware reinforcement learning (RL) with Bayesian estimation via particle filtering. To achieve this, we addressed the challenge of integrating the many non-differentiable steps of the estimation process, such as measurements and particle filter resampling, into the training routine. Our RL-based approach is suitable for optimizing both non-adaptive and adaptive strategies using a neural network. We provide an implementation of this technique in the form of a Python library called qsensoropt, along with several pre-built applications for relevant physical platforms, including NV centers, photonic circuits, and optical cavities. Using our method, we have achieved results that surpass the current state-of-the-art in experimental design for numerous tasks. Beyond Bayesian estimation, by leveraging model-aware RL, it is also possible to find optimal controls for minimizing the Cramér-Rao bound, based on Fisher information.
\end{abstract}

\maketitle

\section{Introduction}
\label{sec:introduction}
In recent years, the synergy between machine learning and quantum information has attracted increasing attention. These two technological fields can complement each other in various ways. On the one hand, quantum technologies, particularly quantum computers, have the potential to address classical machine learning challenges, such as classification and sampling, using both classical and quantum data~\cite{flamini_photonic_2020, broughton_tensorflow_2021, bergholm_pennylane_2022}. On the other hand, traditional machine learning can enhance quantum information tasks, including state preparation~\cite{bukov_reinforcement_2018, zhang_when_2019, niu_universal_2019, porotti_deep_2022}, optimal quantum feedback~\cite{porotti_gradient-ascent_2023}, error correction~\cite{fosel_reinforcement_2018}, device calibration~\cite{cimini_calibration_2019, ban_neural-network-based_2021, nolan_machine_2021, nolan_frequentist_2021}, characterization~\cite{nguyen_deep_2021}, and quantum tomography~\cite{palmieri_experimental_2020, quek_adaptive_2021, hsieh_direct_2022}. This work falls within the latter category, utilizing model-aware reinforcement learning~\cite{marquardt_machine_2021, marquardt_online_2021, krenn_artificial_2023, porotti_gradient-ascent_2023} (RL) to identify optimized adaptive and non-adaptive control strategies for application-relevant tasks in quantum metrology and estimation~\cite{belliardo_applications_2024}. The problem of optimal experimental design~\cite{fisher_design_1935} has already been addressed using machine learning techniques~\cite{foster_variational_2021,baydin_toward_2021,ballard_machine_2021,ivanova_implicit_2021,sarra_deep_2023,foster_deep_2021}. In this manuscript, we advance further by proposing a tool applicable to a broader range of problems, providing control solutions that are either superior to or easier to implement than their model-free counterparts. We present the theory, along with two relevant examples, and the details of the mathematical approach, while in~\cite{PhysRevA.109.062609}, the full range of optimization problems we have solved with this technique is discussed.

\subsection{Model-based and model-free reinforcement learning.}
The purpose of this section is to briefly review the key definitions in the field of reinforcement learning (RL) to better convey the novelty of our approach and clarify the associated terminology. Reinforcement learning is a mathematical framework for modeling decision-making in environments where outcomes are random but can be influenced by the actions of an agent, which is the mechanism making decisions. RL provides a systematic approach to optimizing the actions of the agent, taking into account both immediate and future consequences.

The objective of the actions and observations performed by the RL agent is to execute a task within the environment. After each execution, a loss function is computed, and the agent's strategy is updated based on this loss value. This process is referred to as training. If the model of the environment is known, it can be incorporated into the gradient of the loss function to update the agent's parameters. This forms the basis of model-based RL. Conversely, if the dynamic of the environment is completely unknown, the agent must implicitly learn how the environment responds to its actions in order to minimize the loss. While several studies have applied RL to quantum metrology~\cite{cimini_deep_2023, fallani_learning_2022, xu_generalizable_2021, xiao_parameter_2022, fiderer_neural-network_2021}, the innovation introduced here is the use of a model-aware approach, which, as we demonstrate through comparison with the results of~\cite{fiderer_neural-network_2021}, offers a more effective method for training optimized strategies. For further clarification on terminology and the distinction between RL and other decision-making frameworks, we refer to~\cite{moerland_unifying_2022}. While the term ``model-based reinforcement learning'' typically refers to scenarios where the model of the system and the optimization of the strategy occur simultaneously, in this manuscript, we rely on a pre-characterized model of the system, hence we adopt the term ``model-aware RL''. Our use of model-aware RL is akin to the concept of digital twins in industry, where a complete model of a product or system is simulated on a computer to optimize its performance. We provide evidence that incorporating the system model into the training process enhances efficiency, enabling better control solutions than those derived from model-free RL. With our tool, we achieve high-performance Bayesian experimental design, where high performance refers to the ability of our methods to surpass strategies obtained through other techniques.

\subsection{Review of the literature}
In the following paragraph, we review the field of optimization in quantum metrology, discussing the strengths and limitations of various approaches in comparison to the framework we propose. Similar challenges to those addressed by our method have been studied previously. These works on optimizing quantum metrology can be broadly categorized into four classes, which we will now present one by one.

The first class includes well-established competitor frameworks, such as the toolbox proposed by Meyer \textit{et al.}, which employs a variational approach for optimizing measurements and states~\cite{meyer_variational_2021}. In this work, the authors introduce a scheme to optimize probe state preparation and measurements to maximize the classical Fisher information obtained from the process. They demonstrate the success of this approach by applying it to multiphase estimation with GHZ states and to the problem of triangulating the position of a spin with three NV centers. In contrast to our approach, the variational toolbox introduced by Meyer \textit{et al.} does not allow for Bayesian estimation or consider adaptive strategies. A similar approach can be found in the library QuantEstimation~\cite{zhang_quanestimation_2022}, which implements different bounds for quantum metrological tasks (Fisher information, Holevo-Cramér-Rao bound, quantum Ziv-Zakai bound, and Bayesian estimation), along with various optimization algorithms that we have not considered (particle swarm optimization and differential evolution). However, both QuantEstimation and the variational toolbox studied by Meyer \textit{et al.} do not account for the use of neural networks for adaptive experiments, unlike our framework. We now turn to the two libraries, QInfer~\cite{granade_qinfer_2017} and Optbayesexpt~\cite{mcmichael_optbayesexpt_2021}, which are similar tools that optimize Bayesian experimental design for a range of experimental situations, but only consider greedy optimizations, i.e., one measurement at a time, via an approximation of the information gain per measurement. In contrast, the approach we present in this manuscript can plan measurements several steps ahead, potentially for the entire duration of the estimation. Another important tool recently introduced in the domain of optimal quantum metrology is a quantum comb-based approach for the simultaneous optimization of states, measurement, and estimator in one-shot Bayesian experiments, as proposed in~\cite{bavaresco_designing_2023}. Here, the authors consider a single encoded probe and a single measurement, from which they perform Bayesian estimation. Using the formalism of higher-order quantum operations, they develop a mechanism based on semidefinite programming to find the optimal probe, quantum measurement, and estimator function in both single- and multi-parameter scenarios. The main limitation of this approach is its extension to the multi-shot scenario, where the complexity of quantum comb optimization grows exponentially.

The second class of papers concerns those based on the optimization of Fisher information~\cite{liu_quantum_2017, xu_generalizable_2019, rembold_introduction_2020, schuff_improving_2020, xu_generalizable_2021, liu_optimal_2022, xiao_parameter_2022, qiu_efficient_2022}. Most of these approaches rely on some form of reinforcement learning or GRAPE, but the Fisher information analysis is limited to local estimation, and these works typically lack coverage of adaptive measurements or are only applicable to specific platforms (e.g., NV centers).

The third class includes theoretical works that advocate for the necessity of optimal control in quantum metrology and conceptually shape the working principles of our approach, though without presenting any concrete implementation~\cite{baydin_toward_2021, ballard_machine_2021, krenn_artificial_2023, vedaie_framework_2023, gebhart_learning_2023}. In~\cite{baydin_toward_2021}, the authors introduce MODE (Machine-learning Optimized Design of Experiments), a collaborative research program aimed at leveraging modern machine learning and statistical programming tools across different scientific domains.

The fourth class consists of applications of variational quantum circuits to specific platforms and tasks. These are generally non-adaptive (with two exceptions~\cite{ma_adaptive_2021,PhysRevApplied.22.044058}) and can be Bayesian~\cite{kaubruegger_quantum_2021, marciniak_optimal_2022, kaubruegger_optimal_2023,PRXQuantum.3.020350,Cao2024} or based on the quantum Fisher information~\cite{kose_superresolution_2023, heras_photonic_2023, yang_variational_2022}. The work~\cite{kose_superresolution_2023} is particularly noteworthy, as it studies the optimization of super-resolution imaging for observing Earth.

Among all the works discussed in this literature review, none present a framework capable of addressing both Bayesian and frequentist estimations, adaptive and non-adaptive metrology, across different platforms. This manuscript introduces a new framework designed to fill this gap through an innovative combination of techniques from statistics and machine learning. We have chosen an approach to Bayesian estimation based on particle filtering (or sequential Monte Carlo) due to the speed of the estimation performed in this manner and the ability to parallelize multiple simulations of the experiment. This approach involves many non-differentiable steps, such as simulating measurements and resampling from the posterior distribution, which we needed to account for when computing gradients.

\section{Quantum metrology with reinforcement learning}
In this section, we discuss how reinforcement learning (RL) is applied to quantum metrology. After providing a general overview of our scheme's working principle, we discuss how parameters of interest are encoded in quantum metrology, followed by a description of the Bayesian inference technique we employ. We then introduce two fundamental concepts in our approach: the measurement loop and the resources-loss paradigm. Finally, we define the loss function for various estimation tasks and comment on the technical aspects of the training process.

Given a specific physical platform and metrological task, the set of tunable parameters in the experiment is identified. These are the parameters that can be adjusted using various experimental controls, such as knobs and dials. The agent’s role is to decide the optimal settings for these controls before each measurement on the system. We train the agent to optimally control these parameters, minimizing the error metric through a gradient descent optimization procedure, using backpropagation to compute the derivatives throughout the entire history of the estimation process. The loss function minimized during the training is related to the final estimation error obtained after completing the experiment and performing Bayesian inference. In the examples presented in this paper, the agent is a small neural network, while the environment it interacts with encompasses the entire experimental setup, including the system that stores the estimation results and performs the Bayesian inference. In other words, the processed information extracted from the experiment is considered part of the environment.

This estimation procedure has been abstracted and decoupled from the specifics of any particular sensor or physical platform, allowing our tool to function as a versatile optimization method for quantum sensors. We demonstrate the broad applicability of our methodology by optimizing a wide range of estimation tasks on the nitrogen-vacancy (NV) center platform~\cite{chen_quantum_2018, rembold_introduction_2020}, for both single- and multi-parameter metrology, including DC magnetometry~\cite{fiderer_neural-network_2021}, AC magnetometry, decoherence estimation~\cite{arshad_real-time_2024}, and hyperfine coupling characterization~\cite{joas_online_2021}. In the domain of photonic circuits, we studied tasks such as multiphase discrimination, the agnostic Dolinar receiver~\cite{zoratti_agnostic-dolinar_2021}, and coherent state classification, both for cases where the states are classically known and where they must be learned from a quantum training set. For frequentist estimation, we explored the sensing of detuning frequency in a driven optical cavity~\cite{fallani_learning_2022}. In this paper, we present the applications to DC magnetometry and quantum communication, while other applications to NV centers (AC-field, decoherence, hyperfine coupling estimation) and photonic circuits (quantum machine learning with photonic circuits, multiphase estimation, coherent state classification) are discussed in~\cite{PhysRevA.109.062609}.

\subsection*{Encoding of the probe}
\label{subsec:probe_encoding_intro}
In quantum metrology, we deal with an environment or process characterized by a fixed number of parameters, denoted as $\vtheta \in \Theta$. These parameters are unknown, and our goal is to estimate them. To accomplish this, a \textit{quantum probe} with known dynamics interacts with the environment or undergoes the process of interest. By measuring the state of the probe, which now depends on $\vtheta$, we can extract information about these parameters, assuming the dynamic of the interaction is completely understood. In this context, quantum probes are systems that are well-characterized, easily manipulable, and often quite simple. Further details on the encoding of the probe can be found in the Supplementary Information~\cref{sec:phys_system}.

For the purpose of optimizing control strategies, the evolution of the probe and the measurement outcomes are simulated. The training phase is separated from the deployment, where the measurement process takes place on the actual sensor during in the experiment.

\subsection*{Bayesian estimation and particle filter}
\label{subsec:bayesian_update_intro}
Bayesian estimation is a step-by-step method for updating information about the unknown parameters of a system we are measuring, by refining a probability distribution after each measurement. The process begins with the definition of a \textit{prior distribution} $\pi (\vtheta)$ over the parameters $\vtheta$, which encapsulates our initial belief about the value of the unknown parameters before any measurements are taken. After the first measurement, this prior is updated to form the \textit{posterior distribution}, denoted as $P(\vtheta)$. To represent the posterior distribution, we use the particle filter method~\cite{del_moral_nonlinear_1997, arulampalam_tutorial_2002, liu_sequential_1998} (PF), which approximates it as an ensemble of points $\lbrace \vtheta_j \rbrace_{j=1}^N$ in the parameter space $\Theta$, with each point assigned a weight $\lbrace w_j \rbrace_{j=1}^N$, where $N$ is the number of particles. Essentially, we approximate the posterior distribution with a sum of $\delta$-functions, as follows:
\begin{equation}
	P(\vtheta) \simeq \sum_{j=1}^{N} w_j^t \delta (\vtheta - \vtheta_j) \; ,
\end{equation}
Initially, the particles are sampled from $\pi(\vtheta)$ and the weights are set to $w_j = \frac{1}{N}$. As new information becomes available, the weights are updated accordingly. From the particle filter, we derive an estimator $\hvtheta \in \Theta$ for $\vtheta$, which in our application is either the mean of the posterior or the most likely value for $\vtheta$. When the measurements on the quantum probe are weak (as opposed to projective), it is also necessary to account for the measurement backaction for each possible value of the unknown parameters $\vtheta$. For further details, refer to the Supplementary Information \cref{sec:particle_filter}.

\subsection*{Input and output of the controlling agent}
\label{subsec:controlling_agent_intro}
A ``summary'' of the information encapsulated in the Bayesian posterior represented by the particle filter (PF)—such as the mean and covariance matrix of the distribution $P(\vtheta)$—is provided as input to the neural network agent, which subsequently outputs the control settings. It is crucial for the agent to be specifically trained on the experiment it is designed to optimize. This necessitates that precise values for parameters like decoherence rates and visibilities are known and integrated into the simulation, unless these parameters are included in the set of $\vtheta$ to be estimated. In this manner, the knowledge gained about $\vtheta$ through measurements can be adaptively utilized by the agent to control both the evolution of the system and the measurements performed on the probe, with the objective of maximizing the final precision of the estimation. We envision conducting experiments using a small, trained agent deployed on fast hardware, such as a Field Programmable Gate Array (FPGA), situated in close proximity to the experimental setup.

\subsection*{The precision-resources paradigm}
\label{subsec:prec_res_paradigm}
In our framework, each measurement conducted on the probe consumes a certain amount $r$ of a specific ``resource'', which is considered costly within the context of the experiment and must be defined by the user based on the limitations of the setup. Once the total available resources $R$ are exhausted, the estimation process concludes, and the final value of the estimator $\hvtheta$ is computed. Examples of resources include the total estimation time, which is relevant for the NV center platform, the average number of photons consumed, or the amplitude of a signal, as seen in the Dolinar receiver. For optimizing the metrological task, defining the resource is as crucial as establishing the precision figure of merit. There is no universally correct or incorrect resource for an estimation task; it ultimately depends on the experimentalist's choices and their understanding of the laboratory limitations in implementing the task.

\onecolumngrid

\begin{figure}[th]
  \centering
  \includesvg[width=0.8\textwidth]{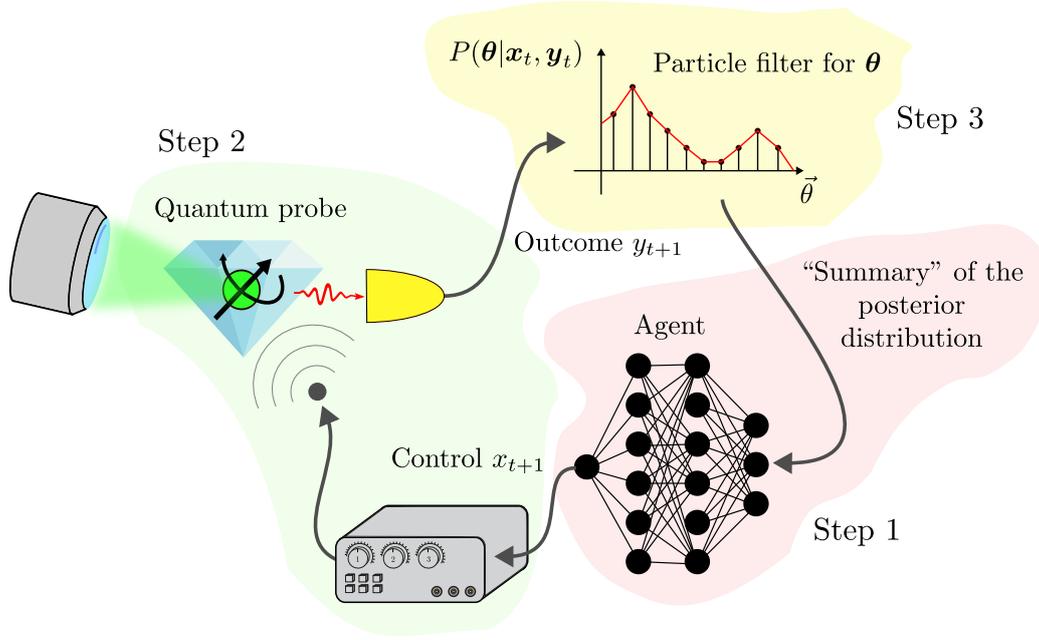}
  \caption{Schematic representation of the three steps of information flow within the measurement loop. The labels correspond to the $(t+1)$-th iteration. In the first step (pink region of the figure), the summary information computed from the particle filter is input into the agent (depicted here as a neural network), which determines the control parameters for both the evolution and measurement of the probe during this iteration, collectively represented by the variable $x_{t+1}$. In the second step (green region), the parameters $\vtheta$ are encoded in the probe state, and the measurement is performed, yielding the outcome $y_{t+1}$. In the third step (yellow region), this outcome is fed into the particle filter, leading to an update of the Bayesian posterior distribution on the parameters $\vtheta$ and the state of the probe (if applicable).
  }\label{fig:pipeline}
  \end{figure}
\twocolumngrid%

\subsection*{The measurement loop}
\label{subsec:meas_loop}
The metrological task is simulated as a sequence of consecutive operations, referred to as the \textit{measurement loop}, as illustrated in \cref{fig:pipeline}. Within this loop, for each iteration numbered from $t=0$ to $M-1$, a single measurement is conducted. We proceed by describing the generic iteration of the loop (specifically the $t+1$-th iteration), which consists of three steps. As indicated in the caption of \cref{fig:pipeline}, we denote the controls generated by the agent for the evolution of the probe and the settings for its measurements as $x_{t+1}$. The outcome of the measurement is denoted by $y_{t+1}$, both obtained at the $t+1$-th iteration of the loop. The objects $\vx_t := (x_0, x_1, \ldots, x_{t})$ and $\vy_t := (y_0, y_1, \ldots, y_{t})$ are tuples containing the controls and measurement outcomes up to time $t$. The distribution $P(\vtheta|\vx_t, \vy_t)$ represents the Bayesian posterior updated with the outcomes up to step $t$ of the measurement loop.
\begin{enumerate}
	\item In the case of the adaptive strategy, the selection of $x_{t+1}$ by the agent can be expressed, without loss of generality, through the mapping
	\begin{equation}
		x_{t+1} := \mathcal{F}_{\vlambda} \lbrace P(\vtheta| \vx_t, \vy_t); \vy_t; R_{t}; t \rbrace \; ,
		\label{eq:controls_computation}
	\end{equation}
	where $r_j$ denotes the resource consumption at the $j$-th step of the protocol, and the total resource consumed up to the $t$-th step is computed as $R_t := \sum_{j=0}^{t} r_j$. Non-adaptive strategies are described by mappings $\mathcal{F}$ that do not depend functionally on $P(\vtheta| \vx_t, \vy_t)$ or $\vy_t$, expressed as
	\begin{equation}
		x_{t+1} := \mathcal{F}_{\vlambda} \lbrace R_{t}; t \rbrace \; .
		\label{eq:controls_computation_static}
	\end{equation}
	The mapping $\mathcal{F}_{\vlambda}$ depends on the trainable parameters of the strategy, collectively denoted as $\vlambda$, which are later optimized. In the context of the non-adaptive strategies discussed here, the agent simply consists of a list of controls that are applied sequentially in the measurement loop, leading to $\vlambda = \vx_{M-1}$. For all the examples discussed in this manuscript, the neural network has five hidden layers with 64 neurons each, and the activation function used is $\tanh$, known for its effectiveness in approximating smooth functions~\cite{de_ryck_approximation_2021}.
	\item Assuming the measurements are projective and the probe's state is reinitialized after each iteration, the probability of observing the outcome $y_{t+1}$ at the $t+1$-th step is given by $p(y_{t+1}|x_{t+1}, \vtheta)$, which is computed using the Born rule according to the known quantum dynamics of the probe coded in the simulation. This probability, henceforth referred to as the ``model'' relies solely on the controls $x_{t+1}$ and the parameters to be estimated, $\vtheta$. At this second step of the measurement loop, the outcome $y_{t+1}$, a stochastic variable, is drawn from the model distribution, expressed as
	\begin{equation}
		y_{t+1} \sim p(y_{t+1}|x_{t+1}, \vtheta) \; .
		\label{eq:outcome_extraction}
	\end{equation}
	If the probe is subjected to a weak measurement, then the outcome probability depends on the entire sequence of previous controls and outcomes because of the measurement backreaction. In this case the sampled outcome is expressed as
	\begin{equation}
		y_{t+1} \sim p(y_{t+1}|\vx_{t+1}, \vy_{t}, \vtheta) \; .
		\label{eq:outcome_extraction_stateful}
	\end{equation}
	\item The observation of $y_{t+1}$ is subsequently incorporated into the posterior using Bayes' rule, formulated as
	\begin{equation}
		P(\vtheta|\vy_{t+1}, \vx_{t+1}) \propto p(y_{t+1}|x_{t+1}, \vtheta) P(\vtheta|\vx_t, \vy_t) \; .
		\label{eq:bayes_rule}
	\end{equation}
	During the first iteration, the prior $\pi(\vtheta)$ is utilized in place of the posterior. If the measurements are weak, the model probability takes the form described in \cref{eq:outcome_extraction_stateful}.
\end{enumerate}
The stopping condition of the measurement loop can be trivial, such as setting a maximum number of iterations $M$, or based on the available resources, for example, imposing a limit on $R_t$.

\subsection*{Training with model-aware reinforcement learning}\label{subsec:agent_training}
The figure of merit for precision depends on the specific metrological task. In the examples concerning the NV center platform, where the parameters $\vtheta$ are continuous, the mean square error (MSE) is employed. The loss for a single estimation is expressed as follows:
\begin{equation}
	\ell (\hvtheta, \vtheta) := \tr \left[ G \cdot (\hvtheta-\vtheta) (\hvtheta-\vtheta)^\intercal \right] \; ,
	\label{eq:loss_mse}
\end{equation}
where $G \ge 0$ represents a positive semidefinite weight matrix, and $\hvtheta$ denotes the mean of the posterior distribution. The weight matrix $G$ determines the contribution of various errors to the loss $\ell (\hvtheta, \vtheta)$ and delineates the parameters of interest from the nuisance parameters, with the latter being assigned corresponding entries of zero in the $G$ matrix.

In discrete estimation tasks, as illustrated subsequently for a photonic platform, both $\vtheta$ and $\hvtheta$ are discrete, such that $\vtheta, \hvtheta \in \Theta = \{ \vtheta_1, \vtheta_2, \ldots, \vtheta_k \}$. The loss for a single instance of this task can be represented using a Kronecker delta:
\begin{equation}
	\ell (\hvtheta, \vtheta) := 1 - \delta(\hvtheta, \vtheta) \; ,
	\label{eq:loss_delta}
\end{equation}
where
\begin{equation}
	\hvtheta := \arg \max_{\vtheta} P(\vtheta|\vx_{t}, \vy_{t}) \; ,
\end{equation}
denotes the maximum \textit{a posteriori} estimator. Optimizing the control strategy involves identifying the agent that minimizes the average loss $\mathbb{E} [ \ell (\hvtheta, \vtheta) ]$, averaged over all potential choices of $\vtheta$ and all stochastic processes involved in the estimation of $\vtheta$, as detailed in the Supplementary Material \cref{sec:comp_diff_loss}.

Each potential agent is characterized by a set of trainable variables, denoted as $\vlambda$, which influence the individual losses of the problem as well as the associated $\mathbb{E} [ \ell (\hvtheta, \vtheta) ]$. The optimal strategy can be abstractly defined by the value $\vlambda^\star := \mbox{argmin}_{\vlambda}\mathbb{E} [ \ell (\hvtheta, \vtheta) ]$ that minimizes the average loss. The training of the agent is implemented as an iterative algorithm aimed at discovering a strategy that closely approximates the performance of the optimal $\vlambda^\star$ through a sequence of recursive updates, denoted as $\text{TS}(1)$, $\text{TS}(2)$, $\cdots$, $\text{TS}(I)$, where ``TS'' stands for ``training step'':
\begin{equation}
	\vlambda_{0} \overset{\text{TS}(1)}{\longrightarrow}
	\vlambda_1 \overset{\text{TS}(2)}{\longrightarrow} \cdots \overset{\text{TS}(I)}{\longrightarrow}\vlambda_{I} \simeq \vlambda^{\star} \; ,
	\label{eq:training}
\end{equation}
with $\vlambda_{0}$ representing the initial weights of the neural network, initialized using the normal Glorot initializer, and the initial bias set to zero.

The construction of the learning trajectory in \cref{eq:training} relies on the computation of an estimation ${\cal L}(\vlambda)$ of the average loss $\mathbb{E} [ \ell (\hvtheta, \vtheta) ]$ associated with an agent $\vlambda$. This is typically accomplished by simulating in parallel $B$ estimations of randomly selected values $\vtheta_1, \cdots, \vtheta_B$ of the parameters $\vtheta$. Consequently, we can express:
\begin{equation}
	{\cal L}(\vlambda) := \frac{1}{B} \sum_{k=1}^B\ell (\hvtheta_k, \vtheta_k) \simeq \mathbb{E} [ \ell (\hvtheta, \vtheta) ] \; ,
	\label{eq:simple_loss}
\end{equation}
where $\ell (\hvtheta_k, \vtheta_k)$ denotes the local loss of the $k$-th estimation, which possesses a functional dependence on $\vlambda$ due to the multiple controlling actions of the agent.

Exploiting this dependence, we can compute the gradient $\mathcal{G}(\vlambda) := \frac{\dd}{\dd \vlambda} \mathcal{L} (\vlambda)$ of ${\cal L}(\vlambda)$ using automatic differentiation (AD), executed in reverse through all operations of the measurement loop. The agent parameters are updated at each training step using stochastic gradient descent, as follows:
\begin{equation}
	\vlambda_{i} \overset{\text{TS}(i+1)}{\longrightarrow} \vlambda_{i+1}  =  \vlambda_i  - \alpha {\mathcal{G}}(\vlambda_i) \; ,
	\label{eq:simple_sgd_update}
\end{equation}
with $\alpha \in (10^{-4}, 10^{-1})$ representing the learning rate. In the reported examples, the Adam optimizer~\cite{kingma_adam_2015} is utilized. This algorithm accumulates past gradients observed during training and uses them to adaptively modify the learning rate for each parameter. The purpose of this modified gradient descent approach is to enhance the training process by smoothing updates, allowing the optimizer to gain momentum in directions with consistent gradients while mitigating oscillations in others. This aids the algorithm in converging more rapidly and prevents it from becoming trapped in areas with noisy or minimal gradients. The algorithm accepts an external learning rate, which serves as a baseline for the adaptive learning rates.

Since the derivatives are propagated through the model for the sensor as described in \cref{eq:outcome_extraction}, this training constitutes a form of model-aware policy gradient reinforcement learning. The gradient descent training of $\vlambda$ will converge to a minimum of the loss; however, there is no guarantee that this minimum will be $\vlambda^\star$. Given that the loss is defined in terms of the stochastic outcomes $\vy_t$, special precautions are required to compute an unbiased estimator for its gradient~\cite{porotti_gradient-ascent_2023}, which entails incorporating the log-likelihood terms $\log p(y_{t+1}|\vx_{t+1}, \vy_t, \vtheta)$ into the loss. For further details regarding the loss definition and its gradient, refer to the Supplementary Material \cref{sec:comp_diff_loss}.

When conducting an estimation with a fixed number of measurements $M\ped{max}$ or a fixed maximum amount of resources $R\ped{max}$, selecting a loss $\mathcal{L}(\vtheta)$ that is sensitive solely to the performance of the estimator $\vtheta$ at the conclusion of the estimation does not necessarily yield optimal strategies for $M < M\ped{max}$ and $R < R\ped{max}$. A straightforward solution would be to repeat the optimization for each smaller $ R\ped{max}$ that requires characterization. However, it is possible to find an approximate solution for all $R \le R\ped{max}$ through a training that optimizes the cumulative loss instead of \cref{eq:simple_loss}, expressed as:
\begin{equation}
	\mathcal{L}_\text{cum} (\vlambda) := \frac{1}{M\ped{max} B} \sum_{t=0}^{M\ped{max}-1} \sum_{k=1}^B \ell (\hvtheta_{k, t}, \vtheta_k) \; .
	\label{eq:simple_cumulative}
\end{equation}
This cumulative loss encourages the agent to learn a strategy that is optimal for all $R \le R\ped{max}$ and has been employed in the examples involving the NV center platform. An alternative version of the loss is the logarithmic loss, which employs the logarithm of the average loss on the batch rather than the average loss in \cref{eq:simple_cumulative}. Further details can be found in \cref{subsec:cum_losses}.

\subsection*{Differentiability of the particle filter}
The primary component of our approach is the integration of particle filter (PF) Bayes updates with model-based reinforcement learning. This presents a challenge, as PF updates encompass steps where the differentiability necessary for gradient computation is not immediately apparent. As the estimation progresses, the weights of the PF become concentrated on a limited number of particles. To optimize memory utilization, we implement a resampling procedure that, when invoked, extracts a new set of particles $\lbrace \vtheta_j' \rbrace_{j=1}^N$ in accordance with the posterior distribution $P(\vtheta)$, resetting the weights to $w_j' = \frac{1}{N}$. This resampling procedure comprises three steps, which can be toggled on and off at discretion. These steps include: resampling from the posterior $P(\vtheta)$, perturbing the newly extracted particles, and proposing new particles. We have optimally integrated these steps through a trial-and-error procedure, as detailed in Supplementary Information \cref{subsec:resampling_scheme}. All these steps entail the extraction of discrete stochastic variables, an operation that, in principle, lacks differentiability and would substantially hinder the subsequent gradient propagation necessary for reinforcement learning.

While the latter two steps can be made differentiable using the reparametrization trick (see Supplementary Information \cref{subsec:diff_pf}), addressing the challenge of resampling the discrete PF ensemble necessitates modifying the loss function by incorporating the log-likelihood of the stochastic outcomes, as we do for the measurements. However, for a large number of particles $N$, this modification would adversely affect the variance of the estimated gradient during simulations. Instead, we utilize importance sampling to extract the new particles from a distribution $Q(\vtheta)$ distinct from the posterior, and we set the new weights proportional to the factor $\frac{P(\vtheta)}{Q(\vtheta)}$, ensuring that the PF consistently represents the posterior~\cite{karkus_particle_2018}. In this manner, the gradient can propagate through a resampling event via the term $P(\vtheta)$ in the weights. Along with the importance sampling, we have implemented the correction introduced by \'Scibior and Wood~\cite{scibior_differentiable_2021} to achieve differentiable resampling, demonstrating its efficacy for the mean square error loss (see Supplementary Information \cref{subsec:scibior_correction}). This correction complements importance sampling and aims to add the fewest possible log-likelihood terms for particle extraction events to the loss, thereby maintaining the stability of the training. The Bayes rule, being the product of the model probability and the previous posterior, is trivially differentiable.

\section{Applications}
\label{sec:results}
In this section, we present two applications of model-aware reinforcement learning (RL) to static field magnetometry using nitrogen-vacancy (NV) centers and to quantum communication utilizing the Dolinar receiver. We provide a brief overview of the two distinct problems to demonstrate the utility of our approach.

\begin{itemize}

	\item \textbf{Magnetometry}. In this example, we address magnetic field estimation employing solid-state quantum sensors, specifically a single qubit in diamond known as an NV center. We estimate a single, continuous parameter, which is the precession frequency of the spin, that is proportional to the external magnetic field we want to measure.

	\item \textbf{Quantum communication}. In this example we evaluate the application of RL to the agnostic Dolinar receiver. This task entails the discrimination of two coherent states transmitted by a distant laboratory. There are two variables to be estimated: one is discrete and contains information regarding the message, while the other is continuous, representing the intensity of the signal to be detected. The latter parameter is a nuisance.

\end{itemize}

For both problems, we successfully identified adaptive and non-adaptive strategies that surpassed the results previously reported in the literature.

\subsection*{Magnetometry with NV centers}
\label{subsec:nv_center}
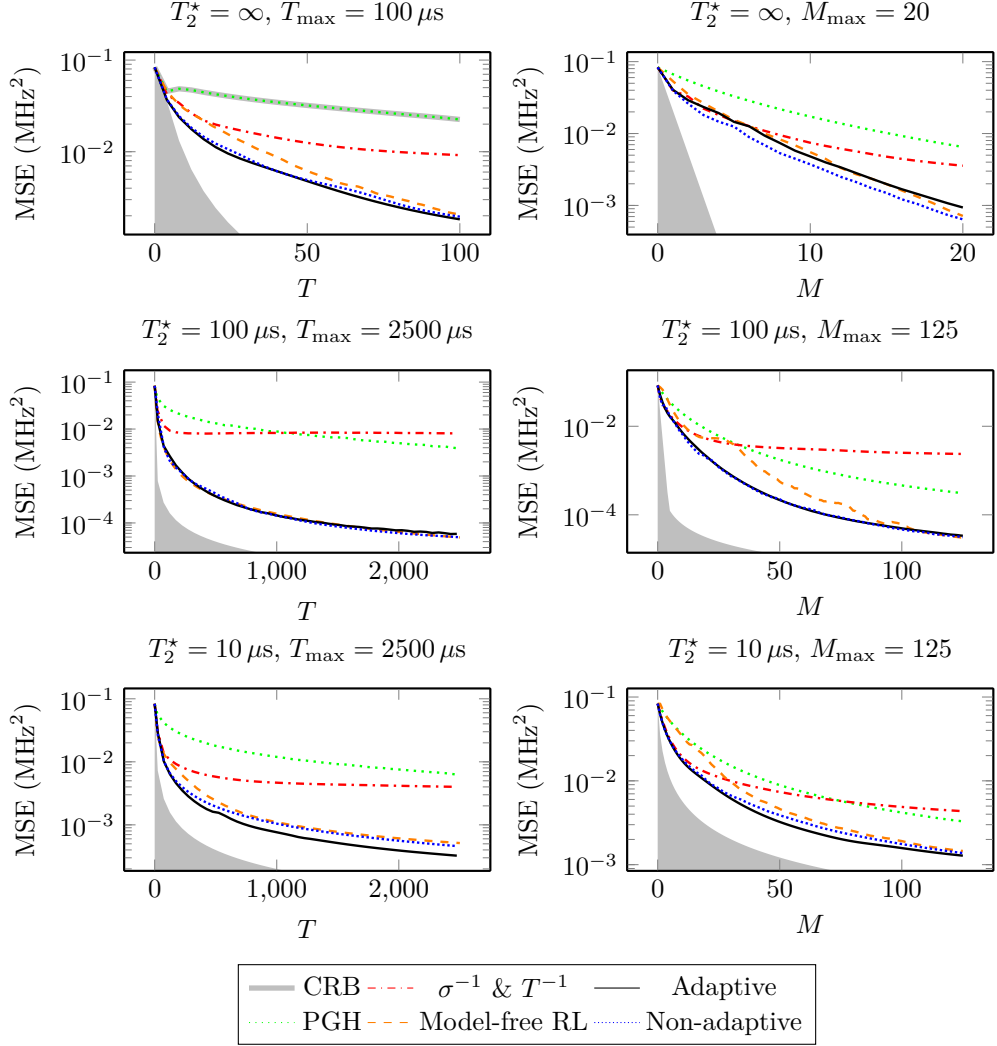
\begin{figure}[th]
  \centering
  \begin{tikzpicture}
    \begin{groupplot}[
      group style={
        group size=2 by 4,
        vertical sep=1.8cm,
        horizontal sep=1.8cm,
        },
      height=4cm, width=6.4cm,]
      \nextgroupplot[
      axis on top,
      title={$T\ped{2}^\star=\infty, \, T\ped{max}=100 \, \mu \text{s}$},
      ylabel={MSE ($\text{MHz}^2$)},
      xlabel={$T$},
      ymode=log,
      line width=.9pt,
      transpose legend,
      legend columns=2,
      legend style={at={(1.1,-4.0)},anchor=north},
      cycle multi list={%
        color list\nextlist
        [2 of]mark list
      }]

	  \addplot[color=gray!50, solid, line width=2pt]
	  table [x=Resources, y=MSE, col sep=comma, draw=none] {nv_center_time_lr_0.01_PGH_batchsize_2048_num_steps_128_max_resources_128.00_ll_False_cl_True_eval.csv};
	  \addlegendentry{CRB}

      \addplot[color=green, dotted]
      table [x=Resources, y=MSE, col sep=comma] {nv_center_time_lr_0.01_PGH_batchsize_2048_num_steps_128_max_resources_128.00_ll_False_cl_True_eval.csv};
      \addlegendentry{PGH}


      \addplot[color=red, dashdotted]
      table [x=Resources, y=MSE, col sep=comma] {nv_center_time_lr_0.01_Sigma_batchsize_2048_num_steps_128_max_resources_128.00_ll_False_cl_True_eval.csv};
      \addlegendentry{$\sigma^{-1}$ \& $T^{-1}$}

      \addplot[orange, dashed]
      table [x=Resources, y=MSE, col sep=comma] {nv_center_time_lr_0.01_batchsize_2048_num_steps_128_max_resources_128.00_ll_False_cl_True_fiderer.csv};
      \addlegendentry{Model-free RL}

      \addplot[color=black, solid]
      table [x=Resources, y=MSE, col sep=comma] {nv_center_time_lr_0.01_batchsize_2048_num_steps_128_max_resources_128.00_ll_False_cl_True_eval.csv};
      \addlegendentry{Adaptive}

      \addplot[color=blue, densely dotted]
      table [x=Resources, y=MSE, col sep=comma] {nv_center_time_lr_0.01_static_batchsize_2048_num_steps_128_max_resources_128.00_ll_False_cl_True_eval.csv};
      \addlegendentry{Non-adaptive}

      \path[name path=axis] (axis cs:0,0.000001) -- (axis cs:128,0.000001);

      \addplot[color=black, solid, update limits=false, name path=bound, draw=none]
      table [x=Resources, y=MSE, col sep=comma] {nv_center_time_lr_0.01_batchsize_2048_num_steps_128_max_resources_128.00_ll_False_cl_True_bound.csv};

      \addplot [gray!50] fill between[of=bound and axis]; 

      \pgfplotsset{
        after end axis/.code={
            \draw[thick] (rel axis cs:0,0) -- (rel axis cs:1,0); 
            \draw[thick] (rel axis cs:0,0) -- (rel axis cs:0,1); 
            \draw[thick] (rel axis cs:1,0) -- (rel axis cs:1,1); 
            \draw[thick] (rel axis cs:0,1) -- (rel axis cs:1,1); 
        }
      }

      \nextgroupplot[
      axis on top,
      title={$T\ped{2}^\star=\infty, \, M\ped{max}=20$},
      ylabel={MSE ($\text{MHz}^2$)},
      ymode=log,
      xlabel={$M$},
      line width=.9pt,
      transpose legend,
      legend columns=2,
      legend style={at={(0.5,-0.4)},anchor=north},
      cycle multi list={%
        color list\nextlist
        [2 of]mark list
      }]

      \addplot[color=green, dotted]
      table [x=Resources, y=MSE, col sep=comma] {nv_center_meas_lr_0.0001_PGH_batchsize_2048_num_steps_24_max_resources_24.00_ll_True_cl_True_eval.csv};
%

      \addplot[color=red, dashdotted]
      table [x=Resources, y=MSE, col sep=comma] {nv_center_meas_lr_0.0001_Sigma_batchsize_2048_num_steps_24_max_resources_24.00_ll_True_cl_True_eval.csv};

      \addplot[orange, dashed]
      table [x=Resources, y=MSE, col sep=comma] {nv_center_meas_lr_0.0001_batchsize_2048_num_steps_24_max_resources_24.00_ll_True_cl_True_fiderer.csv};

      \addplot[color=black, solid]
      table [x=Resources, y=MSE, col sep=comma] {nv_center_meas_lr_0.0001_batchsize_2048_num_steps_24_max_resources_24.00_ll_True_cl_True_eval.csv};

      \addplot[color=blue, densely dotted]
      table [x=Resources, y=MSE, col sep=comma] {nv_center_meas_lr_0.0001_static_batchsize_2048_num_steps_24_max_resources_24.00_ll_True_cl_True_eval.csv};

      \path[name path=axis] (axis cs:0,0.000001) -- (axis cs:24,0.000001);

      \addplot[color=black, solid, update limits=false, name path=bound, draw=none]
      table [x=Resources, y=MSE, col sep=comma] {nv_center_meas_lr_0.0001_batchsize_2048_num_steps_24_max_resources_24.00_ll_True_cl_True_bound.csv};

      \addplot [gray!50] fill between[of=bound and axis]; 

      \pgfplotsset{
        after end axis/.code={
            \draw[thick] (rel axis cs:0,0) -- (rel axis cs:1,0); 
            \draw[thick] (rel axis cs:0,0) -- (rel axis cs:0,1); 
            \draw[thick] (rel axis cs:1,0) -- (rel axis cs:1,1); 
            \draw[thick] (rel axis cs:0,1) -- (rel axis cs:1,1); 
        }
      }

      \nextgroupplot[
      axis on top,
      title={$T\ped{2}^\star=100 \, \mu \text{s}, \, T\ped{max}=2500 \, \mu \text{s}$},
      ylabel={MSE ($\text{MHz}^2$)},
      xlabel={$T$},
      ymode=log,
      line width=.9pt,
      transpose legend,
      legend columns=2,
      legend style={at={(0.5,-0.3)},anchor=north},
      cycle multi list={%
        color list\nextlist
        [2 of]mark list
      }]

      \addplot[color=green, dotted]
      table [x=Resources, y=MSE, col sep=comma] {nv_center_time_invT2_0.0100_lr_0.001_PGH_batchsize_64_num_steps_2560_max_resources_2560.00_ll_False_cl_True_eval.csv};


      \addplot[color=red, dashdotted]
      table [x=Resources, y=MSE, col sep=comma] {nv_center_time_invT2_0.0100_lr_0.001_Sigma_batchsize_64_num_steps_2560_max_resources_2560.00_ll_False_cl_True_eval.csv};

      \addplot[orange, dashed]
      table [x=Resources, y=MSE, col sep=comma] {nv_center_time_invT2_0.0100_lr_0.001_batchsize_64_num_steps_2560_max_resources_2560.00_ll_False_cl_True_fiderer.csv};

       \addplot[color=black, solid]
      table [x=Resources, y=MSE, col sep=comma] {nv_center_time_invT2_0.0100_lr_0.001_batchsize_64_num_steps_2560_max_resources_2560.00_ll_False_cl_True_eval.csv};

      \addplot[color=blue, densely dotted]
      table [x=Resources, y=MSE, col sep=comma] {nv_center_time_invT2_0.0100_lr_0.001_static_batchsize_64_num_steps_2560_max_resources_2560.00_ll_False_cl_True_eval.csv};

      \path[name path=axis] (axis cs:0,0.000001) -- (axis cs:2560,0.000001);

      \addplot[color=black, solid, update limits=false, name path=bound, draw=none]
      table [x=Resources, y=MSE, col sep=comma] {nv_center_time_invT2_0.0100_lr_0.001_batchsize_64_num_steps_2560_max_resources_2560.00_ll_False_cl_True_bound.csv};

      \addplot [gray!50] fill between[of=bound and axis];

      \pgfplotsset{
        after end axis/.code={
            \draw[thick] (rel axis cs:0,0) -- (rel axis cs:1,0); 
            \draw[thick] (rel axis cs:0,0) -- (rel axis cs:0,1); 
            \draw[thick] (rel axis cs:1,0) -- (rel axis cs:1,1); 
            \draw[thick] (rel axis cs:0,1) -- (rel axis cs:1,1); 
        }
      }

      \nextgroupplot[
      axis on top,
      title={$T\ped{2}^\star=100 \, \mu \text{s}, \, M\ped{max}=125$},
      xlabel={$M$},
      ylabel={MSE ($\text{MHz}^2$)},
      ymode=log,
      line width=.9pt,
      transpose legend,
      legend columns=2,
      legend style={at={(0.5,-0.3)},anchor=north},
      cycle multi list={%
        color list\nextlist
        [2 of]mark list
      }]

      \addplot[color=green, dotted]
      table [x=Resources, y=MSE, col sep=comma] {nv_center_meas_invT2_0.0100_lr_0.001_PGH_batchsize_1024_num_steps_128_max_resources_128.00_ll_True_cl_True_eval.csv};


      \addplot[color=red, dashdotted]
      table [x=Resources, y=MSE, col sep=comma] {nv_center_meas_invT2_0.0100_lr_0.001_Sigma_batchsize_1024_num_steps_128_max_resources_128.00_ll_True_cl_True_eval.csv};

      \addplot[orange, dashed]
      table [x=Resources, y=MSE, col sep=comma] {nv_center_meas_invT2_0.0100_lr_0.001_batchsize_1024_num_steps_128_max_resources_128.00_ll_True_cl_True_fiderer.csv};

      \addplot[color=black, solid]
      table [x=Resources, y=MSE, col sep=comma] {nv_center_meas_invT2_0.0100_lr_0.001_batchsize_1024_num_steps_128_max_resources_128.00_ll_True_cl_True_eval.csv};

      \addplot[color=blue, densely dotted]
      table [x=Resources, y=MSE, col sep=comma] {nv_center_meas_invT2_0.0100_lr_0.001_static_batchsize_1024_num_steps_128_max_resources_128.00_ll_True_cl_True_eval.csv};

      \path[name path=axis] (axis cs:0,0.000001) -- (axis cs:128,0.000001);

      \addplot[color=black, solid, update limits=false, name path=bound, draw=none]
      table [x=Resources, y=MSE, col sep=comma] {nv_center_meas_invT2_0.0100_lr_0.001_batchsize_1024_num_steps_128_max_resources_128.00_ll_True_cl_True_bound.csv};

      \addplot [gray!50] fill between[of=bound and axis];

      \pgfplotsset{
        after end axis/.code={
            \draw[thick] (rel axis cs:0,0) -- (rel axis cs:1,0); 
            \draw[thick] (rel axis cs:0,0) -- (rel axis cs:0,1); 
            \draw[thick] (rel axis cs:1,0) -- (rel axis cs:1,1); 
            \draw[thick] (rel axis cs:0,1) -- (rel axis cs:1,1); 
        }
      }

      \nextgroupplot[
      axis on top,
      title={$T\ped{2}^\star=10 \, \mu \text{s}, \, T\ped{max}=2500 \, \mu \text{s}$},
      ylabel={MSE ($\text{MHz}^2$)},
      xlabel={$T$},
      ymode=log,
      line width=.9pt,
      transpose legend,
      legend columns=2,
      legend style={at={(0.5,-0.3)},anchor=north},
      cycle multi list={%
        color list\nextlist
        [2 of]mark list
      }]

      \addplot[color=green, dotted]
      table [x=Resources, y=MSE, col sep=comma] {nv_center_time_invT2_0.1000_lr_0.01_PGH_batchsize_128_num_steps_2560_max_resources_2560.00_ll_False_cl_True_eval.csv};


      \addplot[color=red, dashdotted]
      table [x=Resources, y=MSE, col sep=comma] {nv_center_time_invT2_0.1000_lr_0.01_Sigma_batchsize_128_num_steps_2560_max_resources_2560.00_ll_False_cl_True_eval.csv};

      \addplot[orange, dashed]
      table [x=Resources, y=MSE, col sep=comma] {nv_center_time_invT2_0.1000_lr_0.01_batchsize_128_num_steps_2560_max_resources_2560.00_ll_False_cl_True_fiderer.csv};

      \addplot[color=black, solid]
      table [x=Resources, y=MSE, col sep=comma] {nv_center_time_invT2_0.1000_lr_0.01_batchsize_128_num_steps_2560_max_resources_2560.00_ll_False_cl_True_eval.csv};

      \addplot[color=blue, densely dotted]
      table [x=Resources, y=MSE, col sep=comma] {nv_center_time_invT2_0.1000_lr_0.01_static_batchsize_128_num_steps_2560_max_resources_2560.00_ll_False_cl_True_eval.csv};

      \path[name path=axis] (axis cs:0,0.000001) -- (axis cs:2560,0.000001);

      \addplot[color=black, solid, update limits=false, name path=bound, draw=none]
      table [x=Resources, y=MSE, col sep=comma] {nv_center_time_invT2_0.1000_lr_0.01_batchsize_128_num_steps_2560_max_resources_2560.00_ll_False_cl_True_bound.csv};

      \addplot [gray!50] fill between[of=bound and axis];

      \pgfplotsset{
        after end axis/.code={
            \draw[thick] (rel axis cs:0,0) -- (rel axis cs:1,0); 
            \draw[thick] (rel axis cs:0,0) -- (rel axis cs:0,1); 
            \draw[thick] (rel axis cs:1,0) -- (rel axis cs:1,1); 
            \draw[thick] (rel axis cs:0,1) -- (rel axis cs:1,1); 
        }
      }

      \nextgroupplot[
      axis on top,
      title={$T\ped{2}^\star=10 \, \mu \text{s}, \, M\ped{max}=125$},
      xlabel={$M$},
      ylabel={MSE ($\text{MHz}^2$)},
      ymode=log,
      line width=.9pt,
      transpose legend,
      legend columns=2,
      legend style={at={(0.5,-0.3)},anchor=north},
      cycle multi list={%
        color list\nextlist
        [2 of]mark list
      }]

       \addplot[color=green, dotted]
      table [x=Resources, y=MSE, col sep=comma] {nv_center_meas_invT2_0.1000_lr_0.001_PGH_batchsize_1024_num_steps_128_max_resources_128.00_ll_True_cl_True_eval.csv};


      \addplot[color=red, dashdotted]
      table [x=Resources, y=MSE, col sep=comma] {nv_center_meas_invT2_0.1000_lr_0.001_Sigma_batchsize_1024_num_steps_128_max_resources_128.00_ll_True_cl_True_eval.csv};

      \addplot[orange, dashed]
      table [x=Resources, y=MSE, col sep=comma] {nv_center_meas_invT2_0.1000_lr_0.001_batchsize_1024_num_steps_128_max_resources_128.00_ll_True_cl_True_fiderer.csv};

      \addplot[color=black, solid]
      table [x=Resources, y=MSE, col sep=comma] {nv_center_meas_invT2_0.1000_lr_0.001_batchsize_1024_num_steps_128_max_resources_128.00_ll_True_cl_True_eval.csv};

      \addplot[color=blue, densely dotted]
      table [x=Resources, y=MSE, col sep=comma] {nv_center_meas_invT2_0.1000_lr_0.001_static_batchsize_1024_num_steps_128_max_resources_128.00_ll_True_cl_True_eval.csv};

      \path[name path=axis] (axis cs:0,0.000001) -- (axis cs:128,0.000001);

      \addplot[color=black, solid, update limits=false, name path=bound, draw=none]
      table [x=Resources, y=MSE, col sep=comma] {nv_center_meas_invT2_0.1000_lr_0.001_batchsize_1024_num_steps_128_max_resources_128.00_ll_True_cl_True_bound.csv};

      \addplot [gray!50] fill between[of=bound and axis];

      \pgfplotsset{
        after end axis/.code={
            \draw[thick] (rel axis cs:0,0) -- (rel axis cs:1,0); 
            \draw[thick] (rel axis cs:0,0) -- (rel axis cs:0,1); 
            \draw[thick] (rel axis cs:1,0) -- (rel axis cs:1,1); 
            \draw[thick] (rel axis cs:0,1) -- (rel axis cs:1,1); 
        }
      }

  	\end{groupplot}

  \end{tikzpicture}
  \caption{These plots refer to the static field magnetometry with an NV center, conducted under different conditions. The mean squared error (MSE) on $\omega$ is plotted as a function of the total number of consumed resources, which are either the maximum number of measurements $M_{\text{max}}$ or the maximum total free evolution time of the probe, i.e., $T_{\text{max}} \ge \sum_{t=0}^{M-1} \tau_k$. The adaptive and non-adaptive strategies are optimized using model-aware reinforcement learning. The formula used to compute the evolution time when employing the $\sigma^{-1} \& T^{-1}$ and Particle Guess Heuristic (PGH) strategies can be found in the main text. With the label ``Model-free RL'', we denote the performances obtained in~\cite{fiderer_neural-network_2021} using model-free RL, which are never better than the non-adaptive strategy optimized with our techniques. The shaded grey area represents the (non-tight) ultimate precision bound, computed either from the Cramér-Rao bound (CRB) or from bit-counting arguments, as detailed in the Supplementary Information \cref{subsec:DC_magnetometry_bound}. The title of each plot includes the transverse relaxation time $T_2^\star$, which is the time scale of the dephasing noise, along with the maximum amount of resources utilized in the simulations. The number of particles in the particle filter was $N=480$ for the first and third rows, and $N=1024$ (left) and $N=1536$ (right) for the second row, which are significantly smaller numbers than those used in~\cite{fiderer_neural-network_2021}.}
  \label{fig:nvcenter_comparison}
\end{figure}

The nitrogen-vacancy (NV) centre in diamond is a point defect that enables initialization, detection, and control of its electronic spin, featuring very long quantum coherence time, even at room temperature. As such, it has been used in applications such as magnetometry, thermometry, and stress sensing~\cite{gali_ab_2019, doherty_quantum_2022,chen_quantum_2018, maze_rios_quantum_2010, barry_sensitivity_2020}. The electronic spin is sensitive to magnetic fields; for example, static fields determine the electron Larmor frequency, which can be measured as an accumulated phase by a Ramsey experiment. These experiments are realized by applying two $\pi/2$ pulses to the spin, followed by illumination with green light and detection of the photoluminescence. A single measurement has a binary outcome, yielding $\pm 1$ with probabilities
\begin{align}
	p(\pm1|\omega, T_2^\star, \tau) := \frac{1}{2} \pm \frac{1}{2} e^{-\tau/T_2^\star} \cos \left(\omega \tau \right) \; .
	\label{eq:nv_center_model}
\end{align}
The free evolution time $\tau$ is controlled by a trainable agent, while $\omega := \gamma B$ represents the unknown precession frequency to be estimated, which is proportional to the static magnetic field $B$ with $\gamma \simeq 28 \, \text{MHz}/\text{mT}$. The parameter $T_2^\star$ denotes the transverse relaxation time, serving as the time scale for the dephasing induced by magnetic noise. The optimization of the NV center as a magnetometer has been extensively studied in the literature with analytical tools~\cite{schmitt_optimal_2021,ferrie_how_2013}, with numerical methods~\cite{dushenko_sequential_2020,mcmichael_sequential_2021,granade_robust_2012,oshnik_robust_2022,craigie_resource-efficient_2021,bonato_optimized_2016,santagati_magnetic-field_2019,zohar_real-time_2023,nusran_high-dynamic-range_2012,wang_experimental_2017,dinani_bayesian_2019,bonato_adaptive_2017,ferrie_adaptive_2012}, and with machine learning~\cite{liu_repetitive_2020,fiderer_neural-network_2021,tsukamoto_machine-learning-enhanced_2022}. We conducted multiple estimations over the same parameter ranges chosen in the work of Fiderer \textit{et al.}~\cite{fiderer_neural-network_2021}, in order to facilitate an easy comparison of the results. The prior for the frequency $\omega$ is uniform in $(0, 1) \, \text{MHz}$. \cref{fig:nvcenter_comparison} compares the performances of the optimized adaptive (NN) and non-adaptive strategies against the Particle Guess Heuristic (PGH)~\cite{wiebe_hamiltonian_2014}, which is a commonly referenced strategy in the literature. According to this strategy, the evolution time is then computed as $\tau = \left( ||\boldsymbol{\theta}_1-\boldsymbol{\theta}_2||_{2} + \varepsilon \right)^{-1}$ with $\varepsilon := 10^{-5} \, \mu \text{s}^{-1}$. The concept behind it is to gauge the width of the probability distribution by extracting two particles from it at random, i.e. $\boldsymbol{\theta}_1$ and $\boldsymbol{\theta}_2$. There is also another common approach to assess this dispersion of the posterior, which is the use of the $\sigma^{-1}$ strategy, i.e., the evolution time $\tau$ is selected adaptively as the inverse of the standard deviation of the posterior distribution. Additionally, we introduced a variant of the $\sigma^{-1}$ strategy~\cite{ferrie_how_2013}, named $\sigma^{-1}\&T^{-1}$, which accounts for the finite coherence time. According to the $\sigma^{-1} \& T^{-1}$ strategy, the next evolution time $\tau$ is computed from the covariance matrix $\Sigma$ of the current posterior distribution as $\tau = \left[ \text{tr} (\Sigma)^{\frac{1}{2}} + 1/T_2^{\star} \right]^{-1}$.

\onecolumngrid%

\twocolumngrid%
For each plot, the better performance between our optimized adaptive and non-adaptive strategies outperforms all the other approaches, as illustrated in \cref{fig:nvcenter_comparison}. There are two comparisons to be made: on one hand, we have the optimized adaptive versus the non-adaptive strategies, which are both original results of this work; on the other hand, we have model-free versus model-aware reinforcement learning (RL), where the application of the latter to NV center magnetometry has been studied in~\cite{fiderer_neural-network_2021}. We shall begin with the first comparison. Notably, the optimal results for extended coherence times ($T_2^\star = 100 \, \mu \text{s}, \infty$) are achieved using non-adaptive strategies, which offer several practical advantages in experimental implementation. Primarily, since the controls are fixed \emph{offline} before the experiment, there is no requirement for real-time feedback via rapid electronics. Furthermore, there is also no need to update the Bayesian posterior on the fly due to the absence of adaptivity. Instead, the measurement outcomes can be processed offline, post-measurement, using more powerful hardware. This would significantly reduce online memory usage as there is no need for real-time updates to the particle filter. In the third row of \cref{fig:nvcenter_comparison}, we observe a gap between the performances of ``Adaptive'' and ``Non-adaptive'', and in~\cite{belliardo_applications_2024} we provide additional examples of the usefulness of adaptivity for NV centers. Regarding the second comparison of model-aware and model-free RL, we observe that no strategy trained with model-free RL can surpass even the non-adaptive strategy, indicating that the results of~\cite{fiderer_neural-network_2021}, although similar to ours, cannot prove that the neural network has been trained to exploit adaptivity and that it has not simply learned an optimal non-adaptive sequence of measurement times $\tau$. Moreover, we notice that our ``Adaptive'' strategy and the ``Model-free'' approach yield results that are closer toward the end of the estimation, while they differ for intermediate times. This is attributed to our use of cumulative loss. In the simulation with $T_2^\star = \infty, \, M_{\text{max}} = 20$, the NN strategy performs worse than the non-adaptive one because it becomes stuck in a local minimum during training. In \cref{fig:example_controls_nv_dc}, we present five examples of optimal adaptive trajectories for the estimation of $\omega=0.2 \, \text{MHz}$ corresponding to $T\ped{2}^\star = 10$, along with the optimal non-adaptive strategy. We observed also that multiple runs of the agent training will yield consistent performance but not necessarily the same optimized agent. In conclusion, we aim to provide an explanation for why adaptive control appears to offer limited advantage compared to the optimized non-adaptive strategy. For adaptivity to be beneficial, the phase $\omega \tau$ must be known to some extent. As the error on $\omega$ decreases, the evolution time increases, resulting in the uncertainty on $\omega \tau$ not approaching zero even after many measurements, which leaves very little room for adaptivity to enhance estimation precision. We also wish to point out that the Cramér-Rao (CR) bound (the gray area in the plot of \cref{fig:nvcenter_comparison}) is not achievable without entanglement in the probes and measurements, and even then, the gap cannot be fully closed.

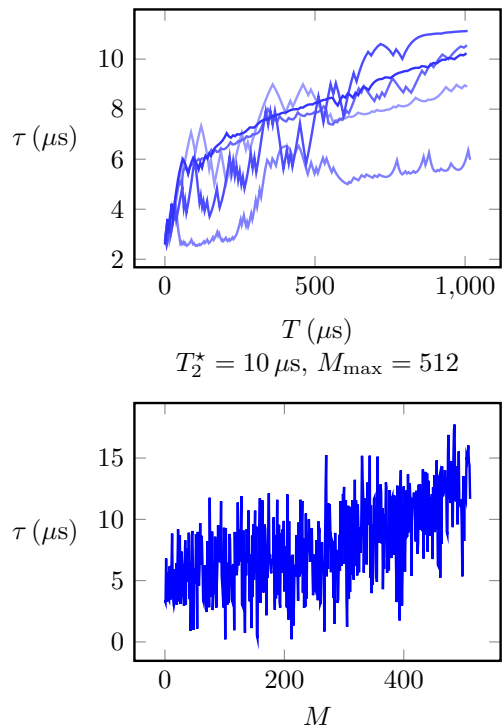
\begin{figure}[th]
  \centering
  \begin{tikzpicture}
    \begin{groupplot}[group style={group size=1 by 2, vertical sep=1.8cm,
      horizontal sep=1.2cm,}, height=5cm, width=6.4cm]
      \nextgroupplot[
      title={$T\ped{2}^\star = 10 \, \mu \text{s}, \, \omega=0.2  \, \text{MHz}, \, T\ped{max}=1024 \, \mu \text{s}$},
      xlabel={$T \, (\mu \text{s})$},
      y label style={rotate=-90},
      ylabel={$\tau \, (\mu \text{s})$},
      line width=.9pt,
      transpose legend,
      cycle multi list={%
        color list\nextlist
        [2 of]mark list
      }]

      \addplot[color=blue!40!white, solid]
      table [x=Resources, y=Tau, col sep=comma] {nv_center_time_invT2_min_0.0900_invT2_max_0.1100_lr_0.01_batchsize_128_num_steps_1024_max_resources_1024.00_ll_False_cl_True_traj_0.csv};

      \addplot[color=blue!50!white, solid]
      table [x=Resources, y=Tau, col sep=comma] {nv_center_time_invT2_min_0.0900_invT2_max_0.1100_lr_0.01_batchsize_128_num_steps_1024_max_resources_1024.00_ll_False_cl_True_traj_1.csv};

      \addplot[color=blue!60!white, solid]
      table [x=Resources, y=Tau, col sep=comma] {nv_center_time_invT2_min_0.0900_invT2_max_0.1100_lr_0.01_batchsize_128_num_steps_1024_max_resources_1024.00_ll_False_cl_True_traj_2.csv};

      \addplot[color=blue!70!white, solid]
      table [x=Resources, y=Tau, col sep=comma] {nv_center_time_invT2_min_0.0900_invT2_max_0.1100_lr_0.01_batchsize_128_num_steps_1024_max_resources_1024.00_ll_False_cl_True_traj_3.csv};

      \addplot[color=blue!80!white, solid]
      table [x=Resources, y=Tau, col sep=comma] {nv_center_time_invT2_min_0.0900_invT2_max_0.1100_lr_0.01_batchsize_128_num_steps_1024_max_resources_1024.00_ll_False_cl_True_traj_4.csv};

      \addplot[color=blue, solid, update limits=false, name path=static_ghost, draw=none]
      table [x=Resources, y=Tau, col sep=comma] {nv_center_meas_invT2_min_0.0900_invT2_max_0.1100_lr_0.001_static_batchsize_256_num_steps_512_max_resources_512.00_ll_True_cl_True_traj_0.csv};

      \nextgroupplot[
      title={$T\ped{2}^\star=10 \, \mu \text{s}, \, M\ped{max}=512$},
      ylabel={$\tau \, (\mu \text{s})$},
      xlabel={$M$},
      y label style={rotate=-90},
      line width=.9pt,
      transpose legend,
      legend columns=2,
      legend style={at={(0.5,-0.3)},anchor=north},
      cycle multi list={%
        color list\nextlist
        [2 of]mark list
      }]

      \addplot[color=blue, solid]
      table [x=Resources, y=Tau, col sep=comma] {nv_center_meas_invT2_min_0.0900_invT2_max_0.1100_lr_0.001_static_batchsize_256_num_steps_512_max_resources_512.00_ll_True_cl_True_traj_0.csv};

    \end{groupplot}
  \end{tikzpicture}
  \caption{Control strategies for the estimation of $\omega$ in direct current (DC) magnetometry. For the time-limited estimation (in the upper panel), five instances of the trajectory of the control $\tau$ produced by the neural network (NN) are plotted for $\omega=0.2 \, \text{MHz}$. The optimized strategies proposed by the NN start from the same point but gradually diverge over time as more data accumulates due to the stochastic nature of the measurement outcomes. The observed ''diffusion`` is attributed to the NN following the estimation adaptively. In the lower panel, the optimal non-adaptive strategy for the measurement-limited estimation is presented. This prescribes an increasing $\tau$ with a random pattern superimposed, which is interpreted as beneficial for compensating for the non-adaptivity of the strategy compared to the upper panel, with the multiple adaptive trajectories.}
  \label{fig:example_controls_nv_dc}
\end{figure}

\subsection*{Agnostic Dolinar receiver}
\label{subsec:dolinar}
\begin{figure}[th]
  \centering
  \includesvg[width=0.9\textwidth]{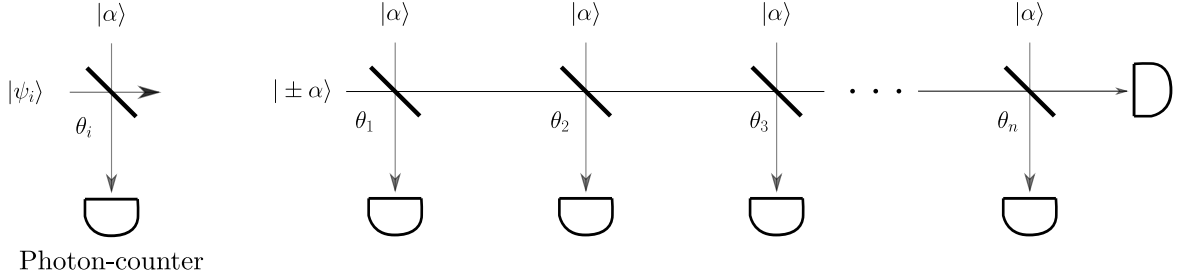}
  \caption{Schematic representation of the agnostic Dolinar receiver: each thick diagonal line symbolizes a beam splitter with programmable reflectivity $\theta_i$. Each ''D`` device denotes a photon counter. On the left side of the figure, the fundamental building block of this apparatus is illustrated. Here, the input state at step $i$, denoted as $\ket{\psi_i}$, is combined with one of the $n$ training states $\ket{\alpha}$. One of the two ports undergoes a measurement via photon counting. At the device's end, the second output port is also measured, ensuring that no information is left unused. It is important to note that the values of the control $\theta_i$ for the $i$-th measurement are determined based on the outcomes of all previous measurements.}
  \label{fig:dolan_physical}
\end{figure}

In this section, we address the challenge of distinguishing between two known coherent states, $\ket{-\alpha}$ and $\ket{\alpha}$, where $\alpha \in \mathbb{R}$ and $\alpha>0$, using a single copy of the signal $\ket{\pm \alpha}$. The Dolinar receiver optimally addresses this problem through linear optics and photon counting~\cite{dolinar_processing_1973,geremia_distinguishing_2004, izumi_displacement_2012,assalini_revisiting_2011, cook_optical_2007, pozza_adaptive_2014,takeoka_implementation_2005}. For this device, multiple machine learning approaches can be found in the literature~\cite{bilkis_real-time_2020,cui_quantum_2022}. In some recent studies~\cite{zoratti_agnostic-dolinar_2021, sentis_quantum_2015}, a variant of this device was introduced that does not require a local oscillator (LO) on the receiver side, which must be in phase with the sender's laser. This is the agnostic Dolinar receiver, in which, instead of the LO, $n$ copies of $\ket{\alpha}$, referred to as the reference states, are sent to the receiver from the sender, alongside the signal $\ket{\pm \alpha}$. We furthermore assume that classical knowledge about the state $\ket{\alpha}$ is absent, i.e., $\alpha$ is an unknown parameter of the estimation. In \cref{fig:dolan_physical}, we schematically represent this device, which leverages the states $\ket{\alpha}^{\otimes n}$ to perform the discrimination task on the sign of the signal $\ket{\pm \alpha}$. The signal $\ket{\pm \alpha}$ enters from the left and is sequentially combined with one of the reference states $\ket{\alpha}$ on a programmable beam splitter with adjustable reflectivity $\theta_i$. At each beam splitter, one of the two ports undergoes measurement by a photon counter, while the residual signal $\ket{\psi_i}$ from the other port is fed forward to the subsequent beam splitter. The photon counting result is used to update the Bayesian posterior on $\alpha$ and on the signal's sign, from which the reflectivity for the upcoming beam splitter is determined via a neural network. In this task, which combines estimation and discrimination, there are two undetermined parameters: one continuous, i.e., the signal's amplitude $\alpha \in \mathbb{R}$, and one discrete, i.e., the signal's sign. The receiver's performance is assessed based on the error probability in the task of signal classification, with the loss being defined in \cref{eq:loss_delta}, while the amplitude $\alpha$ is a nuisance parameter. See~\cite{belliardo_applications_2024} for details regarding the loss and the input to the NN.

\onecolumngrid%

\onecolumngrid%

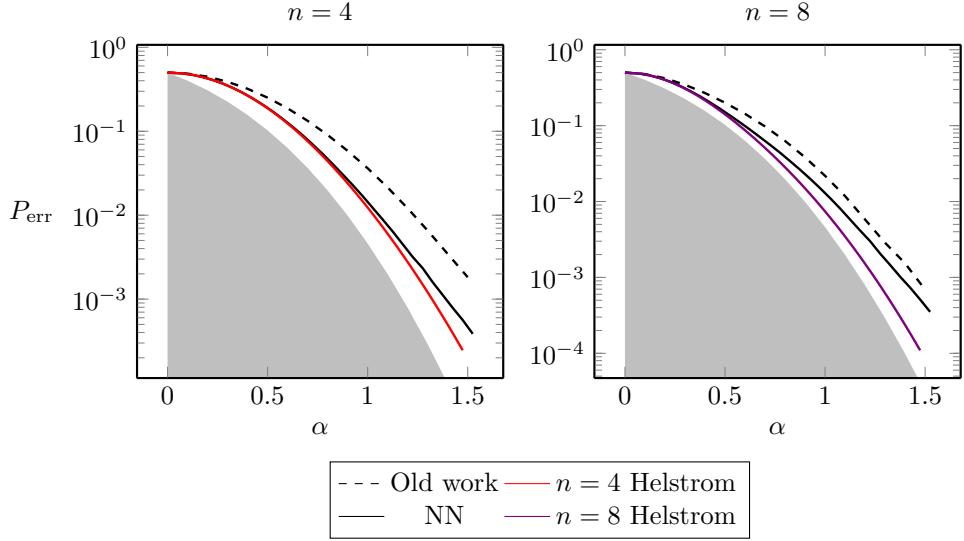
\begin{figure}[th]
  \centering
  \begin{tikzpicture}
    \begin{groupplot}[group style={group size=2 by 1, vertical sep=1.8cm,
      horizontal sep=1.2cm,}, height=6cm, width=6.4cm]
      \nextgroupplot[
      axis on top,
      title={$n=4$},
      xlabel={$\alpha$},
      y label style={rotate=-90},
      ylabel={$P\ped{err}$},
      ymode=log,
      line width=.9pt,
      transpose legend,
      legend columns=2,
      legend style={at={(1.1,-0.25)},anchor=north},
      cycle multi list={%
        color list\nextlist
        [2 of]mark list
      }]

      \addplot[color=black, dashed]
      table [x=alpha, y=ph_4, col sep=comma] {p_err_fabio.csv};
      \addlegendentry{Old work}

      \addplot[color=black, style=solid]
      table [x=Resources, y=ProbError, col sep=comma] {dolinar_batchsize_4096_num_steps_5_max_resources_1.55_ll_False_cl_False_loss_0_eval.csv};
      \addlegendentry{NN}

      \addplot[color=red, style=solid]
      table [x=Resources, y=ProbError, col sep=comma] {helstrom_4.csv};
      \addlegendentry{$n=4$ Helstrom}

       \addplot[color=violet, style=solid, update limits=false, name path=ghost_helstrom_8, draw=none]
      table [x=Resources, y=ProbError, col sep=comma] {helstrom_8.csv};
      \addlegendentry{$n=8$ Helstrom}

      \path[name path=axis] (axis cs:0.0,0.000001) -- (axis cs:1.55,0.000001);

      \addplot[color=black, solid, update limits=false, name path=bound, draw=none]
      table [x=alpha, y=helstrom, col sep=comma] {p_err_fabio.csv};

      \addplot [gray!50] fill between[of=bound and axis];

      \pgfplotsset{
        after end axis/.code={
            \draw[thick] (rel axis cs:0,0) -- (rel axis cs:1,0); 
            \draw[thick] (rel axis cs:0,0) -- (rel axis cs:0,1); 
            \draw[thick] (rel axis cs:1,0) -- (rel axis cs:1,1); 
            \draw[thick] (rel axis cs:0,1) -- (rel axis cs:1,1); 
        }
      }

      \nextgroupplot[
      axis on top,
      title={$n=8$},
      xlabel={$\alpha$},
      y label style={rotate=-90},
      ymode=log,
      line width=.9pt,
      transpose legend,
      legend columns=2,
      legend style={at={(0.5,-0.3)},anchor=north},
      cycle multi list={%
        color list\nextlist
        [2 of]mark list
      }]

      \addplot[color=black, dashed]
      table [x=alpha, y=ph_8, col sep=comma] {p_err_fabio.csv};

      \addplot[color=black, style=solid]
      table [x=Resources, y=ProbError, col sep=comma] {dolinar_batchsize_4096_num_steps_9_max_resources_1.55_ll_False_cl_False_loss_0_eval.csv};

      \addplot[color=violet, style=solid]
      table [x=Resources, y=ProbError, col sep=comma] {helstrom_8.csv};

      \path[name path=axis] (axis cs:0.0,0.000001) -- (axis cs:1.55,0.000001);

      \addplot[color=black, solid, update limits=false, name path=bound, draw=none]
      table [x=alpha, y=helstrom, col sep=comma] {p_err_fabio.csv};

      \addplot [gray!50] fill between[of=bound and axis];

      \pgfplotsset{
        after end axis/.code={
            \draw[thick] (rel axis cs:0,0) -- (rel axis cs:1,0); 
            \draw[thick] (rel axis cs:0,0) -- (rel axis cs:0,1); 
            \draw[thick] (rel axis cs:1,0) -- (rel axis cs:1,1); 
            \draw[thick] (rel axis cs:0,1) -- (rel axis cs:1,1); 
        }
      }

    \end{groupplot}
  \end{tikzpicture}
  \caption{Comparison of error probabilities for various strategies with different numbers of copies of $\ket{\alpha}$, specifically $n=4$ and $n=8$. The shaded gray area represents the region excluded by the Helstrom bound~\cite{helstrom_quantum_1969, holevo_statistical_1973}, which is the lowest error probability theoretically achievable when assuming access to an infinite number of reference states ($n=\infty$). The solid red and violet lines are the Helstrom bound calculated for a finite number of copies of $\ket{\alpha}$~\cite{zoratti_agnostic-dolinar_2021}, specifically $n=4$ and $n=8$. For details on the computation of the Helstrom bound, see~\cite{belliardo_applications_2024}. The black dashed line showcases the lowest error found in the earlier work~\cite{zoratti_agnostic-dolinar_2021}, without machine learning, while the black solid line represents the performance achieved using the neural network (NN). The performances of the optimal non-adaptive strategies have not been reported as they cannot rival those of the NN. For both the training and the performance evaluation, we used $N=512$ particles. The weights and biases of the NN have been initialized randomly.}
  \label{fig:dolinar_confronto}
\end{figure}

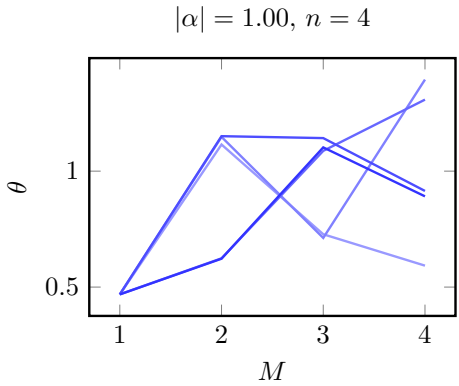
\begin{figure}[th]
	\centering
	\begin{tikzpicture}
		\begin{groupplot}[group style={group size=1 by 2, vertical sep=1.8cm,
				horizontal sep=1.2cm,}, height=5cm, width=6.4cm]
			\nextgroupplot[
			title={$|\alpha|=1.00, \, n=4$},
			ylabel={$\theta$},
			xlabel={$M$},
			line width=.9pt,
			transpose legend,
			cycle multi list={%
				color list\nextlist
				[2 of]mark list
			}]

			\addplot[color=blue!40!white, solid]
			table [x=MeasStep, y=Theta, col sep=comma] {traj_0.csv};

			\addplot[color=blue!50!white, solid]
			table [x=MeasStep, y=Theta, col sep=comma] {traj_1.csv};

			\addplot[color=blue!60!white, solid]
			table [x=MeasStep, y=Theta, col sep=comma] {traj_2.csv};

			\addplot[color=blue!70!white, solid]
			table [x=MeasStep, y=Theta, col sep=comma] {traj_3.csv};

			\addplot[color=blue!80!white, solid]
			table [x=MeasStep, y=Theta, col sep=comma] {traj_4.csv};

		\end{groupplot}
	\end{tikzpicture}
	\caption{Examples of trajectories for the control sequence in the Dolinar receiver are presented. The plots illustrate the selected values for $\theta$ as a function of the measurement step $M$. The total number of measurements is $n=4$, and the intensity of the signal is $|\alpha|=1$. The different shades of blue represent the five distinct trajectories that are being considered.}
	\label{fig:examples_dolinar}
\end{figure}

\twocolumngrid%
The simulation results are presented in \cref{fig:dolinar_confronto}, where we compared the performances of our adaptive procedure with the current state-of-the-art solution for this problem~\cite{zoratti_agnostic-dolinar_2021}. In each scenario, we achieved superior results with the neural network (NN). Notably, we nearly reached the theoretical bound in our primary area of interest, relevant for long-distance communications, which is the full quantum limit with $\alpha \lesssim 1$, and a small number of reference states (specifically, $n = 4$). For large $\alpha$, the error probability is already very small, placing us in the classical limit. See \cref{fig:examples_dolinar} for an example of series of trajectories for the control of the beam splitter phase, as it is determined by the network. For completeness, we mention that we have chosen a prior on $\alpha$ that is uniform in the interval $[0.05, 1.50]$, and the sign is also uniform in $\pm 1$. Additionally, we report the trajectory for five executions of the state discrimination.

\subsection*{Choice of the hyperparameters}
In this section, we provide a brief commentary on the selection of hyperparameters for the examples presented. These parameters include the batch size $B$, the number of particles in the particle filter $N$, and the initial learning rate $\alpha_0$. The selection of these parameters must align with the memory limitations of the computer during training. Specifically, we first empirically determine the number of particles to a value sufficient to ensure that the discretization of the posterior does not compromise the precision of the estimation. This choice subsequently influences the batch size, i.e., the number of simulations that can be executed in parallel. The batch size, in conjunction with the type of loss function, then determines the initial learning rate. For instance, in \cref{fig:nvcenter_comparison}, we utilized $B = 128$ and $\alpha_0 = 10^{-2}$ for the cumulative loss, and $B = 1024$ and $\alpha_0 = 10^{-3}$ for the logarithmic loss. The batch size can also be increased through gradient accumulation, which involves averaging the gradients from multiple executions of a batch of simulations for the update in \cref{eq:simple_sgd_update}. For the Dolinar receiver, we employed $\alpha = 10^{-2}$ and $B = 4096$. Refer to \cref{sec:miscellanea} for further information.

\section{Technical remarks}
\label{sec:technical_remarks}
We emphasize that our technique for optimizing sensors is fundamentally based on two steps. First, a model for the system must be constructed; this model can be analytical if the physics of the system is well understood and characterized, or numerical if necessary. Subsequently, based on the constructed model, the optimal strategy for estimation can be derived through reinforcement learning, as demonstrated in this work. An online approach that involves learning the model concurrently with the training of the strategy exceeds the scope of our current achievements. In both examples presented, we provided the neural network with a summary of the information extracted via Bayesian estimation. A straightforward extension of this approach would involve supplying the network with higher moments of the posterior distribution, allowing the network to learn more intricate details of the optimal estimation strategy. In theory, it is feasible to input the entire posterior distribution into the NN in the form of all particles and their corresponding weights. However, this is rarely practical if the parameters are continuous, although this method is employed in those examples discussed in~\cite{PhysRevA.109.062609}, which involve the estimation of discrete parameters only.

\section{Conclusions}
In this section, we summarize the results obtained from applying model-aware reinforcement learning to metrology and draw conclusions regarding the utility of this approach. Overall, our research underscores the advantages of integrating machine learning with modern quantum technologies. We have introduced a framework, complemented by a versatile library, capable of addressing a wide range of quantum parameter estimation and metrology challenges within both Bayesian and frequentist frameworks, applicable to various platforms. Our methods possess the potential to expedite the development of practical applications in quantum metrology. The ability to accurately estimated physical parameters through quantum systems could revolutionize multiple sectors, including biology, fundamental physics, and quantum communication. Through the application of model-aware reinforcement learning, we aim to facilitate progress in these domains, easing the transition of quantum-based metrology from proof-of-principle experiments to industrial applications.
This work aims to accelerate the search for optimal control strategies in quantum sensors, potentially speeding up their widespread industrial adoption. The technique of model-aware RL for agent optimization can, in principle, be applied to a broad spectrum of problems in quantum information, including quantum error correction and entanglement distillation, though this would necessitate engineering different loss functions. The primary challenge in extending this approach to other fields of quantum information beyond metrology is the rapidly increasing dimensionality of the quantum state spaces that would need to be simulated.

\section{Methods}
\label{sec:methods}
The library qsensoropt has been implemented in Python 3 on the Tensorflow framework. All of the simulations have been done on the High-Performance Computing cluster of Scuola Normale Superiore. The simulations ran on an NVIDIA Tesla GPU with 32GB of dedicated VRAM. The training and evaluation of each strategy took $\mathcal{O} (1)$ hours.

\section{Data availability}

All the data referenced in this work can be found in the GitLab repository, in the folders \texttt{qsensoropt/examples/nv\_center\_dc} and \texttt{qsensoropt/examples/dolinar}. The repository can be found at the URL \url{https://gitlab.com/federico.belliardo/qsensoropt}.

\section{Code availability}

The open source library qsensoropt can be found in the GitLab repository and can be installed with \texttt{pip} following the instruction in the README file. The repository can be found at the URL \url{https://gitlab.com/federico.belliardo/qsensoropt}. The package is also available on PyPI at \url{https://pypi.org/project/qsensoropt/}.

\section{Acknowledgments}
We gratefully acknowledge the computational resources of the Center for High Performance Computing (CHPC) at SNS. F. B. thanks T. Shah for useful discussions. We aknowledge finantial support by MUR (Ministero dell'Istruzione, dell'Università e della Ricerca) through the following projects: PNRR MUR project PE0000023-NQSTI, PRIN 2017 Taming complexity via Quantum Strategies: a Hybrid Integrated Photonic approach (QU-SHIP) Id. 2017SRN-BRK.

\section{Author Contributions}
F. Belliardo conceived of the presented idea under the supervision of F. Marquardt and V. Giovannetti. Federico B. and Fabio Z. have programmed the library and performed the simulations, all the authors have discussed the results and contributed to the final manuscript, with F. Belliardo being the main contributor.

\section{Competing interests statement}
The authors declare no competing interests.

\bibliographystyle{quantum}
\bibliography{bibliografia.bib}

\newpage

\onecolumngrid

\begin{appendices}
\appendix

\section{Schematization of physical systems in qsensoropt}
\label{sec:phys_system}
\subsection*{Encoding of the probe}
Following the standard nomenclature in quantum metrology, we define a quantum \emph{probe} as a quantum system initialized in a reference state $\rho$. This probe encodes the $d$-dimensional vector of parameters $\vtheta \in \Theta$ of interest, undergoing a controllable evolution determined by the controls $x$, i.e., $\rho \to \rho_{x, \vtheta} = \mathcal{E}_{x, \vtheta}(\rho)$, where $\mathcal{E}_{x, \vtheta}$ is a general linear completely positive trace-preserving (LCPT) map. A control is a tunable parameter that can be adjusted during the experiment, which may include factors such as measurement duration, laser frequency detuning in a cavity, or a tunable phase in an interferometer. In a figurative sense, control parameters encompass all the controls on the experimental electronics. A control can be continuous if it takes values within an interval, or discrete if it assumes only a finite set of values (e.g., on and off). The encoded parameters $\vtheta$ may represent properties of the environment, such as a magnetic field acting on a spin in the NV center platform, or certain degrees of freedom of the probe's initial state, such as the parameter $\alpha$ of a coherent state of light $\ket{\alpha}$ in the agnostic Dolinar receiver.

This scheme can also be framed as a \emph{communication protocol}, wherein Alice transmits the state $\rho_{\vtheta}$ to Bob, who is tasked with decoding the vector $\vtheta$. For the sake of generality, we will continue to refer to the quantum system as a probe in these contexts. The objective is to perform a measurement on $\rho_{x, \vtheta}$ to extract information regarding $\vtheta$. It is essential to distinguish that the term quantum parameter estimation in the literature pertains to situations where the encoded probe $\rho_{\vtheta}$ is provided, meaning we start from this state and cannot act on the encoding. In contrast, quantum metrology provides access to the encoding process $\mathcal{E}_{x, \vtheta}$, not merely to the final result. The implicit assumption in parameter estimation~\cite{hayashi_asymptotic_2005} is that the encoding occurs externally to the observed framework.

In both metrology and parameter estimation, we assume that the encoding $\mathcal{E}_{x, \vtheta}$ is applied multiple times or that we are given numerous copies of $\rho_{\vtheta}$, allowing us to gather statistically relevant data by measuring all copies from which we infer the value of $\vtheta$. Quantum metrology represents a more general setting than parameter estimation, and since the techniques developed here apply to quantum metrology, they are also applicable to parameter estimation. An example of parameter estimation is receiving radiation emitted by a distribution of currents on a plane, which depends on the properties of the source, such as temperature~\cite{kose_superresolution_2023}. In this scenario, the quantum probe is the radiation. As the radiation is assumed to be emitted from a distant and inaccessible region, we lack direct access to the quantum channel performing the encoding, receiving only the encoded states, which is the state of the radiated field upon detection.

An example of a quantum metrological task is the estimation of an environmental magnetic field using a spin, where we can select the initial state and the interaction duration. Parameters can be continuous or discrete. Continuous parameters include the magnetic field and temperature, while examples of discrete parameters encompass signal sign and the type of interaction between two quantum systems~\cite{gentile_learning_2021}. When discrete parameters are involved, we operate within the realm of value discrimination. In a metrological task, we may encounter a mixture of continuous and discrete parameters, as observed in the agnostic Dolinar receiver in \cref{subsec:dolinar}. A parameter can also be classified as a nuisance parameter, which is an unknown variable that requires estimation, though we do not evaluate the precision of the procedure with respect to it, as we are not directly interested in its value. An example of a nuisance parameter is the fluctuating optical visibility of an interferometer when our primary interest lies in the phase. Estimating nuisance parameters is often necessary and beneficial for estimating parameters of interest.

\subsection*{Measurement on the probe}
To obtain information on $\vtheta$, it is necessary to perform a measurement on the encoded probe $\rho_{x, \vtheta}$, represented by a positive operator-valued measure (POVM) $\mathcal{M} := \{ M_y^x \}$, where $x$ denotes the control parameters and $y$ signifies the measurement outcome. For simplicity, we will use the notation $x$ to refer to both the controls of the evolution and those of the measurement. The probability of obtaining outcome $y$ can be computed using the Born rule, yielding

\begin{equation}
	p(y|\vtheta, x) := \tr \left( M_{y}^x \rho_{\vtheta, x} \right) \; .
	\label{eq:meas_res_strong}
\end{equation}

If the measurement is projective, the system transitions to a known state, thereby extracting the maximum possible amount of information from $\rho_{x, \vtheta}$. The probe is subsequently reinitialized in the reference state $\rho$, encoded, and measured again, with the outcome probability given by the same expression in \cref{eq:meas_res_strong}. In the case of weak measurement (non-projective), information about $\vtheta$ remains encoded in the probe state, and reinitialization is not performed. The probe may or may not undergo evolution $\mathcal{E}_{x^\prime, \vtheta}$ again, potentially with different controls $x^\prime$. Following this, the probe is measured again using a different POVM $\mathcal{M}^\prime := \{ M^{x^\prime}_{y^\prime} \}$, leading to the outcome $y^\prime$. This procedure can be iterated multiple times until a projective measurement is performed on the probe, at which point its state is reinitialized. For weak measurements, the Born rule prescribes an outcome probability that depends on the entire trajectory of previous controls and measurement outcomes. We denote by $\vx_t := (x_0, x_1, \ldots, x_t)$ and $\vy_t := (y_0, y_1, \ldots, y_t)$ the tuples containing the controls and outcomes up to the $t$-th iteration, respectively. The probability of obtaining $y_{t+1}$ at the $(t+1)$-th step is given by

\begin{align}
	p(y_{t + 1} | \vx_{t+1}, \vy_t, \vtheta) := \tr \left( M_{y_{t + 1}}^{x_{t + 1}} \rho_{\vx_t, \vy_t, \vtheta} \right) \; .
\end{align}

The case of continuous measurement can be simulated by taking the appropriate limits, though it is beyond the scope of this work.

\section{Implementation of the particle filter}
\label{sec:particle_filter}
\subsection{Bayesian update}
\label{subsec:bayesian_update}

If we perform a projective measurement on the probe at each step, the probability of observing outcome $y$, given the control $x$ and the true value $\vtheta$ of the unknown parameters, is reported in \cref{eq:outcome_extraction}. To recover the value of $\vtheta$, we apply the principles of Bayesian estimation. Starting from the prior distribution $\pi(\vtheta)$ on $\vtheta$, we calculate the posterior probability distribution using Bayes' rule, expressed as

\begin{equation}
	P(\vtheta|x, y) = \frac{p(y|x, \vtheta) \pi(\vtheta)}{P(y)} = \frac{p(y|x, \vtheta) \pi(\vtheta)}{\int p(y|x, \vtheta) \pi(\vtheta) \dd \vtheta} \; .
\end{equation}

The denominator serves as the normalization required for $P(\vtheta|x, y)$ to qualify as a probability density. For a series of measurements, we apply Bayes' rule iteratively, using the posterior computed at the previous step as the prior for the next one. Given the tuples of controls $\vx_{t+1}$ and outcomes $\vy_{t+1}$, we compute the posterior at the $(t+1)$-th step from the posterior at the $t$-th step using the formula

\begin{equation}
	P(\vtheta|\vx_{t+1}, \vy_{t+1}) = \frac{p(y_{t+1}|x_{t+1}, \vtheta) P(\vtheta|\vx_{t}, \vy_{t})}{\int p(y_{t+1}|x_{t+1}, \vtheta) P(\vtheta|\vx_{t}, \vy_{t}) \dd \vtheta} \; .
	\label{eq:bayesian_update}
\end{equation}

It is important to note that for each measurement, the probability of obtaining $y_{t+1}$ is independent of the outcomes and controls up to that point and depends solely on $x_{t+1}$. This independence arises from the reinitialization of the probe following projective measurements. To efficiently perform the Bayesian update on a computer, we utilize the particle filter method (PF). This method represents the posterior distribution using a discrete set of points in the parameter space $\Theta$, each associated with a weight. Essentially, this approximates the posterior distribution as a sum of $\delta$-functions, expressed as

\begin{equation}
	P(\vtheta|\vx_t, \vy_t) \simeq \sum_{j=1}^{N} w_j^t \delta (\vtheta - \vtheta_j) \; ,
\end{equation}

where the values $\{ \vtheta_j \}_{j=1}^N$ are referred to as particles and $\{ w_j^t \}_{j=1}^N$ are the weights at step $t$. The values of the particles are sampled from the initial prior $\pi (\vtheta)$, while the weights are initialized uniformly across all particles, i.e., $\omega_{j}^0 := \frac{1}{N}$. The weights vary with each step due to the Bayesian update of the posterior in \cref{eq:bayesian_update}, which corresponds to the transformation

\begin{equation}
	w_j^{t+1} =  \frac{p(y_{t+1}|x_{t+1}, \vtheta_j) w_j^t}{\sum_{j=1}^N p(y_{t+1}| x_{t+1}, \vtheta_j) w_j^t} \; .
	\label{eq:bayes_rule_w}
\end{equation}

The particles $\{ \vtheta_j \}_{j=1}^N$ may also vary with the step $t$. We will introduce a resampling procedure that, when triggered, generates a new set of particles, meaning they do not necessarily remain constant throughout the estimation process. For notational simplicity, we will not include a time index on $\vtheta_j$. We denote the set of particles and their weights as $\mathfrak{p}_t := \{ \vtheta_j, w_j^t \}_{j=1}^N$, referred to as the PF ensemble.

\subsection{Moments of the posterior}
\label{subsec:moments_posterior}
Computing the first moments of the posterior, namely the mean value and the covariance matrix, corresponds to straightforward linear algebra operations on the PF ensemble, expressed as

\begin{equation}
	\hvtheta_t := \int \vtheta P(\vtheta| \vx_t, \vy_t) \dd \vtheta \simeq \sum_{j=1}^{N} w_j^t \vtheta_j \; ,
	\label{eq:mean_pf}
\end{equation}

and

\begin{equation}
	\cov_t := \int (\vtheta - \hvtheta_t) (\vtheta - \hvtheta_t)^\intercal P(\vtheta|\vx_t, \vy_t) \dd \vtheta \simeq \sum_{j=1}^N w_j^t (\vtheta_j - \hvtheta_t) (\vtheta_j - \hvtheta_t)^\intercal \; .
	\label{eq:Sigma_pf}
\end{equation}

The mean value of the posterior $\hvtheta$ serves as our estimator for all continuous parameters throughout this paper. As the estimation progresses, the weights typically concentrate on a limited number of particles, while the remaining particles become negligible for the estimation and may only consume memory. The precision of the estimation is limited by the average distance between the points $\vtheta_j$, which depends on the prior distribution $\pi(\vtheta)$ and the number of particles $N$. The next section will demonstrate how the introduction of a resampling scheme can address this issue by generating a new set of particles $\{ \vtheta_j' \}_{j=1}^N$, which should be situated in regions where the posterior distribution is concentrated. This leads to an increased density of particles in that region, enhancing the resolution in distinguishing close values of $\vtheta$. Throughout the paper, we utilize the same symbols for the ``theoretical'' moments of the posterior (which are not directly accessible) and the approximations of these quantities computed from the PF. The context will clarify which quantities are being referenced at any given time.

\subsection{Resampling scheme}
\label{subsec:resampling_scheme}
While for a small number of unknown parameters we could still obtain good performances even if no resampling procedure is performed, it is essential for larger dimensions. Indeed the density of particles, i.e. the resolution in $\vtheta$, after the initialization, is inversely proportional to the volume of the parameter space, which grows exponentially in the number $d$ of dimension of the $\Theta$ space. To solve this problem it is typical to perform during the estimation a resampling of the particles according to the posterior distribution, which is triggered by the condition
\begin{equation}
	N_{\text{eff}} := \frac{1}{\sum_{j=1}^N (w_j^i)^2} < r_t N \; ,
	\label{eq:resampling_condition}
\end{equation}
where $r_t$ is the resampling threshold that is kept fixed at $r_t=0.5$ in all the simulations of the paper. The left-hand side of \cref{eq:resampling_condition} is sometimes called the effective number of particles $N_{\text{eff}}$.

\subsubsection*{Soft resampling}
The simplest resampling scheme prescribes the extraction of $N$ samples with repetitions from the set of indexes $J=\lbrace 1, \cdots, N \rbrace$, each weighted with the corresponding $w_j$, $j \in J$. We will call $\phi(j): J \rightarrow J$ the map that gives the outcome of the $j$-th extraction event. The indexes $j \in J$ corresponding to the particles $\vtheta_j$ that have large weights are extracted more frequently, while the particles with small weights tend to disappear. In our implementation, we considered a slightly more general version of this procedure which goes under the name of soft resampling~\cite{karkus_particle_2018}, that is, we mix the probability distribution represented by the weights $\lbrace w_j \rbrace_{j=1}^N$ with a uniform distribution on $\lbrace \vtheta_j \rbrace_{j=1}^N$ by constructing the soft-weights $q_j$ defined as
\begin{equation}
	q_j := \alpha w_j + (1-\alpha)\frac{1}{N} \; ,
	\label{eq:mixed_q_j}
\end{equation}
where $\alpha \in [0, 1]$ is a parameter characterizing the effectiveness of the resampling. With $\alpha=1$ we have the traditional procedure, while with $\alpha=0$ no actual resampling is performed, because we extract the new particles from a uniform distribution, just like at the beginning. With $\alpha=0$ the particles with low weights are not cut away from the ensemble but persist after the process. With an intermediate value of $\alpha$ (by default we set $\alpha=0.5$) only a fraction $\alpha$ of the particles are effective for the resampling, because the other fraction $(1-\alpha)$ is expected to be distributed uniformly. We call $\vtheta_j^\prime$ the new particles extracted from $q_j$, i.e.
\begin{equation}
	\vtheta_j^\prime = \vtheta_{\phi(j)} \; .
	\label{eq:theta_j'}
\end{equation}
Their corresponding weights are chosen, so that the ensemble of the PF represents the same distribution as before the resampling. These are
\begin{equation}
	w'_j \propto \frac{w_{\phi(j)}}{q_{\phi(j)}} = \frac{w_{\phi(j)}}{\alpha w_{\phi(j)} + (1-\alpha)\frac{1}{N}} \; ,
	\label{eq:new_weights_to_norm}
\end{equation}
that still need to be normalized. With this choice for $w'_j$ the PF represents the correct posterior even though the particles have been sampled from a different distribution. The probability density function represented by the PF is, roughly speaking, proportional to the product of the weights $w_j'$ and the density of particles at the position $\vtheta_j^\prime$, i.e. $q_{\phi(j)}$, which with our choice for $w_j'$ is exactly $w_{\phi(j)}$, i.e. the weight of the particle $\vtheta_j^\prime$ prior to the resampling step. In the next section, we detail this relation. The reader that is interested in the successive steps of the resampling can however skip it. The soft resampling scheme, which is based on importance sampling~\cite{li_resampling_2015}, will be crucial in making the PF differentiable~\cite{zhu_towards_2020, ma_particle_2020}. We might want, in general, to perform a subsampling of the particles, that is, we sample from the distribution in \cref{eq:mixed_q_j} not $N$ but $\gamma N$ particles, with $0 < \gamma \le 1$. We will later in the resampling routine propose $(1-\gamma) N$ new particles that will help us in representing the posterior better, so that we have in total after the resampling step $N$ particles again. In this case the weights in \cref{eq:new_weights_to_norm} will be normalized as $w_j' \rightarrow \frac{w_j'}{\mathcal{C}}$, where $\mathcal{C}$ is such that
\begin{equation}
	\frac{1}{\mathcal{C}} \sum_{j=1}^{\gamma N} w_{j} = \gamma \; ,
\end{equation}
By default we set $\gamma = 0.99$, that is, only $1\%$ of the particles after the resampling are new.

\subsubsection*{Particle filters and importance sampling}
In this section we review the core ideas underlying the functioning of a particle filter and the principle of importance sampling, as it is applied in our implementation of the soft resampling. Consider a distribution $P(\vtheta)$, from which we sample $N$ particles $\vtheta_j$ with $j=1, \cdots, N$. Let us define an hypercube $\mathcal{C}$ of volume $\dd \vtheta$ centred around $\vtheta$, and let us call $n(\vtheta, d^N \vtheta)$ the number of particles in the said hypercube, i.e.
\begin{equation}
	n(\vtheta, \dd \vtheta) := \sum_{j=1}^N \chi_{\mathcal{C}} (\vtheta_j) \;
\end{equation}
with $\chi_{\mathcal{C}}$ being the characteristic function of the hypercube. We can write
\begin{equation}
	\frac{1}{N} n(\vtheta, d^N \vtheta) \rightarrow P(\vtheta) \dd \vtheta \quad \text{for} \; N \rightarrow \infty \; ,
	\label{eq:convergence_particle_filter}
\end{equation}
that is in the limit of large $N$ the fraction of particles in the hypercube tends to the probability in such volume element. In a PF we associate to each particle $\vtheta_i$ a weight $w_i$ and we can define the total weight in the hypercube $\mathcal{C}$ as
\begin{equation}
	P(\vtheta) \dd \vtheta \simeq w(\vtheta, d^N \vtheta) := \sum_{j=1}^N w_j \chi_{\mathcal{C}} (\vtheta_j) \; .
\end{equation}
This total weight is the probability distribution actually represented by the PF. In the limit of large $N$, for a smooth distribution, we can consider the weight a function of the point $w(\vtheta)$, which varies smoothly in space and is approximatively constant in the hypercube $\mathcal{C}$. This leads us to write
\begin{equation}
	P(\vtheta) \dd \vtheta \simeq w(\vtheta) \sum_{j=1}^N \chi_{\mathcal{C}} (\vtheta_j) =  w(\vtheta) n(\vtheta, \dd \vtheta) \; .
\end{equation}
This means that the distribution represented by the particle filter is proportional to the product of the weights and the density of the particles. This is however not the only way to represent $P(\vtheta)$. Suppose that for whatever reason we sample the particles $\vtheta_j$ from $Q(\vtheta)$, but that we actually want to represent the distribution $P(\vtheta)$. Then we can multiply the weights $w_j=1/N$ of each particle $\vtheta_j$ with the corrective factor $P(\vtheta_j)/Q(\vtheta_j)$, which remains approximately constant inside the region $\mathcal{C}$, i.e.
\begin{equation}
	w(\vtheta, d^N \vtheta) = \frac{1}{N} \sum_{j=1}^N \frac{P(\vtheta_j)}{Q(\vtheta_j)} \chi_{\mathcal{C}} (\vtheta_j) \simeq \frac{P(\vtheta)}{Q(\vtheta)} \frac{1}{N} \sum_{j=1}^N \chi_{\mathcal{C}} (\vtheta_j) = \frac{P(\vtheta)}{Q(\vtheta)} \frac{n(\vtheta, d^N \vtheta)}{N} \; ,
\end{equation}
since now the particles are distributed spacially according to $Q(\vtheta)$ the density of particles will tend to $Q(\vtheta)$ for large $N$ that, according to \cref{eq:convergence_particle_filter}, gives $w(\vtheta, d^N \vtheta) \rightarrow Q(\vtheta)$, therefore we have $w(\vtheta, d^N \vtheta) \simeq P(\vtheta)$. In the case of soft resampling the distribution $Q(\vtheta)$ is constructed from $P(\vtheta)$ as
\begin{equation}
	Q(\vtheta) \dd \vtheta :=  \left[ \alpha w(\vtheta) + (1-\alpha) \frac{1}{N} \right] n(\vtheta, \dd \vtheta) \; ,
\end{equation}
and the factor that multiplies the weights is $P(\vtheta)/Q(\vtheta) = w(\vtheta)/\left[ \alpha w(\vtheta) + (1-\alpha) \frac{1}{N}\right]$.
The factor used in \cref{eq:new_weights_to_norm} for the particle at $\vtheta_j^\prime$ contains $w_\phi(j)$ , which is the weight at this point in the original distribution $P(\vtheta)$.

\subsubsection*{Gaussian perturbation}
\label{subsub:gaussian_perturbation}
After the soft resampling, we add a perturbation to the particles as proposed in~\cite{granade_robust_2012}, that is, we define
\begin{equation}
	\vtheta_j^{\prime \prime} := \beta \vtheta_j^\prime + (1-\beta) \hvtheta_t + \vdelta \; ,
	\label{eq:perturbation}
\end{equation}
where $\beta \in (0, 1]$, $\hvtheta_t$ is the mean of the posterior approximated in \cref{eq:mean_pf} and $\vdelta$ is a random variable distributed according to
\begin{equation}
	\vdelta \sim \mathcal{N} (0, (1-\beta^2) \cov_t) \; .
\end{equation}
With this expression we move the particles toward the mean of the posterior, which is our estimator for $\vtheta$ and at the same time we lift the degeneracy of the $\vtheta_j^\prime$, that comes about because the particle $\vtheta_j$ with high weights appear many times in the new particles ensemble. Were the degeneracy not removed, all these copies of the same particle wouldn't contribute much to improving the resolution of the PF. This holds true unless they are perturbed, at which point they can encode the small scale behavior of the posterior. Because of the perturbation in \cref{eq:perturbation} the PF does not represent anymore the posterior $P(\vtheta|\vx_t, \vy_t)$ exactly. We now compute the probability distribution for $\vtheta_j^{\prime \prime}$ after the perturbation step. The particles are distributed in the space according to the $q_j$ weights in \cref{eq:mixed_q_j} and we call this distribution $Q(\vtheta')$. Let us write \cref{eq:perturbation} as $\vtheta_j^{\prime \prime} = \beta \vtheta_j^\prime + \vdelta'$ with
\begin{equation}
	\vdelta' \sim \mathcal{N} ((1-\beta) \hvtheta_t, (1-\beta^2) \cov_t) \; ,
	\label{eq:perturbation_prime}
\end{equation}
being a perturbation with non-null mean value. Then the probability density for $\vtheta_j^{\prime \prime}$ is the convolution of the probability of a particle being at position $\vtheta'$ and the probability of the noise causing a displacement $\vdelta' = \vtheta'' - \beta \vtheta'$, i.e.
\begin{equation}
	\wt{Q}(\vtheta'') = \int Q(\vtheta') g_\beta (\vtheta'' - \beta \vtheta') \dd \vtheta' = \sum_{j=1}^{\gamma N} q_{\phi(j)} g_\beta (\vtheta'' - \beta \vtheta'_j) \; ,
	\label{eq:perturbation_q}
\end{equation}
where $g_\beta$ is the Gaussian probability density associated to $\vdelta'$, i.e.
\begin{equation}
	g_{\beta} (\vtheta) := (2 \pi)^{-\frac{\dd}{2}} (1-\beta^2)^{-\frac{1}{2}} \det (\cov_t)^{-\frac{1}{2}} \exp \left[ -\frac{1}{2 (1-\beta^2)} \left( \vtheta - (1-\beta) \hvtheta_t \right)^\intercal \cov_t^{-1} \left( \vtheta - (1-\beta)\hvtheta_t \right) \right] \; .
	\label{eq:resampling_normal_distr}
\end{equation}
In \cref{eq:perturbation_q} we also substituted the integral with a summation being the probability $Q(\vtheta')$ discrete. According to the principles of importance sampling the distribution represented by a PF is the product of the weights and the density of particles, which reads
\begin{equation}
	\wt{P}(\vtheta_j) \propto \frac{P(\vtheta)}{Q(\vtheta)} \wt{Q}(\vtheta_j) \simeq P(\vtheta) \; ,
	\label{eq:p_after_perturbation}
\end{equation}
In principle, we could correct the distribution for this perturbation by computing exactly \cref{eq:perturbation_q} and accounting for it in the weights $w_j'$, in our implementation we don't do it however, since it would be very small anyway.

\subsubsection*{New particles proposal}
\label{subsub:new_particles}
We still need to produce $(1-\gamma) N$ new particles and we do it by extracting them from the Gaussian distribution with the same two first moments of the PF ensemble, i.e.
\begin{equation}
	\vtheta_j^{''} \sim \mathcal{N} \left( \hvtheta, \cov \right) \; ,
	\label{eq:resampling_normal}
\end{equation}
for $j=\gamma N, \cdots, N$. The mean and the covariance matrix are defined in \cref{eq:mean_pf} and \cref{eq:Sigma_pf} respectively. This is done again to increase the density of particles in the region of high probability, but it works properly only for unimodal distributions. The weights of these new particles are set to $w_j' = \frac{1}{N}$, so that their normalization is
\begin{equation}
	\sum_{j=\gamma N}^N w'_j = 1-\gamma \; .
\end{equation}
This extra particles and weights are concatenated directly to $\lbrace \vtheta_j^{\prime \prime}\rbrace_{j=1}^{\gamma N}$ and $\lbrace w_j'\rbrace_{j=1}^{\gamma N}$. We then rename the new weights and particles, i.e. $w_j' \rightarrow w_j$ and $\vtheta''_j \rightarrow \vtheta_j$, and with that the resampling procedure is concluded. In doing the last step of proposing new particles we are mixing the distribution represented by the PF $P(\vtheta)$ as it comes out of the perturbation step in \cref{eq:p_after_perturbation} with the distribution $g_0$ in \cref{eq:resampling_normal_distr}. At the end the PF ensemble represents the distribution
\begin{equation}
	P'(\vtheta) = \gamma P(\vtheta) + (1-\gamma)g_0 (\vtheta) \; .
\end{equation}
Again we do not correct for this distortion, which could be done by modifying the weights properly.

\subsubsection*{Resampling of the batch}
\label{subsub:batch_resampling}
\begin{figure}[h]
	\centering
	\includegraphics[width=\textwidth]{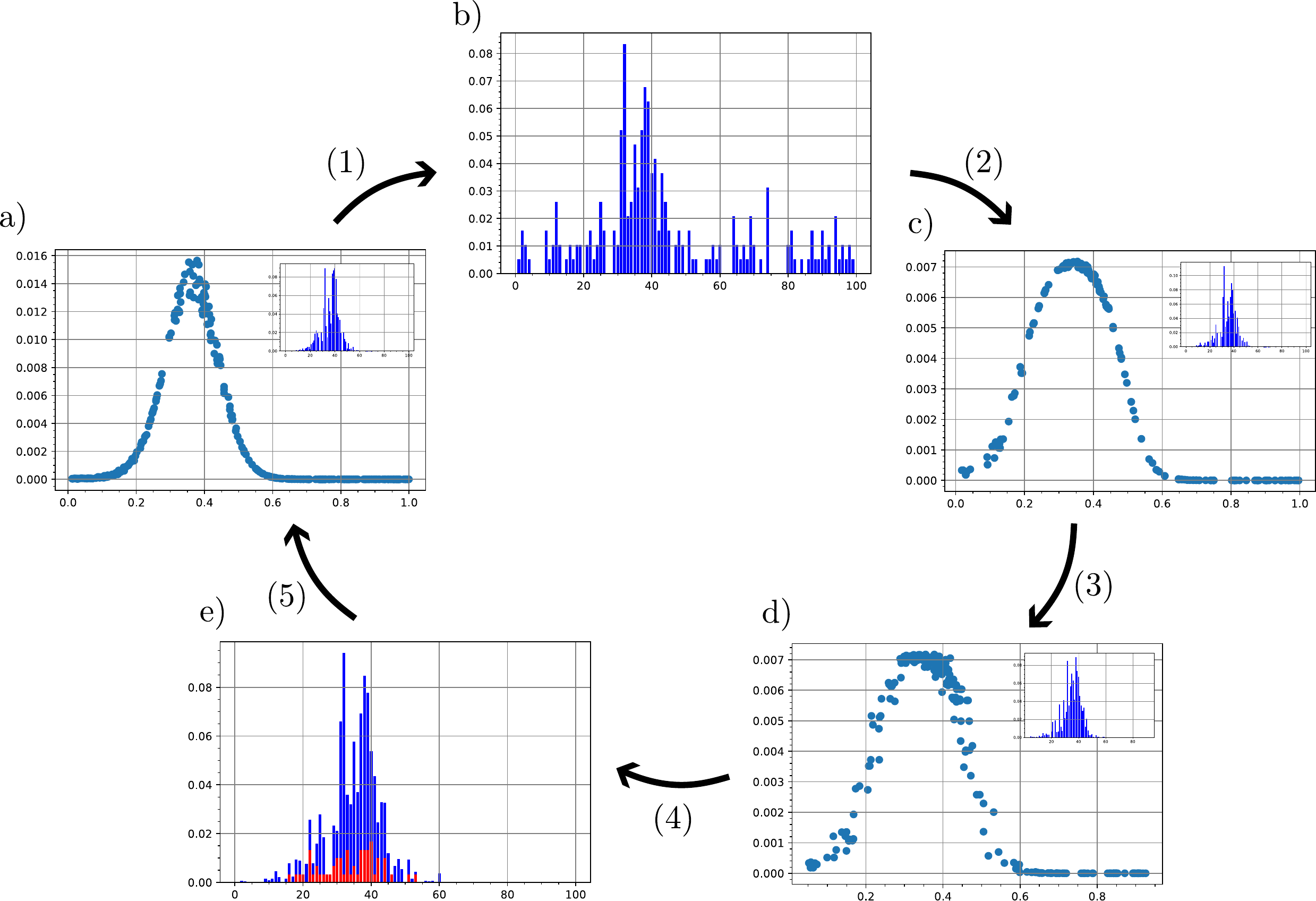}
	\caption{The ensemble of the PF before the resampling is represented in a), the scatter plot are the points $(\vtheta_j, w_j)$. This plot doesn't represent directly the posterior, because it doesn't take into account the density of particles. In all the plots, the inserted histogram is the actual posterior represented by the PF. Once the condition \cref{eq:resampling_condition} is satisfy the first step of the resampling is executed, which is the transformation (1) and corresponds to sampling with repetitions from \cref{eq:mixed_q_j}, the plot b) is the spatial density of particles after this action. The scatter plot c) is the distribution of the PF after the weights have been corrected according to \cref{eq:new_weights_to_norm} (transformation (2)). (3) is the application of the Gaussian noise in \cref{eq:perturbation} and (4) is the sampling of the extra proposed particles in \cref{eq:resampling_normal}. Plot e) is the distribution represented by the PF when the resampling routine is complete, where in red the contribution of the new particles is highlighted. At last (5) is the repeated application of the Bayes rule \cref{eq:bayes_rule_w}, following some new measurement on the probe, that leads to the ensemble of the PF being again in need of resampling. To emphasise the effect of each transformation we have set $\alpha=0.5, \beta=0.9, \gamma=0.8$. The total number of particles was $N=10^3$ and the effective number of particles in a), that is before the resampling, was $N_{\text{eff}}=93.8$. In the histograms the interval $[0, 1]$ of the scatter plots has been mapped to $[0, 100]$.}
	\label{fig:pf_cycle}%
\end{figure}

In order to compute the precision of the estimation, we need the results of many runs of the simulation, possibly executed in parallel on a GPU. In these circumstances the resampling is performed on all the instances of the estimation as soon as the condition \cref{eq:resampling_normal} holds true for at least a fraction $f$ of the estimations in the batch, which by default is set to $f=0.98$. The premature resampling of an estimation run will have a quite strong detrimental effect on the goodness of the posterior represented by the PF, on the contrary a late resampling is much less probable to distort the distribution, this is the reason why we set $f$ so close to one, that is, we want to limit as much as possible the number of simulations that are prematurely resampled. With the current implementation at each step either all the simulations is the batch are resampled or none. An improvement to the PF would be to resample selectively only those runs that are in need of resampling, and leave the other untouched until they satisfy \cref{eq:resampling_normal}, so that whatever number of runs could be resampled at each step. The complete resampling cycle, including the extraction and the new particle and the importance sampling is represented in \cref{fig:pf_cycle}.

\subsection{State particle filter}
\label{subsec:state_pf}
In this section we describe what happens when we are acting with weak (non-projective) measurements on the probe. In this case the probability to observe the outcome $y_{t+1}$ at the step $t+1$ depends on all the string of previous outcomes and controls, that is on the whole trajectory $\tau := (\vx_t, \vy_t)$, as well as on the current control $x_{t+1} $. This means we must substitute $p(y_{t+1}|x_{t+1}, vtheta)$ with $p(y_{t+1}|\vtheta, \vx_{t+1}, \vy_t)$ in \cref{eq:bayes_rule_w}. Since we avoid the reinitialization of the probe, its state depends on all the evolution history. With this change in the outcome probability all the formulas of the previous section remain valid. To compute $p(y_{t+1}|\vtheta, \vx_{t+1}, \vy_t)$, we need to keep track of the state of the probe. In order to do so we introduce the state particle filter. In this data structure we save for each particle $\vtheta_j$ the state of probe had the system evolved under the action of $\mathcal{E}_{\vtheta_j, x}$, with the controls and the outcomes being the ones actually applied/observed in the evolution, we indicate this state with $\rho_{\vtheta_j, \tau_t}$. To this state we associate the weight $w_j$ of the particle $\vtheta_j$. The state particle filter represents the posterior probability distribution for the state of the probe conditioned on the trajectory $\tau_t$. The expression for $\rho_{\vtheta_j, \tau_t}$ reads
\begin{equation}
	\rho_{\vtheta_j, \tau_t} = \mathcal{M}_{y_t}^{x_t} \circ \mathcal{E}_{\vtheta_j, x_t} \circ \mathcal{M}_{y_{t-1}}^{x_{t-1}} \circ \mathcal{E}_{\vtheta_j, x_{t-1}} \circ \cdots \circ \mathcal{M}_{y_1}^{x_1} \circ \mathcal{E}_{\vtheta_j, x_1} (\rho)\; ,
	\label{eq:evolution_rho_j}
\end{equation}
where
\begin{equation}
	\mathcal{M}_{y_t}^{x_t} (\rho) := \frac{M_{y_t}^{x_t} \rho M_{y_t}^{x_t \dagger}}{\tr \left[ M_{y_t}^{x_t} \rho M_{y_t}^{x_t \dagger} \right]} \; ,
\end{equation}
is the backreaction of the measurement on the state of the probe. The estimator for the probe state at the step $t$ is
\begin{equation}
	\widehat{\rho}_{\tau_t} := \sum_{j=1}^N w_j \rho_{\vtheta_j, \tau_t} \; ,
\end{equation}
that is, the mean of the state on the posterior distribution for the parameters. The estimator $\widehat{\rho}_{\tau_t}$ can then be fed to the agent, to contribute to the computation of the next control. When the resampling is performed on the PF ensemble we get a new set of particles $\vtheta_j^\prime$ and their corresponding states must be also updated. This means we have to keep track of the vectors $\vx_t$ and $\vy_t$ in the simulations and recompute the evolution of the whole state particle filter from the beginning, so that we get $\rho_{\vtheta_j^\prime, \tau_t}$. From the computational point of view, the fact that we need these rather memory intensive structures of the PF and the state PF tells us that the optimization loop presented here can be applied only to rather small and simple quantum sensors.

\subsection{Multimodal posterior distributions}
\label{subsec:multimodal_distributions}
The resampling procedure presented in the previous section has some limitation in dealing with multimodal distributions. In this case the mean of the posterior may lay in a region of relatively low probability between two peaks and the accumulation of particles in this region after a resampling would be detrimental to the precision of the estimation. From its own design it would be difficult to modify the PF so that it accounts for multiple maxima. The informations that we can easily extract from the PF are its moments and from them the actual positions of the maxima are not straightforward to obtain. Multimodal posterior distributions are however common in quantum metrology. For example in multiphase estimation, like the measurement of the hyperfine interaction in NV-${}^{13}C$. To promote the preservation of secondary features in the posterior distribution we can use multiple particle filter at once. In this situation a set of PFs, with different priors, are updated in parallel and only together represent the full Bayesian posterior. To reduce the memory requirement of such approach we could consider simple Gaussian distributions instead of full PFs. We start by approximating the prior distribution $\pi (\vtheta)$ with as a sum of $L$ Gaussians:
\begin{equation}
	\pi (\vtheta) \simeq \sum_{l=1}^{L} w_l \mathcal{N} (\boldsymbol{\mu}_l, \vsigma_l) \; .
	\label{eq:gaussian_prior_approximation}
\end{equation}

If the parameters $\boldsymbol{\mu}_l, \vsigma_l$ are fixed then the Bayesian update step can be done by solving a linear regression problem to find the best new values for $\lbrace w_l \rbrace_{l=1}^L$ that represent the posterior. In this way the PF has however a limited resolution, determined by the initial Gaussian. If we also let $\boldsymbol{\mu}_l, \vsigma_l$ change during the Bayesian update step, then we solve the problem of having limited resolution, but we now have to deal with a non-linear regression problem.

\section{Differentiability of the particle filter}
\label{sec:diff_pf_generic}
In this section we discuss what happens when the resampling routine of the particle filter is switched on, and, in particular, what we need to do to assure that the gradient produced by the automatic differentiation is correct.

\subsection{Differentiable PF through reparametrization and soft resampling}
\label{subsec:diff_pf}
\subsubsection*{Differentiability of the soft resampling}
As seen in \cref{subsec:gradient_loss}, the gradient can't be propagated through randomly extracted variables, therefore when the categorical resampling is executed, the particles $\vtheta_j^\prime$ in \cref{eq:theta_j'} don't have any connection with the controls, i.e.
\begin{equation}
	\frac{\dd \vtheta_j^\prime}{\dd \vlambda} = 0\; .
\end{equation}
Similarly, the weights are reinitialized and loose every dependence on the history of the estimation. At the moment of taking the gradient we won't be able to account for anything that happened before the last resampling. This means that the training routine optimizes the agent only for the later steps, although what has been learnt in this context may be useful also for the earlier measurements. The soft resampling with $\alpha < 1$, introduced in \cref{sec:particle_filter}, is able to partially remove this obstacle. With this trick the dependence on $\vlambda$ is passed from the old weights to the new ones through \cref{eq:new_weights_to_norm}. However, the gradient doesn't backpropagate entirely but it is attenuated by the factor $1-\alpha$. The price to pay for propagating the gradient is that the $N$ particles are not all fully effective for the resampling, instead only a fraction $\alpha$ of them participate to it. As discussed in \cref{subsec:gradient_loss}, since we are extracting stochastic variables from the distribution in \cref{eq:mixed_q_j}, we should also add the corresponding log-likelihood terms $\sum_{j} \log q_{\phi(j)}$ to the loss. However adding so many terms would increase too much the variance of the gradient. Either we don't account for these log-likelihoods and we accept the gradient to have a bias, or we use the correction introduced in~\cite{scibior_differentiable_2021}, that prescribes to substitute the definition of the new weights $w_j'$ with an appropriate surrogate expressions. Introducing this correction is the default behaviour of our code but it can be applied only if the loss $\ell (\hvtheta, \vtheta)$ is of a certain form. It can be proved that for the MSE defined in \cref{eq:loss_mse} this gives the correct gradient. See \cref{subsec:scibior_correction} for a complete discussion.

\subsubsection*{Differentiability of the perturbation}
The next transformation on the particles is the perturbation of \cref{eq:perturbation}. Again we would be unable to propagate the gradient through the perturbation $\vdelta'$, if we extracted it directly from the Gaussian $\mathcal{N} (\boldsymbol{\mu}, \cov)$, with $\boldsymbol{\mu} = (1-\beta) \hvtheta_t$ and $\cov = (1-\beta^2) \cov_t$. For this reason we apply the reparametrization trick and write
\begin{equation}
	\vdelta' (\vy_{t}, \vlambda) = \cov (\vy_{t}, \vlambda)\vu + \boldsymbol{\mu} (\vy_{t}, \vlambda) \; ,
	\label{eq:first_perturb_u}
\end{equation}
where $\vu$ is extracted from the multivariate standard Gaussian $\vu \sim \mathcal{N} (0, \id)$. The perturbation is now a differentiable functions of $\vlambda$. For the extraction of $\vu$ we do not need to add a corresponding log-likelihood term, as discussed in \cref{subsec:gradient_loss}, because its probability density function doesn't depend on $\vlambda$.

\subsubsection*{Differentiability of the proposed particles}
For the last step of the resampling, which consists in proposing new particles $\vtheta_j^{\prime \prime}$, extracted from $\mathcal{N} (\hvtheta_t, \cov_t)$, we again exploit the reparametrization trick and write
\begin{equation}
	\vtheta_j^{\prime \prime} (\vy_{t}, \vlambda) = \cov_t (\vy_{t}, \vlambda)\vu_j + \hvtheta_t (\vy_{t}, \vlambda) \; ,
	\label{eq:second_perturb_u}
\end{equation}
which is again differentiable and doesn't require a log-likelihood term.

\subsubsection*{Differentiability of the state particle filter}
Regarding the differentiability of the state particle filter discussed in \cref{subsec:state_pf}, we observe that its elements are functions of $\vlambda$ through the trajectory $\tau_t (\vlambda) = (\vx_t(\vy_t, \vlambda), \vy_t)$:
\begin{equation}
	\rho_j (\vlambda) := \rho_{\vtheta_j, \tau_t (\vlambda)} \; .
\end{equation}
Under the assumption that the encoding and the measurements appearing in \cref{eq:evolution_rho_j} are differentiable we can propagate the gradient through the evolution of the probe in \cref{eq:evolution_rho_j} . When the particles are resampled and the new states are computed, the dependence of the new state $\rho_j (\vlambda)$ on the old weights of the PF, and therefore on $\vlambda$, persists through the measurement backreaction operators $\mathcal{M}_{x_t(\vlambda)}^{y_t}$. A new dependence on $\vlambda$ appears in the evolution map $\mathcal{E}_{\vtheta_j (\vlambda), x_{t}(\vlambda)}$, coming from the new particles $\vtheta_j(\vlambda)$, so that we can write
\begin{equation}
	\rho_j' (\vlambda) := \rho_{\vtheta_j (\vlambda), \tau_t (\vlambda)} \; .
\end{equation}

\subsection{Differentiable PF through the correction of \'Scibior and Wood}
\label{subsec:scibior_correction}
In~\cite{scibior_differentiable_2021} a correction was introduced to make the resampling procedure in a PF differentiable. This can be implemented in place of the soft resampling, or alongside with it. The default behaviour of our software is to perform the soft resampling with $\alpha=0.5$ alongside the \'Scibor and Wood correction. With this choice only half of the gradient is backpropagated through soft resampling, the other half is done by the \'Scibor and Wood correction. The former prescribes to modify the normalized weights $w_j'$ of \cref{eq:new_weights_to_norm} to
\begin{equation}
	\wt{w}_j' \leftarrow w_j' \frac{q_{\phi(j)}}{\sg{q_{\phi(j)}}} \; ,
	\label{eq:scibior_correction}
\end{equation}
where the meaning of the symbols is that of \cref{subsec:resampling_scheme}. In this formula we are using the stop gradient operator $\text{sg} \left[ \cdot \right]$, which is an instruction that tells the automatic differentiation frameworks not to compute the derivatives of the expression inside the operator. This correction has no effects in the forward pass, but produces additional gradient terms in the backward pass. We see in this section, that for a MSE loss the extra terms in the gradient appearing because of this surrogate expression are exactly the log-likelihoods that we would have to insert following the conclusions of \cref{subsec:gradient_loss}, although this observation can't be extended to a generic loss. Let us start from the expression for the MSE, when one and only one resampling is performed in the whole experiment, at step $t$, i.e.
\begin{equation}
	\Delta^2 \hvtheta = \int \ell (\hvtheta, \vtheta) P(\tau_{t+1:M-1} | \vtheta_j', \vtheta) \left( \prod_{j=1}^N P(\vtheta_j'|\tau_{0:t}) \right) P(\tau_{0:t} | \vtheta) \pi (\vtheta) \dd \tau_{M-1} \left( \prod_{j=1}^N \dd \vtheta_j' \right) \dd \vtheta \; .
	\label{eq:mse_resampling}
\end{equation}
We assume for clarity that the perturbation and the extraction of the extra particles, in \cref{eq:first_perturb_u} and \cref{eq:second_perturb_u} respectively, are turned off, i.e. $\beta=\gamma=1$. The object $\tau_{\alpha:\beta}$ with $\alpha, \beta$ integers in $[0, M-1]$ is the trajectory between the steps $\alpha$ and $\beta$ (extrema included), i.e. $\tau_{\alpha:\beta} = (\vx_{\alpha:\beta}, \vy_{\alpha: \beta})$ with $\vx_{\alpha:\beta} = (x_\alpha, x_{\alpha+1}, \ldots, x_\beta)$ and $\vy_{\alpha:\beta} = (y_\alpha, y_{\alpha+1}, \ldots, y_\beta)$. Reading \cref{eq:mse_resampling} from right to left we encounter the probability densities for all the random variable extractions in chronological order. First the extraction of the true values $\vtheta$ for the simulation instance, then the trajectory up to the resampling point, then the extraction of the new particles $\vtheta_j'$, and finally the measurements after the resampling, i.e. the trajectory after the $t$-th step until the end. This last probability depends also on the values of the new particles, through the posterior distribution momenta, that are passed to the agent that decides the next control. We now insert the expression for $\ell(\hvtheta, \vtheta)$ found in \cref{eq:loss_mse} in \cref{eq:mse_resampling}. We postpone the computation of the trace to the end and expand the error matrix for the estimator in \cref{eq:mean_pf}, i.e.
\begin{equation}
	(\hvtheta-\vtheta) (\hvtheta-\vtheta)^\intercal = \sum_{i, j=1}^N w_i' w_j' {\vtheta_i'} \vtheta_j'^\intercal - \vtheta \left( \sum_{j=1}^N w_j' \vtheta_j'^\intercal \right) - \left( \sum_{j=1}^N w_j' \vtheta_j' \right) \vtheta^\intercal  + \vtheta \vtheta^\intercal \; .
	\label{eq:mse_expanded}
\end{equation}

Each term in the first summation gives a contribution to $\Delta^2 \hvtheta$ equal to
\begin{equation}
	\int w_i' w_j' \vtheta_i' \vtheta_j'^\intercal P(\tau_{t+1:M-1}|\vtheta) P(\vtheta_i'|\tau_{0:t}) P(\vtheta_j'|\tau_{0:t}) P(\tau_{0:t} | \vtheta) \pi(\vtheta) \dd \tau_{M-1} \dd \vtheta_i' \dd \vtheta_j' \dd \vtheta \; ,
\end{equation}
where we neglect all the integrals on the variables $\vtheta_\alpha'$ with index $\alpha \neq i, j$, because they do not appear in the integrand. The gradient of this term with respect to $\vlambda$ gives rise to the usual likelihood terms for the measurement plus the following extra terms coming from the resampling:
\begin{equation}
	\sum_{i, j}^N w_i' w_j' \vtheta_i' \vtheta_j'^\intercal \left( \frac{\dd \log P(\vtheta_i'|\tau_{0:t})}{\dd \vlambda} + \frac{\dd \log P(\vtheta_j'|\tau_{0:t})}{\dd \vlambda} \right) \; ,
	\label{eq:second_order_loglike}
\end{equation}
for $i, j = 1, \dots, N$. Similarly the linear terms in \cref{eq:mse_expanded} give the following likelihood terms:
\begin{equation}
	- \sum_{j=1}^N w_j' \left( \vtheta_j' \vtheta^\intercal + \vtheta \vtheta_j'^\intercal \right) \frac{\dd \log P(\vtheta_j'|\tau_{0:t})}{\dd \vlambda} \; ,
	\label{eq:first_order_loglike}
\end{equation}
plus the same terms with $\vtheta_j'^\intercal$. There is no likelihood terms associated to the constant $\vtheta \vtheta^\intercal$ in \cref{eq:mse_expanded}. Now we shall see that deriving the surrogate expression gives the same terms in the gradient. Let us write the total derivative of the error matrix:
\begin{eqnarray*}
	\frac{\dd}{\dd \vlambda} (\hvtheta-\vtheta) (\hvtheta-\vtheta)^\intercal &=& \frac{\dd \hvtheta} {\dd \vlambda} (\hvtheta-\vtheta)^\intercal + (\hvtheta-\vtheta) \frac{ d \hvtheta^\intercal }{\dd \vlambda} \\
	&=&  \left( \sum_{i=1}^N \frac{\dd \wt{w}_i'}{\dd \vlambda} \vtheta_i' \right) \left( \sum_{j=1}^N {w}_j' \vtheta_j' -\vtheta \right)^\intercal + \left( \sum_{i=1}^N {w}_i' \vtheta_i' - \vtheta \right) \left( \sum_{j=1}^N \frac{\dd \wt{w}_j'}{\dd \vlambda} \vtheta_j'^\intercal \right) \; .
\end{eqnarray*}
Where the derivative doesn't act the weights $\wt{w}_j'$ become $w_j'$. From the definition of $\wt{w}_j'$ in \cref{eq:scibior_correction} we compute the derivative of the surrogate expression, that is
\begin{equation}
	\frac{\dd \wt{w}_j'}{\dd \vlambda} = \frac{d w_j'}{\dd \vlambda} + w_j' \frac{\dd \log q_{\phi(j)}}{\dd \vlambda} \; .
\end{equation}
We know keep track only of the extra likelihood terms coming from the surrogate part of the weights $\wt{w}_j'$ and organize the terms according to the order in $\vtheta_j'$. We have the second order terms:
\begin{equation}
	\sum_{i, j}^N w_i' w_j' \vtheta_i' \vtheta_j'^\intercal \left ( \frac{\dd \log q_{\phi(i)}}{\dd \vlambda} + \frac{\dd \log q_{\phi(j)}}{\dd \vlambda} \right) \; ,
\end{equation}
and the first order ones:
\begin{equation}
	\sum_{j=1}^N w_j'\left( \vtheta_j' \vtheta^\intercal + \vtheta \vtheta_j'^\intercal \right) \frac{\dd \log q_{\phi(j)}}{\dd \vlambda} \; ,
\end{equation}
which correspond respectively to \cref{eq:second_order_loglike} and \cref{eq:first_order_loglike}, once we realize that $P(\vtheta_j'|\tau_{0:t}) = q_{\phi(j)}$. One of the advantages of this approach is the reduce variance of the gradient estimator, which would explode, where we to insert all the likelihood terms for the new particle extractions at the end of the estimation. The correction, however, doesn't produce always the correct gradient for the loss, but only when $\ell (\hvtheta, \vtheta)$ is a polynomial function of the weights $\wt{w}_j'$. Consider the estimation of single parameter $\theta \in [0, 2 \pi)$, we might want to use a loss functions $l(\widehat{\theta}, \theta)$ that respect the circular nature of the parameter, like
\begin{equation}
	l(\widehat{\theta}, \theta) := \sin(\widehat{\theta}-\theta)^2 \; .
\end{equation}
The gradient with respect to $\vlambda$, when the correction is implemented, is
\begin{eqnarray}
	\frac{\dd}{\dd \vlambda} l(\widehat{\theta}, \theta) &=& 2 \sin (\widehat{\theta}-\theta) \cos (\widehat{\theta}-\theta) \frac{\dd \widehat{\theta}}{\dd \vlambda} \\
	&=& \sin (2\widehat{\theta}-2\theta) \sum_{j=1}^N \frac{\dd \wt{w}_j'}{\dd \vlambda} \theta_j' \\
	&=& \sin (2\widehat{\theta}-2\theta) \sum_{j=1}^N \left( \frac{d w_j'}{\dd \vlambda} \theta_j' + w_j' \theta_j' \frac{\dd \log q_{\phi(j)}}{\dd \vlambda} \right) \; ,
\end{eqnarray}
so that the likelihood term in the gradient is
\begin{equation}
	\sin (2\widehat{\theta}-2\theta) \sum_{j=1}^N w_j' \theta_j' \frac{\dd \log q_{\phi(j)}}{\dd \vlambda} \; ,
\end{equation}
while it should be
\begin{equation}
	\sin(\widehat{\theta}-\theta)^2 \sum_{j=1}^N \frac{\dd \log q_{\phi(j)}}{\dd \vlambda} \; .
\end{equation}
%
%
%
Another example where the correction fails in the loss of \cref{eq:loss_delta}. Let us take $\hvtheta$ to be the maximum likelihood estimator, then, since a small perturbation in the posterior distribution won't change it we have
\begin{equation}
	\frac{\dd \hvtheta}{\dd \vlambda} = \frac{\dd \hvtheta}{\dd w_{j}} \frac{\dd w_j}{\dd \lambda} = 0\; ,
\end{equation}

therefore the correction is useless to backpropagate the gradient through the resampling step and we must rely only on the importance sampling. Incidentally we notice that the loss for \cref{eq:loss_delta} is a pure-likelihood expression, analogous to the loss in regular Policy Gradient RL.

\section{Computation and differentiation of the loss function}
\label{sec:comp_diff_loss}
As in most optimization problems, the trainable variables of the agent are updated with a version of the stochastic gradient descent. In this section, we define the loss function for this training, compute its gradient, and comment on the computational resources required by the training.

\subsection{Definition of the loss function}
\label{subsec:general_loss}
The two scalar losses that we used in this work are the MSE, defined in \cref{eq:loss_mse}, used for continuous parameters, and the discrimination loss of \cref{eq:loss_delta}, used for discrete parameters, that converges to the error probability when averaged. If the parameter to be estimated is a phase we might want to take as loss the circular variance~\cite{holevo_probabilistic_2011}. In the following section we adopt the symbol $\ell(\hvtheta, \vtheta)$ for the loss and keep the discussion completely general. We mention that this analysis would apply also to a more general class of losses, being functions of the of the PF ensemble, i.e. $\ell(\mathfrak{p}, \vtheta)$, provided they are well-behaved as functions. The expected value of the loss on the trajectory is
\begin{align}
  \Delta^2 \hvtheta_\tau := \int \ell (\hvtheta, \vtheta) P(\hvtheta| \tau_{M-1}, \vtheta) \dd \hvtheta \; ,
  \label{eq:single_trajectory_mse}
\end{align}
with $\tau_{M-1} := (\vx_{M-1}, \vy_{M-1})$ indicating the complete trajectory. This definition presumes a stochastic dependence of the estimator $\hvtheta$ computed from the PF on the outcomes and the controls of the measurements, collectively denominated $\tau_{M-1}$. This is codified by the probability density $P(\hvtheta| \tau_{M-1}, \vtheta)$. This stochasticity can be due to the resampling routine or, in general, to the construction of the estimator $\hvtheta$, which could entail the sampling from a distribution, which is however never the case in our examples. The quantity $\Delta^2 \hvtheta_\tau$ refers to a single trajectory of the PF, the one indicated with $\tau_{M-1}$. We wish however to consider the average of the MSE over all the possible trajectories $\tau_{M-1}$ weighted appropriately. The expectation value over $\tau_{M-1}$ is expressed by the following operator
\begin{align}
  \expval[\tau]{\cdot} := \int \cdot \, P(\vx_{M-1}, \vy_{M-1} | \vtheta) \dd \vx_{M-1} \dd \vy_{M-1} \; ,
\end{align}
which applied to \cref{eq:single_trajectory_mse} gives
\begin{align}
  \expval[\tau]{\Delta^2 \hvtheta_\tau} = \int \ell (\hvtheta, \vtheta) P(\hvtheta | \vtheta) \dd \hvtheta \; ,
  \label{eq:exp_mse}
\end{align}
where we have defined
\begin{align}
  P(\hvtheta|\vtheta) := \int P(\hvtheta| \tau_{M-1}, \vtheta)  P(\tau_{M-1} | \vtheta) \dd \vx_{M-1} \dd \vy_{M-1} \; .
\end{align}
We also want to take the expectation value of $\hvtheta$ on the prior $\pi(\vtheta)$ through the operator
\begin{align}
  \expval[\vtheta]{\cdot} := \int \cdot \, \pi(\vtheta) \dd \vtheta \; ,
\end{align}
which applied to $\expval[\tau]{\Delta^2 \hvtheta_\tau}$ gives the figure of merit for the error
\begin{align}
  \Delta^2 \hvtheta: = \expval[\vtheta]{ \expval[\tau]{\Delta^2 \hvtheta_\tau}} =  \int \ell (\hvtheta, \vtheta) P(\hvtheta) \dd \hvtheta \; ,
  \label{eq:mse_no_theta}
\end{align}
with
\begin{align}
  P(\hvtheta) := \int P(\hvtheta| \tau_{M-1})  P(\tau_{M-1} | \vtheta) \pi (\vtheta) \dd \vx_{M-1} \dd \vy_{M-1} \dd \vtheta \; .
  \label{eq:prob_hvtheta}
\end{align}
This is the probability density for the final estimator $\hvtheta$, given that the true value $\vtheta$ is extracted from the prior $\pi(\vtheta)$ at the beginning and we average over the trajectory $\tau_{M-1}$ that is stochastically generated in the simulation, through the actions of the agent and the measurements. The expression in \cref{eq:mse_no_theta} suggests us a straightforward way to approximate the error from the numerical simulation (\textit{à la} Monte Carlo), i.e.
\begin{align}
  \Delta^2 \hvtheta \simeq \frac{1}{B}\sum_{k=1}^B \ell (\hvtheta_k, \vtheta_k) \; ,
  \label{eq:mse_montecarlo}
\end{align}
where $\vtheta_k$ is the true value of the parameters in the $k$-th simulation and $\hvtheta_k$ is the corresponding final estimator. By carrying out the complete estimation in a batch of $B$ simulated experiments, with each $\vtheta_k$ extracted from $\pi(\vtheta)$, we are effectively sampling $\hvtheta$ from $P(\hvtheta)$ so that by the law of large number we can approximate the expectation value of the loss function in \cref{eq:mse_no_theta} with the empirical mean on the batch. Each simulation in the batch follows its particular trajectory, which will be different from the ones of the other simulations, because the randomly extracted measurement outcomes are different. Notice that in distinction with the notation of the previous sections the subscript in \cref{eq:mse_montecarlo} doesn't refer to the step of the measurement cycle, but to the index of the simulation in the batch: the estimators $\hvtheta_k$ are always evaluated at the last step $t=M-1$. We call $B$ the \textit{batchsize} of the simulation. The right-hand side of \cref{eq:mse_montecarlo} will be the loss to be minimized by the training procedure. A natural question that arises here, is why aren't we using the covariance matrix as estimated from the PF in the computation of the MSE? The answer is that the PF may be imprecise for the evaluation of the variance, in particular, it tends to underestimate it, because some tails of the distribution $P(\hvtheta | \tau_{M-1})$ may not be very well represented. We prefer to estimate the MSE empirically from the sampled $\hvtheta_k$, extracted from the true distribution $P(\hvtheta)$, in order to avoid biases. The loss of \cref{eq:mse_montecarlo} is the closest it can be to the precision we would observe in an experiment.

\subsection*{Definition of the loss for limited resources}
In the previous paragraph we have implicitly assumed that the stopping condition of the estimation was based on the number of measurement $M$, i.e. we had a fixed number of measurement in each instance of the estimation. If, however, the resources are not simply related to the number of measurement steps, since each estimation in the batch follows its own trajectory, we may have different termination times, which correspond to the sensor employing a different number of measurement steps to consume all the available resources. In this section we introduce the notation $\hvtheta_{k, t}$, where the first subscript $k$ is the index in the batch, and the second $t$ is the measurement step. Whatever the nature of the resource chosen, to avoid having infinite loops we always fix a maximum number of measurement steps $M$ in the simulation, that should to be much larger than the expected number of iterations before the resources run out. At each step only the PF ensemble of those estimations which haven't terminated yet are updated with the Bayes rule, all the others, which have already consumed the available amount of resources, remain ``freezed'', since no measurement is performed and therefore no update is applied. Nevertheless all the quantities computed from the PF ensemble, e.g. $\hvtheta_{k, t}$ and $\Sigma_{k, t}$ are defined potentially for all the estimation steps $t=0,\cdots, M-1$. To put it simply if $t_k^\star$ was the index of the last measurement for the $k$-th estimation in the batch before it running out of resources, then $\hvtheta_{k, t} = \hvtheta_{t_k^\star, k}$, $\Sigma_{k, t} = \Sigma_{t_k^\star, k}$ for $t \ge t_k^\star$. In general, the PF ensemble remains the same if no new measurement outcomes are incorporated, i.e. $\mathfrak{p}_{k, t} = \mathfrak{p}_{t_k^\star, k}$ for $t \ge t_k^\star$. The simplest stopping condition for the measurement cycle is now that all the $B$ estimations in the batch have concluded, but to reduce the simulation time we only ask for at least a fraction $\nu=0.98$ of estimations to have terminated. These would exclude those simulations that are taking too long to terminate. We define $M'$ the realized number of iterations in the measurement loop determined by this condition, so that the loss in \cref{eq:mse_montecarlo} becomes
\begin{equation}
	\Delta^2 \hvtheta \simeq \frac{1}{B'}\sum_{k=1}^{B'} \ell (\hvtheta_{k, M'}, \vtheta_k) \; ,
	\label{eq:mse_montecarlo_resources}
\end{equation}
where the summation is taken only on those $B' = \lceil \nu B \rceil$ estimations in the batch that have terminated.

\subsection{Dependence of the loss on the trainable variables}
\label{subsec:particular_loss}
We go on by deriving from \cref{eq:mse_no_theta} an expression for the MSE, that is more directly related to the quantities simulated, under the hypotheses that the resampling has been turned off, i.e. $r_t = 0$, and that the computation of $\hvtheta$ from the ensemble of the PF doesn't require any stochastic operation. These are working hypotheses, which will allow to make useful observations and generalizations, whose domain of applications is however not limited by the said hypotheses. We begin observing that the controls $\vx_{M-1}$ produced by the agent are deterministic functions of the ensemble of the PF, for example through the mean and the covariance matrix. Therefore, the weights of the PF are in turn deterministic functions of the measurement outcomes, as they are computed with \cref{eq:bayes_rule_w}, so that we can write the controls $\vx_{t}$ and the estimator $\hvtheta_{t}$ at step $t$ as
\begin{align}
  \vx_{t} = g_1 (\vy_{t-1}, \vlambda) \quad \text{and} \quad \hvtheta_{t} = g_2 (\vy_{t}, \vlambda) \; ,
  \label{eq:deterministic_theta_x}
\end{align}
for two appropriate functions $g_1$ and $g_2$. Beside the outcomes both the controls and the estimators depend on the trainable variables of the agent, indicated with $\vlambda$, for the aforementioned reasons. Under these hypotheses the probabilities appearing in \cref{eq:prob_hvtheta} can be rewritten as
\begin{equation}
	P (\vx_{M-1}, \vy_{M-1}| \vtheta) = \delta (\vx_{M-1} - g_1 (\vy_{M-2}, \vlambda)) p(\vy_{M-1}|\vtheta, \vlambda)\; , \\
 	\label{eq:simplified_probabilities_1}
\end{equation}
\begin{equation}
	P(\hvtheta|\vx_{M-1}, \vy_{M-1}) = \delta (\hvtheta - g_2 (\vy_{M-1}, \vlambda)) \; .
	\label{eq:simplified_probabilities_2}
\end{equation}
Solving the integrals in $\dd\vx_{M-1}$ and in $\dd \hvtheta$ in \cref{eq:mse_no_theta}, we get the following expression for the MSE
\begin{align}
  \Delta^2 \hvtheta = \int \ell (\hvtheta(\vy_{M-1}, \vlambda), \vtheta) p(\vy_{M-1}|\vtheta, \vlambda) \pi(\vtheta) \dd \vy_{M-1} \dd \vtheta \; .
  \label{eq:final_mse_exp}
\end{align}
This is an expectation value on the probability distribution of the tuple of outcomes $\vy_{M-1}$. We introduce $\omega := (\vy_{M-1}, \vtheta)$ and redefine the loss for the next sections as
\begin{align}
  \ell (\omega, \vlambda) := \ell (\hvtheta(\vy_{M-1}, \vlambda), \vlambda) \; .
\end{align}
The object $\omega$ contains all the variables that depend on the specific instance of the simulation so that the empirical approximation of $\Delta^2 \hvtheta$ from \cref{eq:final_mse_exp} is
\begin{align}
  \mathcal{L} (\vlambda) := \frac{1}{B}\sum_{k=1}^B \ell (\omega_k, \vlambda) \; ,
  \label{eq:loss_omega_k}
\end{align}
with $\omega_k := (\vy_{k, M-1}, \vtheta_k)$. The true values $\vtheta_k$ are sampled from $\pi(\vtheta)$ at the beginning of the run. In case the agent is a NN the trainable variables are the weights and the biases, while for the non-adaptive strategy the variables are directly the tuple of all the controls, i.e.  $\vlambda = (x_1, x_2, \ldots, x_{M-1})$. The average loss in \cref{eq:loss_omega_k} will be also named the scalar loss, in contrast to $\ell (\omega_k, \vlambda)$, which is the individual loss or the vector loss, since it has a free index $k$.

\subsection{Gradient of the loss}
\label{subsec:gradient_loss}
The simulation of the quantum sensor, the particle filter, and the evaluation of the NN are implemented in the chosen automatic differentiation (AD) environment, i.e. TensorFlow (TF), so that at the end of the simulation we can take the gradient of the loss in \cref{eq:loss_omega_k} with respect to $\vlambda$ with no effort and obtain
\begin{align}
  \frac{\dd \mathcal{L} (\vlambda)}{\dd \vlambda} = \frac{1}{B} \frac{\dd}{\dd \vlambda} \sum_{k=1}^B \ell (\omega_k, \vlambda) \; .
  \label{eq:gradient_first_attempt}
\end{align}
The automatic differentiation framework does all the derivatives automatically, that we would need to evaluate analytically or numerically otherwise. Even if the outcomes $\vy_{k, M-1}$ are extracted from a probability distribution that depends on $\vlambda$, as it is because each of them is sampled from $p(y_{k, t+1}|\vx_{k, t+1}, \vy_t, \vtheta_k)$ and the controls $\vx_{k, t+1}$ depend on $\vlambda$, in TF and other similar frameworks their derivatives with respect to $\vlambda$ are always null by construction, i.e.
\begin{align}
  \frac{\dd}{\dd \vlambda} y_{k, t+1} = 0 \; ,
  \label{eq:derivative_random}
\end{align}
in other words, the gradient cannot propagate through the extraction of random variables. This is a consistent behaviour of automatic differentiation frameworks, and has to do with the fact that the sampled variables are considered constant tensors in the construction of the graph, on the same level as other numerical constants fixed by the programmer. We will show now, that much like in~\cite{porotti_gradient-ascent_2023}, the gradient of the loss produced by AD in \cref{eq:gradient_first_attempt} is not correct and will lead to a suboptimal training routine. Another term must be added that keeps track of the sampled variables during the evolution. To understand why this is so, let us start from the theoretical definition of $\Delta^2 \hvtheta$ in \cref{eq:final_mse_exp} and take its gradient with respect to $\vlambda$. The two terms $p(\vy_{M-1}|\vtheta, \vlambda)$ and $\hvtheta(\vy_{M-1}, \vlambda)$ both depend on $\vlambda$. The first one can be expanded as follows:
\begin{align}
  p(\vy_{M-1}|\vtheta, \vlambda) = \prod_{t=0}^{M-1} p(y_{t}|\vx_{t}, \vy_{t-1}, \vtheta) \; ,
\end{align}
and the dependence on the controls $\vx_{t}$ is a dependence on the trainable variables of the agent $\vlambda$. The second term $\hvtheta(\vy_{M-1}, \vlambda)$ depends on $\vlambda$ through the PF weights, which are updated with the Bayes rule \cref{eq:bayes_rule_w}, that features the term $p(y_{t+1}|\vx_{t+1}, \vy_{t}, \vtheta)$, where again the controls $\vx_{t+1}$ are $\vlambda$-dependent. The complete gradient of the right-hand term of \cref{eq:final_mse_exp} reads therefore
\begin{align}
  \int \frac{\dd}{\dd \vlambda} \ell (\hvtheta(\vy_{M-1}, \vlambda), \vtheta) p(\vy_{M-1}|\vtheta, \vlambda) \dd \vy_{M-1} + \int \ell (\hvtheta(\vy_{M-1}, \vlambda), \vtheta) \frac{\dd p(\vy_{M-1}|\vtheta, \vlambda)}{\dd \vlambda} \dd \vy_{M-1}\; .
\end{align}
The first term is in the form of an expectation value and can be straightforwardly approximated in a Monte Carlo simulation. It corresponds exactly to the naïve gradient of the loss in \cref{eq:gradient_first_attempt} computed by the AD framework. The second term can be written as
\begin{align}
  \int \ell (\hvtheta(\vy_{M-1}, \vlambda), \vtheta) \frac{\dd \log p(\vy_{M-1}|\vtheta, \vlambda)}{\dd \vlambda} p(\vy_{M-1}|\vtheta, \vlambda) \dd \vy_{M-1} \; ,
\end{align}
which is now in the form of an expectation value on the trajectories of the simulation and can be evaluated simultaneously with the first term, provided we keep track of $\log p(\vy_{M-1}|\vtheta, \vlambda)$. This second contribution to the gradient can be approximated as
\begin{align}
  \frac{1}{B} \sum_{k=1}^B \ell (\omega_k, \vlambda) \frac{\dd \log p(\vy_{k, M-1}|\vtheta_k, \vlambda)}{\dd \vlambda} \; ,
  \label{eq:second_term_gradient}
\end{align}
on a batch of $B$ simulations. The term $\log p(\vy_{k, M-1}|\vtheta_k, \vlambda)$ is the sum
\begin{align}
  \log p(\vy_{k, M-1}|\vtheta_k, \vlambda) = \sum_{t=0}^{M-1} \log p(y_{k, t}|\vy_{k, t-1}, \vtheta_k, \vlambda) \; ,
\end{align}
where we exchanged the dependence on $\vx_{k, t}$ of the factors $p(y_{k, t}|\vx_{k, t}, \vy_{k, t-1}, \vtheta_k)$ for the dependence on $\vlambda$. This logarithm can be accumulated step by step in the simulation, after the extraction of each measurement outcome. Notice that for $B$ simulations in the batch, we have to compute $B$ cumulated probabilities, because each trajectory is different. In conclusion, the total gradient is
\begin{align}
  \frac{1}{B} \sum_{k=1}^B \left[ \frac{\dd}{\dd \vlambda} \ell (\omega_k, \vlambda) + \ell (\omega_k, \vlambda) \frac{\dd \log p(\vy_{k, M-1}|\vtheta_k, \vlambda)}{\dd \vlambda} \right] \; .
\end{align}
By introducing the \textit{stop gradient} operation we can write it in the convenient form
\begin{align}
  \frac{1}{B} \frac{\dd}{\dd \vlambda} \sum_{k=1}^B \lbrace \ell (\omega_k, \vlambda) + \sg{\ell (\omega_k, \vlambda), \vtheta)} \log p(\vy_{k, M-1}|\vtheta_k, \vlambda) \rbrace \; ,
  \label{eq:final_gradient}
\end{align}
that requires only one gradient, which makes it more straightforward to implement in the AD framework. In this formula we are using the stop gradient operator $\text{sg} \left[ \cdot \right]$, which is an instruction that tells the automatic differentiation frameworks not to compute the derivatives of the expression inside the operator. We are therefore naturally led to introducing the modified loss $\widetilde{\mathcal{L}}(\vlambda)$, i.e.
\begin{align}
  \widetilde{\mathcal{L}} (\vlambda) := \frac{1}{B} \sum_{k=1}^B \widetilde{\ell} (\omega_k, \vlambda) \; ,
  \label{eq:modified_loss}
\end{align}
with
\begin{align}
  \widetilde{\ell} (\omega_k, \vlambda) := \ell (\omega_k, \vlambda)+\sg{\ell (\omega_k, \vlambda)} \log p(\vy_{k, M-1}|\vtheta_k, \vlambda) \; .
  \label{eq:modified_ell}
\end{align}
which is the correct function to be minimized. For a resource limited estimation the modified loss is the average of the $B'$ simulations that have terminated. In \cref{subsec:tuning_hyperparameters} we comment on the gradient descent process, on the choice of the hyper-parameters, and the typical behaviour of the loss in the training. The second term of the gradient in \cref{eq:final_gradient} is similar to the loss of the Policy Gradient method in reinforcement learning~\cite{sutton_policy_1999}, where the probabilities arise because of the stochastic extraction of the policy, while here the NN produces directly the action and the stochasticity comes from the measurement outcome extraction. The necessity of introducing such terms when dealing with the gradient of expressions involving non-reparametrizable random variables has been known in the machine learning literature for a while~\cite{farquhar_loaded_2019} and the expressions involving stop-gradient operators go under the name of surrogate expressions. However, the first time this has appeared in the physics literature is in~\cite{porotti_gradient-ascent_2023}, applied to quantum feedback. Had we neglected the log-likelihood term of \cref{eq:final_gradient} we would have introduced a bias in the gradient. Not adding the log-likelihood terms means not only a slower convergence in the training but possibly also converging to a worse minimum.

\subsubsection*{Log-likelihood terms in the loss}
%
We can generalize and say that whatever extraction of random variables we perform during the simulation, we need to add a corresponding log-likelihood term, but only if the probability distribution from which they have been extracted depends on $\vlambda$, implicitly or explicitly. With growing number of extracted variables the variance of the gradient grows and if the batchsize is too small this can severely affect the training, then we might loose convergence and end up in a bad local minimum. If possible we advice to reparametrize the random variables and account for the backpropagation of the gradient in a more direct way.  Depending on the quantum sensor we are simulating we might be able to implement the extraction of the measurement outcomes through a differentiable reparametrization. If we can write the measurement outcome $y_t$ as
\begin{equation}
	y_t = g(u_t, x_t, \vtheta) \; ,
\end{equation}
where $u_t$ is a random variable extracted from a probability distribution independent on $\vlambda$, i.e. $\frac{\dd}{\dd \vlambda} p(u_t) = 0$, then we can omit the corresponding log-likelihood term in \cref{eq:final_gradient}. The gradient propagates now directly through the measurement outcome, i.e.
\begin{equation}
	\frac{\dd}{\dd \vlambda} y_{t} \neq 0 \; ,
\end{equation}
and we can differentiate the loss in \cref{eq:loss_omega_k} as it is. The log-likelihood term would be $\log p(u_t)$, which is independent on $\vlambda$. The reparametrization can however be applied only to continuous variables, see~\cite{porotti_gradient-ascent_2023} for more details. In \cref{sec:diff_pf_generic} we apply this technique in the resampling step of the particle filter.

\subsubsection*{Adding a baseline}
%
We can also try to add a baseline to \cref{eq:final_gradient}, as suggested in~\cite{weaver_optimal_2001} for RL with Policy Gradient. This means modifying the gradient to
\begin{equation}
	\frac{1}{B}\frac{\dd}{\dd \vlambda}  \sum_{k=1}^B \left[ \ell (\omega_k, \vlambda) + \sg{\ell (\omega_k, \vlambda) - \mathcal{B}} \log p(\vy_{k, M-1}|\vtheta_k, \vlambda) \right] \; ,
\end{equation}
with the standard choice for the baseline $\mathcal{B}$ being
\begin{equation}
	\mathcal{B} := \frac{1}{B} \sum_{k=1}^B \ell (\omega_k, \vlambda) \; ,
	\label{eq:loss_with_baseline}
\end{equation}
that is, inside the stop gradient we subtract to each loss in the batch the mean value of the loss on the batch. It is important for $\mathcal{B}$ to be a constant across the simulations indexed by $k$. We briefly see in the following that the introduction of $\mathcal{B}$ doesn't change the expected value gradient, while it can be proved that it reduces the variance of the gradient~\cite{weaver_optimal_2001}. Consider the following chain of equalities
\begin{equation}
	0 =	\frac{\dd}{\dd \vlambda} \int p(\vy_{M-1}|\vtheta, \vlambda) \dd \vy_{M-1} = \int \frac{1}{p(\vy_{M-1}|\vtheta, \vlambda)} \frac{\dd p(\vy_{M-1}|\vtheta, \vlambda)}{\dd \vlambda} p(\vy_{M-1}|\vtheta, \vlambda) \dd \vy_{M-1} \; ,
	\label{eq:exp_value_log}
\end{equation}
where the first one comes from the normalization of $p(\vy_{M-1}|\vtheta_k, \vlambda)$ and in the second we divided and multiplied for $p(\vy_{M-1}|\vtheta_k, \vlambda)$ upon swapping the integral and the derivate. The rightmost term of \cref{eq:exp_value_log} is now in the form of an expectation value, that can be approximated by the following summation in the simulation:
\begin{equation}
	\int \frac{1}{p(\vy_{M-1}|\vtheta, \vlambda)} \frac{\dd p(\vy_{M-1}|\vtheta, \vlambda)}{\dd \vlambda} p(\vy_{M-1}|\vtheta, \vlambda) \dd \vy_{M-1} \simeq \frac{1}{B} \sum_{k=1}^B \frac{1}{p(\vy_{k, M-1}|\vtheta_k, \vlambda)} \frac{\dd p(\vy_{k, M-1}|\vtheta_k, \vlambda)}{\dd \vlambda} \; .
\end{equation}
Where the term in the right-hand summation is the derivative of the log-likelihood, so that we expect
\begin{equation}
	\frac{\dd}{\dd \vlambda} \sum_{k=1}^B \log p(\vy_{k, M-1}|\vtheta_k, \vlambda) \simeq 0 \; ,
	\label{eq:sum_log_null}
\end{equation}
for large $B$. Adding the baseline in \cref{eq:loss_with_baseline}, means adding a terms proportional to the derivative of the log-likelihood, with the proportionality constant being $\mathcal{B}$, which has null expectation value.

\subsubsection*{Discrete control space}
%
In this section we briefly comment on the case in which the control space is discrete, that is, $x_t$ can be chosen only among finitely many elements, i.e $x_t \in \chi = \lbrace x_1, x_2, \ldots, x_R \rbrace$. This happens for example in the experiment presented in~\cite{cimini_experimental_2023}, where the control parameter was the topological charge of the q-plate. In this case the agent produces a probability distribution on the set $\chi$ as outcome, just like in Policy Learning, and a random $x_t$ is extracted from this categorical distribution. In this scenario we need to revisit the \cref{eq:simplified_probabilities_1} and \cref{eq:simplified_probabilities_2}, that now need to accommodate also for the probability of extracting a particular $x_{t}$:
\begin{equation}
	 P (\vx_{M-1}, \vy_{M-1}|\vtheta) = \prod_{t=0}^{M-1} p(y_t|\vx_t, \vy_{t-1}, \vtheta) g(x_t|\vy_{t-1}, \vx_{t-1}, \vlambda) \\
	\label{eq:simplified_probabilities_x_stoc_1}
\end{equation}
\begin{equation}
	 P(\hvtheta|\vx_{M-1}, \vy_{M-1}) = \delta (\hvtheta - g_2 (\vy_{M-1}, \vx_{M-1})) \; .
	\label{eq:simplified_probabilities_x_stoc_2}
\end{equation}
Substituting this expressions in \cref{eq:mse_no_theta} we get
\begin{equation}
	\Delta^2 \hvtheta = \int \ell (\hvtheta(\vy_{M-1}, \vx_{M-1}), \vtheta) \prod_{t=0}^{M-1} p(y_t|\vx_t, \vy_{t-1}, \vtheta) g(x_t|\vy_{t-1}, \vx_{t-1}, \vlambda) \dd \vx_{M-1} \dd \vy_{M-1} \; .
	\label{eq:loss_stoch_x}
\end{equation}
By repeating the derivation of the loss, considering that $p(y_t|\vx_t, \vy_{t-1}, \vtheta)$ doesn't depend on $\vlambda$ anymore, the log-likelihood term of \cref{eq:final_gradient} becomes
\begin{equation}
	 \sum_{t=0}^{M-1} \sg{\ell (\omega_k, \vlambda)} \log g(x_{k, t}|\vy_{t-1, k}, \vx_{t-1, k}, \vlambda) \; .
\end{equation}

\subsubsection*{Stochastic estimator}
%
In all the applications of this work the estimator $\hvtheta$ is always  a deterministic function of the PF ensemble. Even for the case of values discrimination, this is computed as the most likely hypothesis at the end of the experiment, which doesn't require any sampling. A perfectly valid estimator for $\theta$ would be one sample extracted from the Bayesian posterior $P(\vtheta|\vy_{M-1}, \vlambda)$. Doing so requires adding a term to the log-likelihood term in the loss in \cref{eq:final_gradient}, which now becomes
\begin{equation}
	\sum_{t=0}^{M-1} \sg{\ell (\omega_k, \vlambda)} \left[ \log p(\vy_{k, M-1}|\vtheta_k, \vlambda) + \log P(\hvtheta_k|\vy_{k, M-1}, \vlambda) \right] \; .
\end{equation}
A differentiable expression for the posterior distribution at $\hvtheta$ might not be accessible if the parameters are continuous, but if they are discrete, the term $P(\hvtheta|\vy_{M-1}, \vlambda)$ is just the weight corresponding to the discrete values $\hvtheta$.

\subsection{Definition of the cumulative and logarithmic losses}
\label{subsec:cum_losses}
When doing a simulation for a certain $M$, if we want the result of the training to give us the optimal strategy also for $M_2 < M$ we can introduce the cumulative loss, that also takes into account the loss at intermediate steps. A naïve approach is to extend the MSE to all steps between $t=0$ and $t=M-1$, and write
\begin{equation}
	\mathcal{L}\ped{cum}(\vlambda) := \frac{1}{MB} \sum_{t=0}^{M-1} \sum_{k=1}^B \ell (\hvtheta_{k, t}, \vtheta_k) \; ,
	\label{eq:all_time_loss}
\end{equation}
where $\hvtheta_{k, t}$ is the estimator at step $t$ of the $k$-th simulation. With this loss the agent is incentivized to make the estimator $\hvtheta$ converge to the value $\vtheta$ as soon as possible. However the error on the first time steps of the estimation dominates the later errors in the summation, and this puts pressure on the agent to optimize the first steps of the procedure at the expense of the later precision. To solve this problem we divide each terms in the sum \cref{eq:all_time_loss} by a function $\eta (\vtheta, t)$, i.e.
\begin{equation}
	\mathcal{L}\ped{cum}(\vlambda) := \frac{1}{MB} \sum_{t=0}^{M-1} \sum_{k=1}^B \frac{\ell (\hvtheta_{k, t}, \vtheta_k)}{\eta (\vtheta_k, t)} \; ,
	\label{eq:all_time_loss_h}
\end{equation}
where $\eta (\vtheta_k, t)$ is the expected precision of the estimation at step $t$ given the true value $\vtheta_k$, or an approximation to it, in the form of a lower bound for example, like the Cramér-Rao bound. This new loss measures the relative variation of the error from the reference value. Even if $\eta (\vtheta_k, t)$ is a rigorous lower bound on the MSE we can't expect the inequality
\begin{equation}
	\ell (\hvtheta_{k, t}, \vtheta_k) \ge \eta(\vtheta_k, t) \; ,
\end{equation}
to hold exactly for every $t$ and $k$, as there will be fluctuations due to the finite batchsize. From the practical point of view this means that it is possible for the loss of some training steps to be $\mathcal{L} (\vlambda) < 1$, which doesn't necessarily point toward a bug in the implementation of the training. With \cref{eq:all_time_loss_h} we still incentivise the agent to be as fast as possible in reaching a good precision, and not wait until the end, because then it will be rewarded by the reduced loss for all the duration of the experiment. Another possibility to account fairly for the MSE at intermediate times is to take the logarithm of the mean error on the batch and write the cumulative loss as
\begin{equation}
	\mathcal{L}\ped{log}(\vlambda) := \frac{1}{M} \sum_{t=0}^{M-1} \log \left[ \frac{1}{B} \sum_{k=1}^B \ell (\hvtheta_{k, t}, \vtheta_k) \right] \; .
	\label{eq:log_loss}
\end{equation}
The advantage of this approach is that it doesn't require any prior known reference value for the error. Notice that this loss is not in the form of an expectation value of $\ell (\hvtheta, \vtheta)$ over a batch.

\subsection{Cumulative and logarithmic losses for a resource limited estimation}
In this section we comment on the form taken by the cumulative and logarithmic losses in the case of a limited number of resources. Given $M'$ the realized number of iterations of the measurement loop \cref{eq:all_time_loss_h} becomes
\begin{equation}
	\mathcal{L}\ped{cum}(\vlambda) := \frac{1}{M'B} \sum_{t=0}^{M'-1} \sum_{k=1}^B \frac{\ell (\hvtheta_{k, t}, \vtheta_k)}{\eta (\vtheta_k, t)} \; ,
	\label{eq:all_time_loss_h_resources}
\end{equation}
notice, that at difference with \cref{eq:mse_montecarlo_resources} all the simulations in the batch are considerer, not only those $B'$ that were already ended as the measurement loop stopped. If one estimation in the batch is ended prematurely with respect to all the other, with all the resources consumed to obtain a bad estimator for $\vtheta$ this will have a huge weight in the loss, since the squared error will appear multiple times, until all the other estimations are ended. This means that an unwise use of the resources, which are consumed early to reach a poor result will be strongly penalized. One may think, that since the number of iterations $M'$ is stochastic then teh cumulative loss is a form of ``existential loss'' which would put pressure on th agent to terminate with the smallest number of measurement step possible, this would be at odd with the actual goal of optimizing with fixed resources irrespective of the number of measurements, but indeed the loss is normalized according to $M'$, so that having a short or a long cycle doesn't matter for the computation of $\mathcal{L} (\vlambda)$. Similarly to the cumulative loss, the logarithmic loss for an estimation with a limited number of resources can be expressed as
\begin{equation}
	\mathcal{L}\ped{log}(\vlambda) := \frac{1}{M'} \sum_{t=0}^{M'-1} \log \left[ \frac{1}{B} \sum_{k=1}^B \ell (\hvtheta_{k, t}, \vtheta_k) \right] \; ,
	\label{eq:log_loss_resources}
\end{equation}
where again $M'$ is the actual number of executed iterations of the loop.

\subsection{Gradients of the cumulative and logarithmic losses}
In this section we comment on the expression of the gradient of the cumulative and logarithmic losses, and of the role of the log-likelihood terms that we had inserted in \cref{eq:modified_ell}. The modified cumulative loss, from which the AD framework can directly compute the gradient, reads
\begin{equation}
	\widetilde{\mathcal{L}}\ped{cum}(\vlambda) :=  \frac{1}{MB}\sum_{k=1}^B \Big \lbrace \sum_{t=0}^{M-1} \ell (\hvtheta_{k, t}, \vtheta_k) + \sg{\sum_{t=0}^{M-1} \ell (\hvtheta_{k, t}, \vtheta_k)} \sum_{t=0}^{M-1} \log p(y_{k, t}|\vtheta_k, \vy_{t-1, k}, \vlambda) \Big \rbrace \; .
	\label{eq:all_time_loss_loglike}
\end{equation}
Given that the stop gradient operator is linear, we now make the important observation that the gradient of the log-likelihood terms in the form
\begin{equation}
	 \sg{\ell (\hvtheta_{\alpha, k}, \vtheta_k)} \log p(y_{\beta, k}|\vtheta_k, \vy_{\beta-1, k}, \vlambda) \; .
\end{equation}
with $\beta > \alpha$ have null expectation value on the batch of simulations, that is
\begin{equation}
	\frac{1}{B} \sum_{k=1}^B \ell (\hvtheta_{\alpha, k}, \vtheta_k) \frac{\dd \log p(y_{\beta, k}|\vtheta_k, \vy_{\beta-1, k}, \vlambda)}{\dd \vlambda} \simeq 0 \, ,
	\label{eq:exp_value_alphabeta}
\end{equation}
The expression in \cref{eq:exp_value_alphabeta} is an approximation of the true expectation value
\begin{equation}
	\int \ell (\hvtheta_{\alpha}, \vtheta) \frac{\dd \log p(y_\beta|\vtheta, \vy_{\beta-1}, \vlambda)}{\dd \vlambda}  \pi(\vtheta) \\d \vtheta \prod_{t=0}^{\beta}  p(y_t|\vtheta, \vy_{t-1}, \vlambda) \dd y_{t} \; .
\end{equation}
All the integral for $y_t$ for $t>\beta$ can be simplified in the above formula, since the integrand doesn't depend on these variables. Let us first solve the integral for $\dd y_\beta$. The loss term doesn't depend on this variable, so that we can pull it out of the integral and write
\begin{equation}
	\ell (\hvtheta_{\alpha}, \vtheta) \int \frac{\dd \log p(y_\beta|\vtheta_k, \vy_{\beta-1}, \vlambda)}{\dd \vlambda} p(y_\beta|\vtheta, \vy_{t-1}, \vlambda) \dd y_{\beta} \; ,
\end{equation}
which is equal to
\begin{equation}
	\int \frac{\dd  p(y_\beta|\vtheta, \vy_{t-1}, \vlambda)}{\dd \vlambda} \dd y_{\beta} \\ = \frac{\dd}{\dd \vlambda} \int p(y_\beta|\vtheta, \vy_{t-1}, \vlambda) \dd y_\beta = 0 \; .
\end{equation}
Since the summation \cref{eq:exp_value_alphabeta} tends to zero for large $B$, we can write the loss as following
\begin{equation}
	\widetilde{\mathcal{L}}\ped{cum}(\vlambda) :=  \frac{1}{MB}\sum_{k=1}^B \Big \lbrace \sum_{t=0}^{M-1} \ell (\hvtheta_{k, t}, \vtheta_k) + \sum_{t=0}^{M-1} \sg{\ell (\hvtheta_{k, t}, \vtheta_k)} \log p(\vy_{k, t}|\vtheta_k, \vlambda) \Big \rbrace \; .
	\label{eq:all_time_loss_loglike_reduced}
\end{equation}
which is the expression implemented in the library. Since we have removed some of the stochastic terms in the loss, which average to zero, but nevertheless contribute to the fluctuations, using expression \cref{eq:all_time_loss_loglike_reduced} we expect to have reduced the variance of the gradient, just like we did with the correction of \'Scibior and Wood for the particle resampling. From this derivation we learn that in general the log-likelihood terms of variables extracted in the future with respect to the terms they multiply can be simplified. Notice that in this derivation we haven't assumed projective measurements, that would have meant $p(y_t|\vtheta_k, \vy_{t-1}, \vlambda) = p(y_t|\vtheta_k, \vlambda)$, instead our derivation works in the most general case of a weakly measured probe. We now turn to the gradient of the logarithm loss of \cref{eq:log_loss_resources}. This is somewhat different from the previous cases since now the mean on the batch is inside the logarithm. The expectation value of the loss is
\begin{equation}
	\frac{1}{M} \sum_{t=0}^{M-1} \log \left[ \int \ell (\hvtheta_{t}, \vtheta) p(\vy_{t}|\vtheta, \vlambda) \dd \vy_t \right] \; ,
	\label{eq:log_loss_exp}
\end{equation}
which has the following gradient
\begin{equation}
	\frac{1}{M} \sum_{t=0}^{M-1} \frac{\int \frac{\dd}{\dd \vlambda}  \ell (\hvtheta_{t}, \vtheta) p(\vy_{t}|\vtheta, \vlambda) \dd \vy_t +  \int  \ell (\hvtheta_{t}, \vtheta) \frac{ \dd \log p(\vy_{t}|\vtheta, \vlambda)}{\dd \vlambda} p(\vy_{t}|\vtheta, \vlambda) \dd \vy_t}{\int \ell (\hvtheta_{t}, \vtheta) p(\vy_{t}|\vtheta, \vlambda) \dd y_t}  \; .
	\label{eq:log_loss_exp_gradient}
\end{equation}
This expression can be obtained on the batch of simulations with the modified loss
\begin{equation}
	\widetilde{\mathcal{L}}\ped{log}(\vlambda) := \frac{1}{M} \sum_{t=0}^{M-1} \frac{\sum_{k=1}^B \ell (\hvtheta_{k, t}, \vtheta_k)+ \sum_{k=1}^B \sg{\ell (\hvtheta_{k, t}, \vtheta_k)} \log p(\vy_{k, t}|\vtheta_k, \vlambda)}{\sg{\sum_{k=1}^B \ell (\hvtheta_{k, t}, \vtheta_k)}} \; .
\end{equation}
To get the results for the resources limited estimation we substitute $M$ with $M'$ in the whole section.

\section{Details on the simulations}
\label{sec:miscellanea}

\subsection{Tuning of the hyperparameters}
\label{subsec:tuning_hyperparameters}
\begin{figure}%
	\centering
	\subfloat[\centering Three phases of the training.]{{\includegraphics[width=8cm]{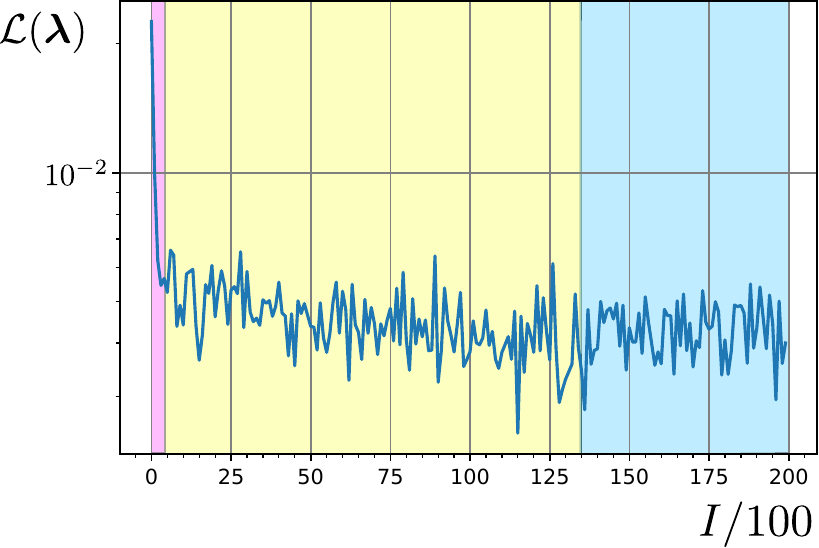} }}%
	\qquad
	\subfloat[\centering Implementations of the trainable variables.]{{\includesvg[width=8cm]{variables_lambdas.svg} }}%
	\caption{On the left picture the three phases of the learning process explained in \cref{subsec:tuning_hyperparameters} are plotted. One tick on the $x$-axis corresponds to $500$ update steps. The agent first learns the rough form of the optimal strategy and later the fine details, before converging. On the right the three agents used in this work are represented, i.e a NN, a ternary decision tree, and a table containing the values of the controls $\vx$, indexed by the measurement step $t$.}%
	\label{fig:three_phases_variables}%
\end{figure}

We mentioned in the main text that the update of the agent's trainable variables is not actually done through \cref{eq:simple_sgd_update}, but via a more sophisticated optimizator called Adam. We observed empirically that a decaying learning rate is beneficial when using the Adam optimizer. This is because the agent first learns the rough features of the optimal solution with a relatively large update step for the variables. Subsequently, with a smaller learning rate, the solution is fine tuned. The Adam optimized, however, already has an internal adaptive update step that is different for every variables, therefore the learning rate should be really only understood as a broad indication of the training speed given to the optimizer. In the original Adam paper~\cite{kingma_adam_2015} the authors consider a learning rate decaying with the inverse square root of the number of update steps. This was also our choice. Let us define $i = 1, 2, \cdots, I$ the index of the update step in the training process, then the learning rate at the $i$-th iteration of the gradient descent is
\begin{equation}
	\alpha_i := \frac{\alpha_0}{\sqrt{i}} \; .
	\label{eq:learning_rate_decrease}
\end{equation}
We observe empirically, that the initial value of the learning rate $\alpha_0$ for a NN should depend on the batchsize $B$. For $B \sim \mathcal{O}(10^3)$ we use $\alpha_0 \sim \mathcal{O} (10^{-2})$, while for $B \sim \mathcal{O}(10^2)$ a value of $\alpha_0 \sim \mathcal{O} (10^{-3})$ is more appropriate. For the non-adaptive strategy we use an initial learning rate that is one order of magnitude larger than the one used for the NN at equal batchsize. The minimum number of training steps $I$ depend strongly on the application, but we observed in all our examples that it should of order $I \sim \mathcal{O}(10^3-10^4)$ to reach convergence. We observed some universal feature in the behaviour of the loss as the training proceeds, which can be associated to three different phases in the training, see \cref{fig:three_phases_variables}. First we have an initial phase of fast learning, which is the shortest one, coloured in pink, followed by the fine tuning phase in yellow and the plateau at the end, with the loss remaining on average constant. As a final note, we mention that when the resampling routine is active we might expect a slow-down of the simulation speed as the training session proceeds. This happens because as the agent is perfected and the loss is reduced, it is more probable that a resampling event is triggered (because the increasing precision means also more concentrated weights in the PF ensemble), which slows down the simulation. In other words, the amount of code that has to be executed in a run is not fixed \textit{a priori}, but depends dynamically on the resampling condition that is checked at run-time.
To end the section we briefly recap the three possible implementations that the trainable variables $\vlambda$ have taken in this work, see \cref{fig:three_phases_variables}. In the case the agent is a NN $\vlambda$ are the weights and the biases of the network When the agent is a decision tree, the controls $x_t$ are associated to each node of the tree, and they are the $\vlambda$ variables. Finally, for a non-adaptive strategy, there is no adaptivity and the controls $\vx$ are codified in a list, indexed by the measurement step $t$. In this case the controls and the training variables coincide, i.e. $\vlambda = \vx_{M-1}$.

\subsection{Fit of the precision}
\label{subsec:fit_precision}
\begin{figure}
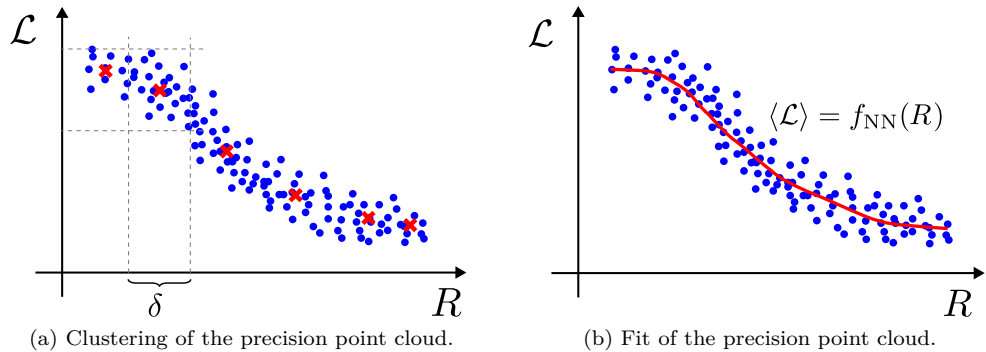
%
	\centering
	\subfloat[\centering Clustering of the precision point cloud.]{{\includesvg[width=6cm]{precision_average.svg} }}%
	\qquad
	\subfloat[\centering Fit of the precision point cloud.]{{\includesvg[width=6cm]{precision_nn.svg} }}%
	\caption{On the left side we represent the precision plot where the cloud of points has been clustered to obtain the red crosses. On the right we use a NN to interpolate and get the average loss at a given value of the resources.}%
	\label{fig:precision_fit}%
\end{figure}

In this section we briefly comment on the way the precision plot are realized throughout the work. The definition of the resources doesn't only impact the stopping condition of the measurement loop, but it defines how the performances of an agent are visualized, since by default we plot the mean loss as a function of the consumed resources. After having trained the agent we simulate many times the estimation and we keep track of the tuples $\mathcal{S} := \lbrace (R_j, \mathcal{L}_j) \rbrace_{j=1}^S$ after each measurement step, containing the consumed resources $R_t$ and the loss $\mathcal{L}$. Since the experiment is a stochastic process we will collect a cloud of points from which a simple relation between the expected precision and the resources must be obtained, see \cref{fig:precision_fit}.
The first possibility is to divide the x-axis of the resources in intervals of size $\delta$, and compute the barycenters of all the points $(R_j, \mathcal{L}_j)$ falling in this interval, these would be the red crosses of \cref{fig:precision_fit}. The second possibility is to fit this cloud of points with a NN. We set the training loss to be the MSE, i.e.
\begin{equation}
	\mathcal{L}\ped{fit} := \frac{1}{S} \sum_{j=1}^S \left[ \mathcal{L}_j - f\ped{NN} (R_j) \right]^2 \; ,
	\label{eq:loss_fit}
\end{equation}
which, for a single value of the resource, i.e. $R_j = R \; \forall j$, would converge to $f\ped{NN} (R) = \frac{1}{S} \sum_{j=1}^S \mathcal{L}_j$, that is a NN will approximate the mean loss. This won't be exactly true for a cloud of points, but with \cref{eq:loss_fit} we incentivise the NN to converge toward the average loss for every value of the resources. All the plots in this paper have been produced with the first method, choosing an appropriate $\delta$, except for those plots on the NV center platform with $T_2^\star = \infty$ and referring to the time-limited estimation. In the PGH line of the first plot in \cref{fig:nvcenter_comparison} there is a non-monotonicity for small $T$, that is a defect in the plot and an artifact on this way of interpolating with a NN.

\subsection{Scaling of the time and memory requirements}
\label{subsec:scaling_resources}
Since the $B$ estimations in a batch can be performed in parallel we will benefit from the use of a GPU or a TPU (Tensor Processing Unit) in the training of the agent. The main difference between the CPU and the GPU is that a CPU has fewer ($\sim \mathcal{O} (10)$) faster cores, while a GPU has many ($\sim \mathcal{O} (10^3)$) slower cores. With a large batchsize the use of hardware acceleration through a GPU will turn out to be essential and we will first examine the resource requirements of model-aware RL assuming that everything that can be parallelized has been is indeed executed in parallel. In this case the time requirement of the simulation is mainly influenced linearly by the number of measurements $M$ in the training loop, that have to be executed necessarily sequentially. The update of the PF and the computation of the distributions moments all require $\mathcal{O}(N)$ multiplications each but can be done in parallel, where $N$ is the number of particles. The memory requirement depends on the batchsize $B$ and the number of particles $N$ in the PF. Nevertheless because of the construction of the gradient, for which we need to keep in memory the results of all the intermediate computations, the number of measurements $M$ has also an almost linear influence on the required memory. Finally, the total time  used in the training is also proportional to the number of update steps $I$. Each update step comprises the complete run of an estimation batch together with the evaluation of the gradient and the update of the controls. The size of the NN has little impact on the training time and memory. We can summarise the above considerations as
\begin{equation}
	\text{Memory} \sim \mathcal{O} (BMN) \;, \quad \; \text{Time}_{\text{Par}} \sim \mathcal{O} (IM) \; .
	\label{eq:scaling_parallel}
\end{equation}
Assuming that nothing can be parallelized (we have a single core) and therefore everything is sequential, if, as usual, the computational time in the CPU is dominated by the number of floating point multiplications, we instead have the time scaling
\begin{equation}
	\text{Time}_{\text{Seq}} \sim \mathcal{O} (IBMN) \; ,
	\label{eq:scaling_sequential}
\end{equation}
while the memory requirement is unchanged. Neither a GPU nor a CPU will perfectly reproduce these theoretical scalings, because the GPU has a limited number of cores, but there is a tendency for a GPU to follow the scaling of $\text{Time}_{\text{Par}}$ and for a CPU $\text{Time}_{\text{Seq}}$. If the batchsize $B$ is very large (or the GPU not very powerful) the simulations in the batch can't all be executed in parallel and $B$ starts to affect also the time requirements. If $B$ and $N$ are small a CPU may complete the training before a GPU, because of the smaller proportionality factor for the time requirement in \cref{eq:scaling_sequential} with respect to \cref{eq:scaling_parallel}, due to the faster cores of the CPU. This analysis applies also to the training of a non-adaptive strategy, which is not resource-saving compared to the training of the NN. In the applications we expect our agent to run on a small controller near the sensor, where most definitely we won't have access to a GPU and lots of memory, which anyway are required only in the training phase. In this situation we have no batch and only one iteration, i.e. $I=B=1$. Furthermore there is no extra $M$ in the memory requirement appearing because of the automatic differentiation, so that the resource scaling in the application will be
\begin{equation}
	\text{Memory} \sim \mathcal{O} (d N+N) \; , \quad \; \text{Time}_{\text{Seq}} \sim \mathcal{O} (M N) \; ,
\end{equation}
where $d$ is the number of parameters. For an estimation limited by the resources instead of the number of measurements, $M$ must be intended as the average number of measurements employed for a fixed amount of resources. In general thanks to the resampling routine we can keep the number of particles fixed while increasing the precision arbitrarily. It is a good practise although to choose $N$ proportional to the number of parameters to estimate, i.e. $N \sim \mathcal{O} (d)$. In the applications the memory requirement of the NN, and the multiplications needed to evaluate it at each step contribute respectively with additional $\mathcal{O} (n_l n_u)$ memory and $\mathcal{O} (n_l n_u^2)$ time (per step), where $n_l$ is the number of layers and $n_u$ the number of units per layer, so that we have
\begin{equation}
	\text{Memory} \sim \mathcal{O} (d N+N+n_l n_u) \; , \quad \; \text{Time}_{\text{Seq}} \sim \mathcal{O} (M N+M n_l n_u^2) \; .
\end{equation}

 If the control is non-adaptive we don't need this extra computations, and if the PF is removed from the picture (because we don't need real time feedback) we have that the memory and time requirements trivialize, i.e they scale respectively as $\mathcal{O} (1)$ and $\mathcal{O} (M)$. Of course the total time of estimation would be influenced also by the time it takes to perform the physical measurement on the probe, but here we are referring only to the computational time.

\section{Optimal strategies for frequentist optimization}
\label{sec:frequentist_optimization}
The qsensoropt library can also optimize the Cramér-Rao bound (based on the Fisher information matrix) for the local estimation of the parameters $\vtheta$. This is frequentist inference instead of Bayesian inference, this last being the main topic of this work. The multiparameter Cramér-Rao bound (CR bound) is a lower bound on the Mean Square Error matrix of the frequentist estimator $\hvtheta$ at the position $\vtheta$, expressed in terms of the FI matrix, i.e.
\begin{equation}
	K := \mathbb{E} \left[ (\hvtheta - \vtheta) (\hvtheta -\vtheta)^\intercal \right] \ge F^{-1} (\vtheta) \; ,
	\label{eq:cr_bound}
\end{equation}
with
\begin{equation}
  F_{ij} (\vtheta) := \mathbb{E}_y \left[ \frac{\partial \log p_{\vtheta} (y)}{\partial \theta_i}  \frac{\partial \log p_{\vtheta} (y)}{\partial \theta_j} \right] \; .
  \label{eq:multiple_parameter_fi}
\end{equation}
This result sets the maximum achievable precision of an estimator around $\vtheta$, in other words, it limits the ability to distinguish reliably two close values $\vtheta$ and $\vtheta+\delta \vtheta$. The expectation value is taken on many realizations of the experiment, i.e. on the probability distribution of the trajectories for the outcomes and the controls. Let us introduce the tuple $\vx$ and $\vy$ containing respectively the entirety of the controls and the outcomes of the measurements done in the experiment, then the FI matrix has the following expression
\begin{equation}
  F_{ij} (\vtheta) := \mathbb{E}_{\vy} \left[ \frac{\partial \log p(\vy|\vx, \vtheta)}{\partial \theta_i} \frac{\partial \log p(\vy|\vx, \vtheta)} {\partial \theta_j} \right] \; ,
\end{equation}
being $p(\vy|\vx, \vtheta)$ the probability of the observed trajectory of outcomes at the point $\vtheta$. Notice that the expectation value is taken on the whole trajectory of the experiment. By contracting \cref{eq:cr_bound} with the weight matrix $G\ge0$ we get the scalar version of the CR bound, i.e.
\begin{equation}
  \text{tr} (G \cdot K) \ge \text{tr} ( G \cdot F^{-1}) := \mathcal{L}(\vlambda) \; ,
\end{equation}
where we have defined the loss to be optimized in the training, as a function of the trainable variables of the agent $\vlambda$. The gradient of the loss is
\begin{equation}
  \frac{\partial \mathcal{L}} {\partial \vlambda} = \text{tr} \left( F^{-1} G F^{-1} \cdot \frac{\partial F}{\partial \vlambda} \right) \; .
\end{equation}
The expectation value in the definition of $F$ is approximated in the simulation by averaging the product of the log-likelihoods derivatives on a batch of estimations, i.e.
\begin{equation}
  F \simeq \widehat{F} = \frac{1}{B}\sum_{k=1}^{B} \frac{\partial \log p(\vy_k|\vx_k, \vtheta)}{\partial \theta_i} \frac{\partial \log p(\vy_k|\vx_k, \vtheta)}{\partial \theta_j} = \frac{1}{B} \sum_{k=1}^B f_k \; .
  \label{eq:def_f_k}
\end{equation}
where $(\vx_k, \vy_k)$ is the trajectory of a particular realization of the experiment in the batch of simulations and $f_k$ is called the observed Fisher information. The unbiased gradient of the loss, that takes into account also the gradient of the probability distribution in the expectation value, can be computed as following
\begin{equation}
  \frac{\partial \mathcal{L}} {\partial \vlambda} \simeq \frac{1}{B} \frac{\partial} {\partial \vlambda} \text{tr} \Big \lbrace \text{sg} \left( \widehat{F}^{-1} G \widehat{F}^{-1} \right) \sum_{k=1}^B \left[ f_k + \text{sg} (f_k) \log p(\vx_k, \vy_k |\vtheta) \right] \Big \rbrace \; .
  \label{eq:grad_cr}
\end{equation}
The $\text{sg} (\cdot)$ is the stop gradient operator, and the probability $p(\vx_k, \vy_k|\vtheta)$ is the likelihood of the particular trajectory, that contains both the probability of the stochastic outcome and that of the control (in case it is stochastically generated). This is the gradient used in the update step of the stochastic gradient descent procedure for the optimization of frequentist estimation. We can also introduce the logarithmic loss, i.e.
\begin{equation}
  \mathcal{L}_{\text{log}} (\vlambda) := \log \text{tr} ( G \cdot F^{-1}) \; ,
  \label{eq:log_loss_cr}
\end{equation}
which is particularly useful to stabilize the training when the FI spans multiple orders of magnitude. If we have a broad prior on $\vtheta$, but we are stile interested in local estimation, we can introduce the average FI, i.e.
\begin{equation}
  \mathcal{F} := \int F (\vtheta) \pi (\vtheta) \dd \vtheta \; ,
\end{equation}
and optimize the loss
\begin{equation}
  \mathcal{L} (\vlambda) := \text{tr} \left[ G \cdot \mathcal{F}^{-1} \right] \le \int \text{tr} \left[ G \cdot F^{-1} (\vtheta) \right] \dd \vtheta \; ,
\end{equation}
which is a lower bound on the expectation value of the CR bound, because of the Jensen inequality applied to the matrix inverse. In the case of a single parameter the training would maximize the expected value of the Fisher information on the prior $\pi(\theta)$. It is possible to use a custom distribution $\widetilde{p}(y|x, \vtheta)$ for extracting the measurements outcomes instead of $p(y|x, \vtheta)$ by using importance sampling. In this case the FI matrix is computed as
\begin{equation}
    F_{ij} (\vtheta) = \mathbb{E}_{\widetilde{p}} \left[ \frac{\partial \log p(\vy|\vx, \vtheta)}{\partial \theta_i} \frac{\partial  \log p(\vy|\vx, \vtheta)}{\partial \theta_j} \cdot \frac{p(\vy|\vx, \vtheta)} {\widetilde{p}(\vy|\vx, \vtheta,)} \right] \; ,
\end{equation}
which can be approximated on a batch as
\begin{equation}
    F \simeq \frac{1}{B} \sum_{k=1}^B f_k \frac{p(\vy_k|\vx_k, \vtheta)} {\widetilde{p}(\vy_k|\vx_k, \vtheta)} \; ,
\end{equation}
with $f_k$ defined in \cref{eq:def_f_k}. Also the gradient of $F$ is changed accordingly. Typically the distribution $\widetilde{p}$ is some perturbation of $p$, for example it can be obtained by mixing $p$ with a uniform distribution on the outcomes. The importance sampling is useful when some trajectories have vanishingly small probability of occurring according to the model $p$ but contribute significantly to the Fisher information. If these trajectories have some probability of occurring sampling with $\widetilde{p}$, then the estimator of the FI might be more accurate. The drawback is that the perturbed probability of the complete trajectory $\widetilde{p}(\vy|\vx, \vtheta)$ might be too different from the real probability (because of the accumulation of the perturbation at each step), so that the simulation might entirely miss the region in the space of trajectories in which the system state moves, thus delivering a bad estimate of the FI and a bad control strategy, upon completion of the training. Whether or not the importance sampling can be beneficial to the optimization should be checked case by case. The derivative with respect to $\vtheta$ in the definition of $f_k$ in \cref{eq:def_f_k} are computed through automatic differentiation. This means there there are two nested automatic differentiation environments, one for the parameter and one for the training variables of the agent.

\section{Precision lower bounds of the examples}
\label{sec:lower_bounds}
In this section we apply the Bayesian Cramér-Rao bound to the estimation of various parameters on the NV center platform. This bounds will be based on the Fisher information~\cite{fisher_design_1935}, which we briefly define in the following. Refer to the literature for more details. Consider a stochastic variable $y$, which is extracted from a probability distribution $p_{\theta} (y)$, where $\theta$ is a parameter we want to estimate. This is a model for an experiment leading to a stochastic outcome. The information on $\theta$ available from $y$ can be measured by the Fisher information (FI), defined as
\begin{equation}
	I(\theta) := \mathbb{E}_y \left[ \left( \frac{\partial \log p_\theta (y)}{\partial \theta} \right)^2 \right] \; ,
	\label{eq:single_parameter_fi}
\end{equation}
where the expectation value is taken over the distribution $p_{\theta} (y)$.  If the experiment allows to be controlled through the parameter $x$, then the outcome probability is $p_\theta(y|x)$ and the FI inherits such dependence, i.e. we write $I(\theta|x)$. If the control parameter $x$ is computed from a strategy $h$, which could be the Particle Guess Heuristic the a neural network, then we indicate it explicitly in the control $x_h$.

\subsection{Bayesian Cramér-Rao bound}
\label{subsec:bcrb}
Given $\theta$ a single parameter to estimate, we call $I(\theta|\vx_h)$ the Fisher information of a sequence of measurements with controls $\vx_h = (x_0^h, x_1^h, \cdots, x_{M-1}^h)$, which are computed from a strategy $h$. The quantity $I(\theta|\vx_h)$, together with the Fisher information of the prior $\pi(\theta)$, i.e. $I(\pi)$, defines a lower bound on the precision $\Delta^2 \widehat{\theta}$ of whatever estimator $\widehat{\theta}$, that contains the expectation value of $I(\theta|\vx_h)$ on $\pi(\theta)$, and is optimized on the strategy $h$. This lower bounds reads
\begin{equation}
	\Delta^2 \widehat{\theta} \ge \frac{1}{\sup_{h} \mathbb{E}_{\theta} \left[ I(\theta|\vx_h) \right] + I(\pi)} \; .
\end{equation}
This definition appears in the work of Fiderer \textit{et al.}~\cite{fiderer_neural-network_2021}. For the NV centers the controls are the evolution time $\tau$ and the phase $\varphi$, this last however doesn't play any role in the computation of the lower bound, and it will be omitted in the following. The Fisher information of a sequence of measurements is always additive, even if the quantum probe is only measured weakly, but in dealing with projective measurements, as it is the case for NV center, the advantage is that the measurements are uncorrelated, and the same expression for the Fisher information applies to all of them, independently on the results of the previous measurements, i.e.
\begin{equation}
	I(\theta|\boldsymbol{\tau}) = \sum_{t=1}^M I(\theta|\tau_t) \le M \sup_{\tau} I(\theta|\tau) \; ,
\end{equation}
where $M$ is the total number of measurements. The optimization of the single measurement FI gives directly the precision bound for the measurement-limited estimation:
\begin{equation}
	\Delta^2 \widehat{\theta} \ge \frac{1}{\sup_{h} \mathbb{E}_{\theta} \left[ I(\theta|\vtau_h) \right] + I(\pi)}
	\ge \frac{1}{M \, \mathbb{E}_{\theta} \left[ \sup_{\tau} I(\theta|\tau) \right] + I(\pi)} \; .
	\label{eq:lower_meas}
\end{equation}
If the total evolution time is the limiting resource, then, the expression for the total FI is
\begin{equation}
	I(\theta|\boldsymbol{\tau}) = T \sum_{t=1}^M \frac{\tau_t}{T} \left[ \frac {I(\theta|\tau_t)}{\tau_t} \right] \le T \sup_{\tau} \frac{I(\theta|\tau)}{\tau} \; ,
\end{equation}
with $\sum_{t=1}^M \tau_t = T$. In this expression the total FI is the weighted sum of the renormalized FI of each measurement, i.e. $\frac{I(\theta|\tau_t)}{\tau_t}$, and can be manifestly upper bounded by concentrating all the weights on the supremum of the renormalized FI. This gives the lower bound for the precision of the time-limited estimation:
\begin{equation}
	\Delta^2 \widehat{\theta} \ge \frac{1}{\sup_{h} \mathbb{E}_{\theta} \left[ I(\theta|\vtau_h) \right] + I(\pi)}
	\ge \frac{1}{T \, \mathbb{E}_{\theta} \left[ \sup_{\tau} \frac{I(\theta|\tau)}{\tau} \right] + I(\pi)} \; .
	\label{eq:lower_time}
\end{equation}
In the following we will apply this general observations to the derivation of the numerical bounds for DC magnetometry.

\subsection{Evaluation of the Fisher information}
\label{subsec:eval_fisher_bound}
Since the measurement outcome in the NV center is binary, we can compute the Fisher information for a parameter $\theta$, given the control $\tau$, as
\begin{equation}
	I(\theta|\tau) = \mathbb{E} \left[ \left( \frac{\partial \log p ({\pm 1}|\theta, \tau)} {\partial \omega} \right)^2 \right] = \frac{\left( \frac{\partial p}{\partial \theta} \right)^2}{p(1-p)} = \frac{\left( \frac{\partial p}{\partial \theta} \right)^2}{\frac{1}{4}-(p-\frac{1}{2})^2} \; ,
	\label{eq:fisher_simplified}
\end{equation}
where we have used the definition in \cref{eq:single_parameter_fi}, and where $p := p(+1|\theta, \tau)$. For example, for a decoherence free estimation of the precession frequency $\omega$ we have $p:=\cos^2 \left( \frac{\omega \tau}{2} \right)$, from which $\frac{\partial p}{\partial \theta} = \tau \sin(\frac{\omega \tau}{2}) \cos(\frac{\omega \tau}{2})$, and finally $I(\omega|\tau) = \tau^2$.

\subsection{DC magnetometry}
\label{subsec:DC_magnetometry_bound}
The lower bounds on the estimation of the frequency $\omega$ are reported in \cref{table:dc_lower_bounds}, and are represented in \cref{fig:nvcenter_comparison} of \cref{subsec:nv_center} as the shaded grey area. The left column of this table contains the bounds for a finite number of measurements $M$, while right column refers to the estimation with a fixed total evolution time $T$. The first row refers to the estimation of $\omega$ with perfect coherence while the second row refers to the estimation of $\omega$ with a finite and know $T_2^\star$. The symbol $I(\omega)$ indicate the FI of the prior for the precession frequency $\omega$.
\begin{table}[htb]
	\centering
	\renewcommand{\arraystretch}{1.5}
	\setlength{\tabcolsep}{6pt}
	\begin{tabular}{|l|cl|cl|}
		\hline
		& Measurement & & Time & \\
		\hline
		\(T_2 = \infty\) &
		\(\frac{2^{-2(M+1)}}{3}\) &
		\refstepcounter{equation}\theequation \label{eq:T2_infty_Meas} &
		\(\frac{1}{T^2+I(\omega)}\) &
		\refstepcounter{equation}\theequation \label{eq:T2_infty_Time} \\
		\hline
		\(T_2 < \infty\) &
		\(\max \Big \lbrace \frac{1}{\mu M (T_2^\star)^2 + I(\omega)}, \frac{2^{-2(M+1)}}{3} \Big \rbrace\) &
		\refstepcounter{equation}\theequation \label{eq:T2_finite_Meas} &
		\(\frac{1}{0.5 \, T T_2^\star + I(\omega)}\) &
		\refstepcounter{equation}\theequation \label{eq:T2_finite_Time} \\
		\hline
	\end{tabular}
	\caption{Lower bounds for the precision of the frequency estimation in DC magnetometry on an NV center.}
	\label{table:dc_lower_bounds}
\end{table}
The numerical values of the quantities appearing in \cref{table:dc_lower_bounds}, for $\omega \in (0, 1) \, \text{MHz}$ are: $\mu = 0.1619$, $I(\omega) = 12 \, \mu \text{s}^2$. In the following we derive these four bounds.
\begin{itemize}
	\item The Fisher information for the decoherence-free precession frequency $\omega$ is given by $I(\omega|\tau) = \tau^2$, so that $\sup_{\tau} I(\omega|\tau) = \infty$ and the analysis based on the Cramèr-Rao bound doesn't gives a useful bound. \cref{eq:T2_infty_Meas} can be found by observing that each measurement gives at most one bit of information about the value of $\omega$, because the measurement has a binary outcomes~\cite{fiderer_neural-network_2021}. This bound is applied also for $T_2^\star < \infty$ in addiction to the one coming from the Fisher information.
	\item With a finite decoherence time $T_2^\star < \infty$ the FI for the frequency $\omega$ is
	\begin{equation}
		I(\omega|\tau, T_2^\star) = \frac{\tau^2 e^{-\frac{2 \tau}{T_2^\star}} \cos^2 \left( \frac{\omega \tau}{2}\right) \sin^2 \left( \frac{\omega \tau}{2}\right) }{\left[ e^{-\frac{\tau}{T_2^\star}} \cos^2 \left( \frac{\omega \tau}{2} \right) + \frac{1-e^{-\frac{\tau}{T_2^\star}}}{2} \right] \left[ e^{-\frac{\tau}{T_2^\star}} \sin^2 \left( \frac{\omega \tau}{2} \right) + \frac{1-e^{-\frac{\tau}{T_2^\star}}}{2} \right]} \; ,
	\end{equation}
	which, by defining $C := \cos^2 \left( \frac{\omega \tau}{2} \right)$, can be bounded in the following way
	\begin{equation}
		I(\omega|\tau, T_2^\star) = \frac{\tau^2 e^{-\frac{2 \tau}{T_2^\star}} C (1-C)}{\left[ \frac{1}{4} - e^{-\frac{2 \tau}{T_2^\star}} (C-\frac{1}{2})^2 \right]}  \le \frac{\tau^2 e^{-\frac{2 \tau}{T_2^\star}}}{1 - e^{-\frac{2 \tau}{T_2^\star}}} = (T_2^\star)^2 \frac{x^2 e^{-2x}}{1-e^{-2x}} \; ,
	\end{equation}
	where $x = \frac{\tau}{T_2^\star}$. The maximization in $x \in \mathbb{R}_+$ gives $\sup_{\tau} I(\omega|\tau, T_2^\star) = \mu (T_2^\star)^2$ with $\mu = 0.1619$. Inserting this expression in \cref{eq:lower_meas} gives the first term in the maximum of \cref{eq:T2_finite_Meas}, the second term was explained in the previous point.
	\item Regarding the time-constrained lower bounds, for $T_2^\star = \infty$, the total FI is maximized by performing a single measurement of time duration $\tau = T$, which gives \cref{eq:T2_infty_Time}, through the application of \cref{eq:lower_time}.
	\item For $T_2^\star < \infty$ we have to maximize the normalized FI in $x \in \mathbb{R}_+$, i.e.
	\begin{equation}
		\frac{I(\omega|\tau, T_2^\star)}{\tau} \le \frac{\tau e^{-\frac{2 \tau}{T_2^\star}}}{1 - e^{-\frac{2 \tau}{T_2^\star}}} \le T_2^\star \frac{x e^{-2x}}{1-e^{-2x}} \le \frac{T T_2^\star}{2} \; ,
		\label{eq:time_fixed_T2_omega}
	\end{equation}
	from which \cref{eq:T2_finite_Time} follows from \cref{eq:lower_time}.
\end{itemize}

\section{Backward recursion method for the optimization of the strategy}
\label{sec:target_f}
In this section we set the stage to understand what function the agent must approximate by formulating the problem in terms of a backward recursion. In this section we will see how the optimal control could theoretically be derived in other ways, so that the training will appear less as an unintelligible black-box and more as the solution to a well-posed problem (though we won't probably have uniqueness). The output of the agent at the $t+1$-th steps is $x_{t+1}$, that is, the control of the current evolutions and measurements. In the following we will define formally the function that the agent must learn to approximate, in doing this we will assume that the control $x$ of the quantum sensor is a continuos real variable. Consider an estimation with $M$ measurements, i.e. $t=0, 1, \dots, M-1$. Let us focus on the last one only and recall the definition of the particle filter ensemble before after the last measure $M-1$, i.e. $\mathfrak{p}_{M} := \lbrace \vtheta^{M-1}_j, w^{M-1}_j \rbrace_{j=1}^N$. Then we can write the ensemble of the PF at the final step $t=M-1$ as
\begin{equation}
	\mathfrak{p}_{M} = \mathfrak{B} (\mathfrak{p}_{M-1}, x_{M-1}, y_{M-1}) \; ,
	\label{eq:pT1}
\end{equation}
where $\mathfrak{B}$ encodes the application of the Bayes rule in \cref{eq:bayes_rule_w}. The ensemble $\mathfrak{p}_{M}$ inherits the stochasticity from the random measurement outcome $y_{M-1}$. Per definition $\mathfrak{p}_{0}$ is the initial PF ensemble that represents the prior $\pi(\vtheta)$. In \cref{subsec:general_loss} we mentioned that the final loss is a scalar function $\ell (\mathfrak{p}_{M}, \vtheta)$ of the final PF ensemble and of the true value $\vtheta$, like the squared derivation of the estimator from the true value. The final loss can be expressed as $\ell (\mathfrak{B} (\mathfrak{p}_{M-1}, x_{M-1}, y_{M-1}), \vtheta)$ and it's expectation value on $y_{M-1}$ (the expected loss), computed with the density in \cref{eq:meas_res_strong}, reads
\begin{equation}
	\overline{\ell} (\mathfrak{p}_{M-1}, x_{M-1}, \vtheta) := \int \ell (\mathfrak{B} (\mathfrak{p}_{M-1}, x_{M-1}, y_{M-1}), \vtheta) p(y_{M-1}| x_{M-1}, \vtheta) d y_{M-1} \; .
\end{equation}
If the outcome probability, the prior and the loss are continuos functions we can also expect $\overline{\ell} (\mathfrak{p}_{M-1}, x_{M-1}, \vtheta)$ to be continuous in its parameters. Without aiming at full rigour, we say that the regularity properties of the probability densities are passed down to the expectation value. Now we look for the minimum of this function, which is realized by solving
\begin{equation}
	\frac{\dd \overline{\ell} (\mathfrak{p}_{M-1}, x_{M-1}^\star, \vtheta)}{\dd x_{M-1}} = 0 \; .
	\label{eq:expected_loss_der_T1}
\end{equation}
This equation defines implicitly the optimal control $x_{M-1}^\star := r_{M-1} (\mathfrak{p}_{M-1}, \vtheta)$, where $x_{M-1}^\star$ realizes the absolute minimum of the expected loss. $r_{M-1}$ inherits some regularity property (at least locally) from $\overline{\ell} (\mathfrak{p}_{M-1}, x_{M-1}, \vtheta)$ thanks to the implicit function theorem. The control $x_{M-1}^\star$ can still have discontinuities in $\mathfrak{p}_{M-1}$ if the expected loss has multiple competing minima. The dependence on $\vtheta$ is rather inconvenient, because it is unknown, but we can think of substituting $\vtheta$ with its estimator $\hvtheta_{M-2}$ to get $x_{M-1}^\star = r_{M-1} (\mathfrak{p}_{M-1}, \hvtheta_{M-2})$. We will however never do explicit optimizations with this approach, the introduction of machine learning in quantum metrology serves precisely to avoid these cumbersome computations. Until now we have only optimized the last control, but fortunately all these operations can be repeated with minor changes for the $t=M-2$ measurement step. Let us start from $\mathfrak{p}_{M-1}$ expressed as function of the ensemble $\mathfrak{p}_{M-2}$:
\begin{equation}
	\mathfrak{p}_{M-1} = \mathfrak{B} (\mathfrak{p}_{M-2}, x_{M-2}, y_{M-2}) \; ,
	\label{eq:pT2}
\end{equation}
we insert this expression in \cref{eq:pT1} to get
\begin{equation}
	\mathfrak{p}_{M} = \mathfrak{B}(\mathfrak{B}(\mathfrak{p}_{M-2}, x_{M-2}, y_{M-2}), x_{M-1}, y_{M-1}) \; .
\end{equation}
By substituting this in the loss $\ell (\mathfrak{p}_{M}, \vtheta)$ we get $\ell (\mathfrak{B}(\mathfrak{B}(\mathfrak{p}_{M-2}, x_{M-2}, y_{M-2}), x_{M-1}, y_{M-1}), \vtheta)$, but the optimal $x_{M-1}$ has been already computed as a function of the PF ensemble. Whatever control we suggest for the step $t=M-2$ it doesn't change the optimal control at the following step. In other words whatever control $x_{M-2}$ gets applied, the optimal action for the last time step is always $x_{M-1}^\star$, we can therefore insert it in the loss at the step $M-2$, i.e.
\begin{align}
	\ell (\mathfrak{p}_{M-2}, x_{M-2}, y_{M-2}, y_{M-1}, \vtheta) := \ell (\mathfrak{B}(\mathfrak{B}(\mathfrak{p}_{M-2}, x_{M-2}, y_{M-2}), r_{M-1} (\mathfrak{B}(\mathfrak{p}_{M-2}, x_{M-2}, y_{M-2}), \vtheta), y_{M-1}), \vtheta) \; ,
\end{align}
where we have also used the expression for $\mathfrak{p}_{M-1}$ of \cref{eq:pT2} in the definition of $x_{M-1}^\star$, and we have redefined the parameters of $\ell$. Again we take the expectation value on the measurement outcomes, i.e.
\begin{equation}
	\overline{\ell} (\mathfrak{p}_{M-2}, x_{M-2}, \vtheta) := \int \ell (\mathfrak{p}_{M-2}, x_{M-2}, y_{M-2}, y_{M-1}, \vtheta) p (y_{M-1}| \vtheta, x_{M-1}^\star) p (y_{M-2}| \vtheta, x_{M-2}) d y_{M-1} d y_{M-2} \; .
	\label{eq:expected_loss_T2}
\end{equation}
By taking the derivative of this expected loss with respect to $x_{M-2}$, we define implicitly the optimal control $x_{M-2}^\star = r_{M-2} (\mathfrak{p}_{M-2}, \vtheta)$ as the solution of
\begin{equation}
	\frac{\dd \overline{\ell} (\mathfrak{p}_{M-2}, x_{M-2}^\star, \vtheta)}{\dd x_{M-2}} = 0 \; ,
	\label{eq:expected_loss_der_T2}
\end{equation}
where $x_{M-2}^\star$ realizes the absolute minimum of $\overline{\ell}$. Notice that this is not a greedy optimization: the value of $x_{M-2}$ is not chosen to optimize the loss one step head in the future but the final loss, knowing what the strategy in the next step will be. We could treat all the previous measurements in the same manner, if it wasn't for the expectation values of the loss, that become more and more complicated. In this way we can inductively proceed in reverse to the start of the estimation $t=0$ and find in the process the family of functions $r_{t} (\mathfrak{p}_t, \vtheta_{t-1})$ that express the optimal controls $x_t^\star$. As discussed previously we can redefine $r_t(\mathfrak{p}_t) = r_t (\mathfrak{p}_t, \hvtheta_{t-1})$, in order to get rid of the dependence on the unknown value of the parameters $\vtheta$. We then introduce $r(\mathfrak{p}_t, t) = r_t(\mathfrak{p}_t)$, which is the function that the agent is trying to approximate in the training, i.e. the map that spits out the optimal control at a given measurement step $t$, provided the ensemble PF at the previous step. Heuristically we expect the optimal control to be inhomogeneous in time because like in many application of RL a good strategy encompasses a phase of ``exploration'' followed by a phase of ``exploitation'' of what has been learned~\cite{marquardt_machine_2021, marquardt_online_2021}.

\section{Backpropagation of the gradient}
\label{sec:gradient_flow}
\begin{figure}[h]
	\centering
	\includesvg[width=0.8\textwidth]{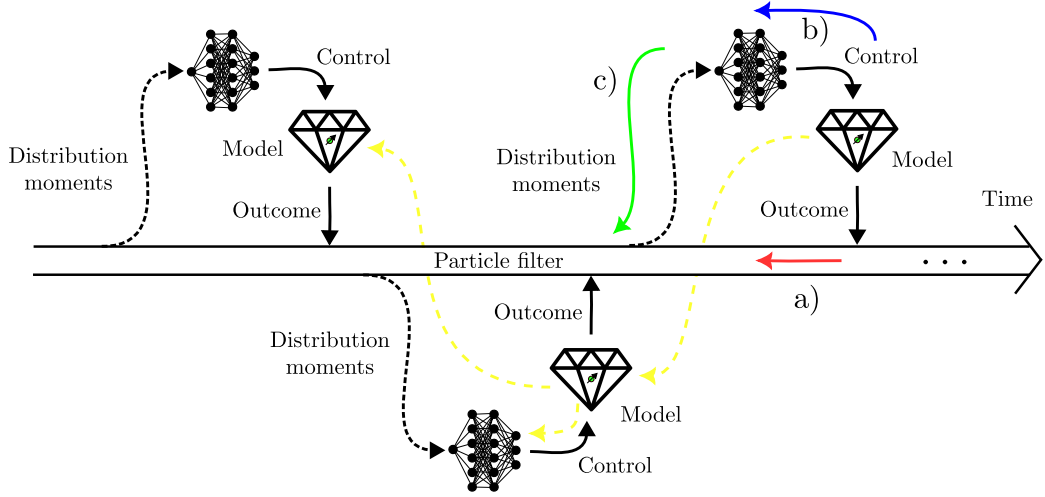}
	\caption{The fat empty arrow in the middle of the picture represents the PF, which is updated via the Bayes rule, indicated with the function $\mathfrak{B}$ in the text, after it receive an outcome from the measurement on the probe. The second term of \cref{eq:total_state}, underlined in red and labelled with a), corresponds to the backpropagation of the gradient along the history of the PF ensemble, whose weights and particles inherits a dependence on $\vlambda$ from the actions of the agent. Visually this corresponds to backpropagation along the red arrow labelled with a). Through the match between the underlined terms in \cref{eq:total_control} and \cref{eq:total_state} and the arrows in the figure can visualize the origin of each term of the gradient. The blue term of \cref{eq:total_control} labelled with b) accounts for the dependence on $\vlambda$ that comes from the controls computed by the agent and propagated through the application of the Bayes rule and the computation of $x_t$. The green term of \cref{eq:total_control} labelled with c) is the gradient propagating from the input of the agent to the previous PF ensemble. This term is responsible for the ``higher-order'' terms of the gradient, that propagate multiple times through the agent. Inserting the stop gradient in \cref{eq:input_stop_gradient} means cutting the green line. The dashed yellow line represents the propagation of the gradient through the state of the probe, when this is not reinitialized between the measurements.}
	\label{fig:gradient_flow_2}
\end{figure}

With the help of an automatic differentiation framework we compute the gradient of the modified loss in \cref{eq:modified_loss} in order to perform the training. Let us for the moment neglect the log-likelihood terms in this expression and concentrate only on the first part. If the loss is computed only from the ensemble at the last measurement step, then it takes the form
\begin{equation}
	\ell_{\vtheta} \circ \mathfrak{B}_{x_{M-1}, y_{M-1}} \circ \mathfrak{B}_{x_{M-1}, y_{M-1}} \circ \cdots \circ \mathfrak{B}_{x_0, y_{0}} (\mathfrak{p}_0) \; .
	\label{eq:gradient_flow_loss}
\end{equation}
where $\ell_{\vtheta} (\mathfrak{p}) := \ell (\mathfrak{p}, \vtheta)$ acts on the ensemble PF and is the individual loss of each simulation, e.g. the square error. The function $\mathfrak{B}_{x_t, y_t} (\mathfrak{p}) := \mathfrak{B} (\mathfrak{p}, x_t, y_t)$ applies the Bayesian update to the PF, which depends on the outcome of the measurement $y_t$ and on the control $x_t$. The initial ensemble of the PF is named $\mathfrak{p}_0$. From a theoretical point of view differentiating this expression means applying repeatedly the rule for the derivation of the composite function and propagating the derivative through the various stages of the estimation, this is called backpropagation. Let us compute explicitly the derivative of the loss for the last two steps of the estimation. We define $\mathfrak{p}_{M} = \mathfrak{B}_{x_{M-1}, y_{M-1}} (\mathfrak{p}_{M-1})$ and $\mathfrak{p}_{M-1} = \mathfrak{B}_{x_{M-2}, y_{M-2}} (\mathfrak{p}_{M-2})$, and apply the chain rule. We will indicate with the partial derivative symbol the derivatives with respect to the parameters of a function (which can also appear in the subscript), while the symbol $\frac{\dd}{\dd \vlambda}$ is reserved to the total derivative with respect to $\vlambda$.
\begin{equation}
	\frac{\dd}{\dd \vlambda} \ell_{\vtheta} ( \mathfrak{p}_{M} ) = \frac{\partial \ell_{\vtheta} (\mathfrak{p}_{M})}{\partial \mathfrak{p}} \frac{\dd}{\dd \vlambda} \mathfrak{B}_{x_{M-1}, y_{M-1}} (\mathfrak{p}_{M-1}) \; .
\end{equation}
The total derivate in the right-hand term an be expanded again with the chain rule:
\begin{equation}
	\frac{\partial \mathfrak{B}_{x_{M-1}, y_{M-1}} (\mathfrak{p}_{M-1} )}{\partial x_{M-1}} \frac{\dd}{\dd \vlambda} x_{M-1} (\vlambda, \mathfrak{p}_{M-1}) +
	\mathunderline{red}{\frac{\partial \mathfrak{B}_{x_{M-1}, y_{M-1}} (\mathfrak{p}_{M-1})}{\partial \mathfrak{p}} \frac{\dd}{\dd \vlambda} \mathfrak{B}_{x_{M-2}, y_{M-2}} (\mathfrak{p}_{M-2})} 	\quad \text{a)} \; .
	\label{eq:total_state}
\end{equation}
Only the parameters $x_{M-1}$ and $\mathfrak{p}_{M-1}$ can carry a dependence on $\vlambda$. Regarding the measurement outcomes $y_{M-1}$ we already discussed their independence on $\vlambda$, expressed by \cref{eq:derivative_random}. The control $x_{M-1}$ has an explicit dependence on $\vlambda$ because it has been produced by the agents, but also has a dependence on $\vlambda$ through the PF ensemble, so that we can expand the total derivative in the following way
\begin{align}
	\frac{\dd}{\dd \vlambda} x_{M-1} (\vlambda, \mathfrak{p}_{M-1}) &= \mathunderline{blue}{\frac{\partial x_{M-1} (\vlambda, \mathfrak{p}_{M-1})}{\partial \vlambda}} \quad \text{b)} \\
	& + \mathunderline{green}{\frac{\partial x_{M-1} (\vlambda, \mathfrak{p}_{M-1})}{\partial \mathfrak{p}} \frac{\dd}{\dd \vlambda} \mathfrak{B}_{x_{M-2}, y_{M-2}} (\mathfrak{p}_{M-2})} \quad \text{c)} \; . \label{eq:total_control}
\end{align}
The first piece of the derivative is the dependence on $\vlambda$ that comes from the last application of the agent, while the second piece represent the backpropagation through the input of the agent and the Bayesian update of the PF ensemble. In general, the gradient is backpropagated through all the applications of the agent until it reaches the beginning. We notice that we can write the total derivative of $\mathfrak{p}_{M}$ as a function of the total derivative of $\mathfrak{p}_{M-1}$, i.e.
\begin{equation}
	\frac{\dd}{\dd \vlambda} \mathfrak{B}_{x_{M-1}, y_{M-1}} (\mathfrak{p}_{M-1}) = Q_{M-1} + H_{M-1} \frac{\dd}{\dd \vlambda} \mathfrak{B}_{x_{M-2}, y_{M-2}} (\mathfrak{p}_{M-2})
\end{equation}
where
\begin{eqnarray}
	Q_{M-1} &:=& \frac{\partial \mathfrak{B}_{x_{M-1}, y_{M-1}} (\mathfrak{p}_{M-1} )}{\partial x_{M-1}} \frac{\partial x_{M-1} (\vlambda, \mathfrak{p}_{M-1})}{\partial \vlambda} \; ,\\
	\label{eq:QT1}
	H_{M-1} &:=& \frac{\partial \mathfrak{B}_{x_{M-1}, y_{M-1}} (\mathfrak{p}_{M-1})}{\partial x_{M-1}} \frac{\partial x_{M-1} (\vlambda, \mathfrak{p}_{M-1})}{\partial \mathfrak{p}} + \frac{\partial \mathfrak{B}_{x_{M-1}, y_{M-1}} (\mathfrak{p}_{M-1})}{\partial \mathfrak{p}} \; .
	\label{eq:HT1}
\end{eqnarray}
We have arrived to a family of recurrence equations, which have the solution
\begin{equation}
	\frac{\dd}{\dd \vlambda} \mathfrak{B}_{x_{M-1}, y_{M-1}} (\mathfrak{p}_{M-1}) = \sum_{t=0}^{M-1} Q_t \prod_{m=t+1}^{M-1} H_m \; ,
	\label{eq:explicit_gradient}
\end{equation}
where $Q_t$ and $H_m$ are defined analogously to $Q_{M-1}$ and $H_{M-1}$. The $H_m$ terms have each two summand, and when multiplied together the number of terms in the gradient grows exponentially in the number of measurement $M$, and generates gradient terms corresponding to multiple repeated backpropagations through the agents. This are a kind of ``higher order'' gradient terms. In our implementation of the training we simplify the gradient by introducing a stop gradient operation before the input of the agent, so that \cref{eq:controls_computation} is actually implemented as
\begin{equation}
	x_{t+1} = \mathcal{F} \lbrace \text{sg} \left[ P(\vtheta| \vx_t, \vy_t); R_{t}; t \right] \rbrace \; ,
	\label{eq:controls_computation_sg}
\end{equation}
This modification doesn't change the forward pass, that is, the results of simulations are the same, but it affects the backpropagation of the gradient, in particular it makes the first term of $H_m$ disappear, because
\begin{equation}
	\frac{\partial x_{m} \left( \vlambda, \sg{\mathfrak{p}_{m}} \right) }{\partial \mathfrak{p}} = 0 \; .
	\label{eq:input_stop_gradient}
\end{equation}
%
Such simplification reduces considerably the training time and doesn't really affect, at least from a theoretical point of view, the ability of the agent to learn the optimal strategy, on the contrary it might even improve it. Before we back up this last assertions we want to recapitulate our analysis of the backpropagation with the help of \cref{fig:gradient_flow_2}.
A control strategy is called myopic if it optimizes the information gained from the next measurement only, while it is non-myopic if it optimizes many steps ahead in the future. It might seem that by cutting the gradient propagation through the green arrows we limit the optimization to the class of myopic strategies, but this is not true because the optimization is always done for the final precision. The input of the agent is basically a constant now and the $t$-th term in the summation of \cref{eq:explicit_gradient} pulls the weights of the NN to minimize the final loss given the PF ensemble at the $t-1$-th step. In order to have non-myopic adaptive strategy backpropagating the gradient through the green arrows is redundant. The gradient of the log-likelihood terms in the loss of \cref{eq:modified_ell} is also simplified when the stop gradient is acting in \cref{eq:input_stop_gradient}. In the model $p(y_{t}|x_{t}, \vtheta)$ the only term that depends on $\vlambda$ is the control, and the gradient propagates through a single application of the agent. Had we not inserted the stop gradient, the gradient of each summands in the log-likelihood would have been propagated through all the past applications of the agent, a scenario that happens anyway if the probe state is not reinitialized between measurements. Before ending the discussion on the gradient backpropagation we want to consider yet another possibility of truncating the gradient. We could indeed image to stop the flow of the derivative through the evolution of the PF ensemble, which means cutting the red line in \cref{eq:total_state}. This eliminates the recurrence equation and trivializes the gradient, which now accounts only for the very last control. That is, with such modification, the agent will learn to optimize only the last measurement. If the loss is cumulative however, like in \cref{eq:all_time_loss}, all the controls will be optimized, but in a myopic way. In this scenario we introduce the modified logarithmic loss, i.e.
\begin{equation}
	\widetilde{\mathcal{L}}_{\text{log}} (\vlambda) := \frac{1}{TB} \sum_{t=1}^{M-1} \log \left[ \frac{\sum_{k=1}^B \ell (\hvtheta_{k, t}, \vtheta_k)}{\sg{ \sum_{k=1}^B  \ell (\hvtheta_{t-1, k}, \vtheta_k)}} \right] \; .
	\label{eq:ig_loss}
\end{equation}
Each term in this summation is the empirical information gain for a Gaussian posterior. In the forward pass the loss is the total empirical information gain, because the stop gradient operators don't play any role, and the series can be resummed. In the computation of the gradient, however, the information gain for each measurement is optimized greedily, as done in~\cite{granade_robust_2012} and in the package optbayesexpt~\cite{mcmichael_optbayesexpt_2021}. If the stop gradient in the denominator is applied only every $n$ measurement, then we are optimizing the information gain planning $n$ steps ahead in the future. It is advisable to put a regularization at the denominator of the loss in \cref{eq:ig_loss} to avoid dividing by zero.

\end{appendices}

\end{document}